\documentclass[structabstract]{aa}

\usepackage{url}
\usepackage[breaklinks=true]{hyperref}
\usepackage{twoopt}
\usepackage[english]{babel}          % English language/hyphenation
\usepackage{placeins}

\usepackage{natbib}
\bibpunct{(}{)}{;}{a}{}{,} %% natbib format for A&A and ApJ

\usepackage[utf8]{inputenc}
\usepackage[normalem]{ulem}
\usepackage{siunitx}
\usepackage{amsmath, amssymb}
\usepackage[farskip=0pt]{subfig}
\usepackage[export]{adjustbox}

\usepackage{multirow,bigstrut,ctable}
\usepackage[normalem]{ulem}
\usepackage{rotating}
\usepackage[varg]{txfonts} % A&A recommended fonts
\usepackage{threeparttable}

\usepackage{xcolor}

\def \kms         {km$\,$s$^{-1}$}
\def \ergs        {erg$\,$s$^{-1}$}

\def \deg         {\text{$^{\circ}$}}
\def \arcmin      {\text{$^\prime$}}
\def \arcsec      {\text{$^{\prime\prime}$}}
\def \hour        {$^{\mathrm{h}}$}
\def \min         {$^{\mathrm{m}}$}
\def \sec         {$^{\mathrm{s}}$}
\def \mjybeam     {mJy\,beam$^{-1}$}
\def \mujybeam    {$\mathrm{\mu}$Jy\,beam$^{-1}$}

\def \mach            {\mathcal{M}}
\def \radmm       {rad~m$^{-2}$}

\newcommand{\hms}[3]{{#1}\hour{#2}\min{#3}\sec}
\newcommand{\dms}[3]{{#1}\deg{#2}\arcmin{#3}\arcsec}
\newcommand{\beam}[2]{{#1}\arcsec$\times${#2}\arcsec}

\newcommand{\Msun}{\text{$\rm M_\odot$}}

\begin{document}

%\title{Linear radio halo, shocks, bays and other wonders observed by MeerKAT in Abell 3667}
\title{MeerKAT view of the diffuse radio sources in Abell 3667 and their interactions with the thermal plasma}
%\subtitle{...}
\titlerunning{MeerKAT view of Abell 3667}

\author{F.~de~Gasperin\inst{1,2}
\and L.~Rudnick\inst{3}
\and A.~Finoguenov\inst{4}
\and D.~Wittor\inst{1,5}
\and H.~Akamatsu\inst{6}
\and M.~Br\"uggen\inst{1}
\and J. O.~Chibueze\inst{7,8}
\and T.~E.~Clarke\inst{9}
\and W.~Cotton\inst{10,11}
\and V.~Cuciti\inst{1}
\and P.~Dom\'inguez-Fern\'andez\inst{12,1}
\and K.~Knowles\inst{13,11}
\and S.~P.~O'Sullivan\inst{14}
\and L.~Sebokolodi\inst{13}
}

\authorrunning{F.~de~Gasperin et al.}

\institute{Hamburger Sternwarte, Universit\"at Hamburg, Gojenbergsweg 112, 21029, Hamburg, Germany, \email{fdg@hs.uni-hamburg.de}
\and INAF - Istituto di Radioastronomia,
via P. Gobetti 101, 40129, Bologna, Italy 
\and Minnesota Institute for Astrophysics, University of Minnesota, 116 Church St. SE, Minneapolis, MN 55455, USA
\and Department of Physics, Gustaf Hällströmin katu 2A, University of Helsinki, Helsinki, FI-00014, Finland
\and Dipartimento di Fisica e Astronomia, Università di Bologna, Via Gobetti 92/3, 40129, Bologna, Italy
\and SRON Netherlands Institute for Space Research, Niels Bohrweg 4, 2333 CA, Leiden, The Netherlands
\and Centre for Space Research, North-West University, Potchefstroom 2520, South Africa
\and Department of Physics and Astronomy, Faculty of Physical Sciences,  University of Nigeria, Carver Building, 1 University Road, Nsukka, Nigeria
\and US Naval Research Laboratory, Code 7213, 4555 Overlook Ave SE, Washington, DC 20375, USA
\and NRAO - National Radio Astronomy Observatory, 520 Edgemont Rd, Charlottesville, VA 22903, USA
\and SARAO - South African Radio Astronomy Observatory, 2 Fir St., Observatory, South Africa
\and Department of Physics, School of Natural Sciences UNIST, Ulsan 44919, Republic of Korea
\and Centre for Radio Astronomy Techniques and Technologies, Department of Physics and Electronics, Rhodes University, P.O. Box 94, Makhanda 6140, South Africa
\and School of Physical Sciences and Centre for Astrophysics \& Relativity, Dublin City University, Glasnevin D09 W6Y4, Ireland
}

\date{Received ... / Accepted ...}

\abstract
%context
{During their lifetimes, galaxy clusters grow through the accretion of matter from the filaments of the large-scale structure and from mergers with other clusters. These mergers release a large amount of energy into the intracluster medium (ICM) through merger shocks and turbulence. These phenomena are associated with the formation of radio sources known as radio relics and radio halos, respectively. Radio relics and halos are unique proxies for studying the complex properties of these dynamically active regions of clusters and the microphysics of the ICM more generally.}
%aim
{Abell 3667 is a spectacular example of a merging system that hosts a large pair of radio relics. Due to its proximity ($z=0.0553$) and large mass, the system enables the study of these sources to a uniquely high level of detail. However, being located at Dec~$=-56.8\deg$, the cluster could only be observed with a limited number of radio facilities.}
%methods
{We observed Abell 3667 with MeerKAT as part of the MeerKAT Galaxy Cluster Legacy Survey. We used these data to study the large-scale emission of the cluster, including its polarisation and spectral properties. The results were then compared with simulations.}
%results
{We present the most detailed view of the radio relic system in Abell 3667 to date, with a resolution reaching 3 kpc. The relics are filled with a network of filaments with different spectral and polarisation properties that are likely associated with multiple regions of particle acceleration and local enhancements of the magnetic field. Conversely, the magnetic field in the space between filaments has strengths close to what would be expected in unperturbed regions at the same cluster-centric distance. Comparisons with magnetohydrodynamic cosmological and Lagrangian simulations support the idea of filaments as multiple acceleration sites. Our observations also confirm the presence of an elongated radio halo, developed in the wake of the bullet-like sub-cluster that merged from the south-east. Finally, we associate the process of magnetic draping with a thin polarised radio source surrounding the remnant of the bullet's cool core.}
%conclusions
{Our observations have unveiled the complexity of the interplay between the thermal and non-thermal components in the most active regions of a merging cluster. Both the intricate internal structure of radio relics and the direct detection of magnetic draping around the merging bullet are powerful examples of the non-trivial magnetic properties of the ICM. Thanks to its sensitivity to polarised radiation, MeerKAT will be transformational in the study of these complex phenomena.}


\maketitle

\section{Introduction}
\label{sec:introduction}
% intro on radio emission from clusters
Galaxy clusters form and grow through mergers of smaller substructures at the intersection of cosmic filaments. 
The energy released during cluster merger events is dissipated through the cluster's volume via shocks and turbulence, heating the intracluster medium (ICM). The ICM is a hot ($\sim 10^7-10^8$ K) and rarefied ($\sim 10^{-3}$ cm$^{-3}$) plasma that shines in the X-rays via thermal bremsstrahlung. Part of the energy of the merger is channelled into the acceleration of particles to relativistic energies and the amplification of magnetic fields. The most spectacular evidence of these non-thermal phenomena in the ICM is observable in the radio band in the form of gigantic ($\sim$1 Mpc scale) diffuse sources called radio relics and radio halos \citep[for a review, see][]{vanWeeren2019}.

%relic&halos
Radio relics have been linked to the presence of merger-induced shock waves propagating in the ICM. These sources are usually located about 1 Mpc from the cluster centre, and they are elongated and polarised, as expected if they trace edge-on shock waves \citep{Ensslin1998}. In some cases, two radio relics are present in the same cluster, usually located along the elongated ICM distribution and oriented perpendicular to it. In these cases, the merger happened in a plane close to the plane of the sky and the merger shocks are seen edge-on. Some examples are CIZA J2242.8+5301 \citep{vanWeeren2010a}, MACS J1752.0+4440 \citep{vanWeeren2012b, Bonafede2012a}, and PSZ1 G108.18-11.53 \citep{deGasperin2015}. Radio relics are believed to be powered through a type-I Fermi mechanism, also known as diffusive shock acceleration (DSA). However, evidence is mounting that the efficiency of the DSA of particles from the thermal pool in weak shocks, such as those generated during cluster mergers, is not enough to explain radio relic luminosities \citep[e.g.][]{Botteon2020}. A modification of standard DSA \citep[e.g.][]{Zimbardo2018} or the shock re-acceleration of seed electrons \citep[e.g.][]{Kang2014, Pinzke2013} are some of the directions that have been explored to mitigate the inconsistency.

Radio halos are believed to arise from turbulent re-acceleration in the magnetised ICM. This is a type-II Fermi mechanism where particles are scattered from magnetic inhomogeneities, resulting in a net increase in their energy \citep{Brunetti2014}. Radio relics and halos are a unique gateway for probing the non-thermal components of the ICM (magnetic fields, cosmic rays, turbulence), which are crucial ingredients for understanding the evolution of clusters \citep{Miniati2015, Ryu2008a, Dolag2008} as well as for high precision cosmology \citep[e.g.][]{Salvati2019}.

\begin{figure}
\centering
\includegraphics[width=.5\textwidth]{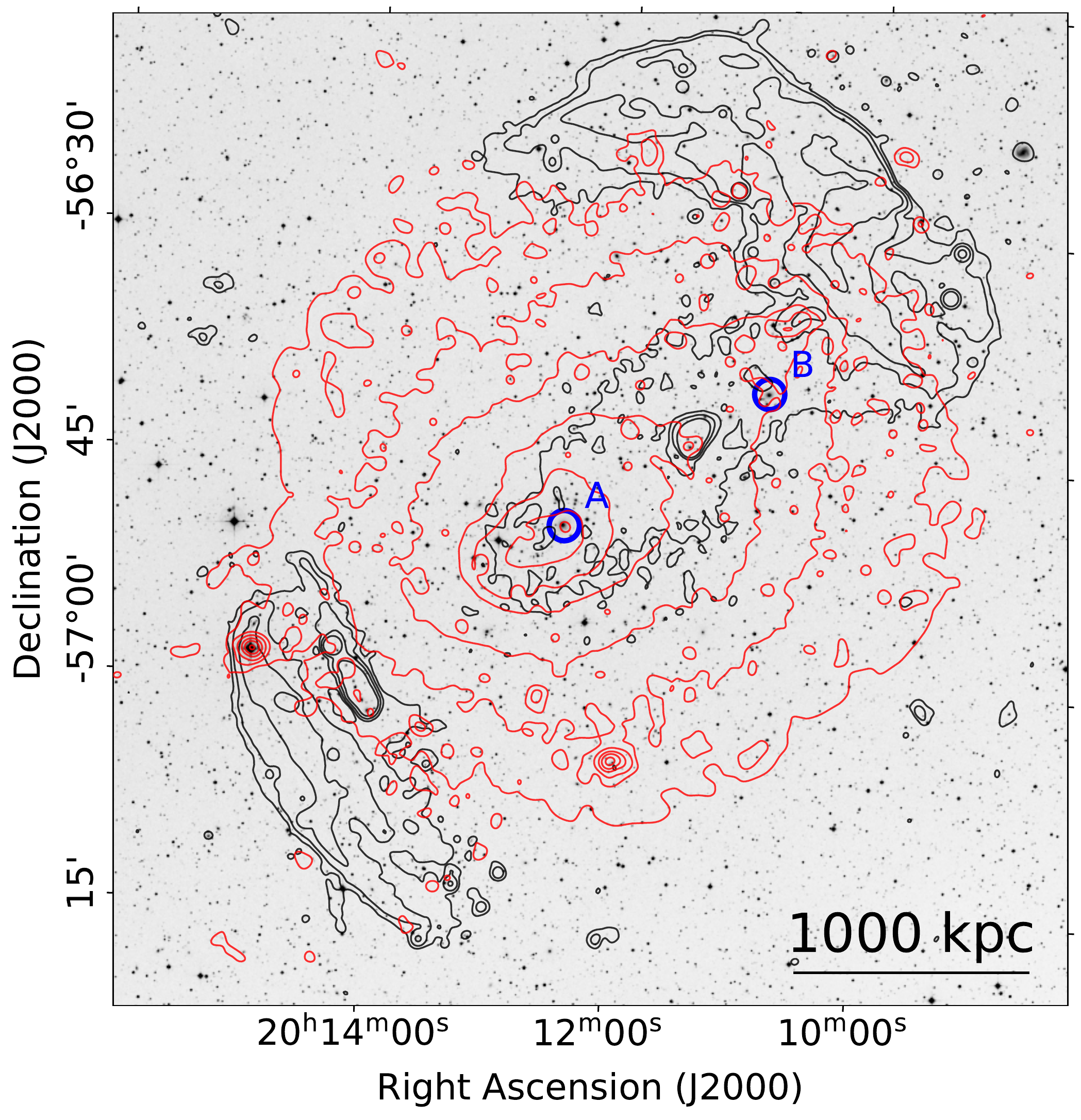}
\caption{Digital Sky Survey r-band image with contours from radio emission in black and from X-ray emission in red. The brightest cluster galaxies (BCG) of the two main sub-clusters are circled in blue and labelled A and B. Contours are taken from the flat-noise map and trace the source-subtracted low-resolution MeerKAT data at $(5,20,80,320)\times \sigma$ with $\sigma=17$~\mujybeam{} and beam \beam{35}{35}.}
\label{fig:opt}
\end{figure}

In this paper we use new MeerKAT data to study the radio emission from extended sources in the galaxy cluster Abell 3667, which hosts a spectacular pair of radio relics and an elongated radio halo.
% M, velocity dispersion, z, subclumps, WL
Abell 3667 (PSZ2 G340.88-33.36) is an outlier in the Abell catalogue \citep{Abell1989}. It is very massive, with $M_{500} = (7.04\pm0.05) \times 10^{14}$~\Msun{} \citep{PlanckCollaboration2016}, and it has a high velocity dispersion, initially estimated to be $\sim 1200-1400$~\kms{} \citep{Proust1988, Sodre1992} and later refined to $\sigma=1056\pm38$~\kms{} \citep{Owers2009}. The projected galaxy density map shows that the cluster consists of a complex blend of substructures \citep[see Fig. 7 of][]{Proust1988}. \cite{Johnston-Hollitt2008} examined 231 spectroscopically confirmed cluster members and found no evidence for dynamical activity in the velocity distribution. With a more detailed analysis of 550 spectroscopically confirmed cluster members, \cite{Owers2009} partitioned Abell 3667 into a main sub-cluster (A) around its cD galaxy, with a velocity dispersion $\sigma = 1073$~\kms{} and a mean peculiar velocity $v_{pec}=-103 \pm 71$~\kms, a north-western (NW) sub-cluster (B) with $\sigma=1039$~\kms{} ($v_{\rm pec}=422 \pm 82$~\kms) centred on a second dominant galaxy, and a south-eastern (SE) subgroup with $\sigma=209$~\kms{} ($v_{\rm pec}=-638$~\kms). The position of the dominant, brightest galaxies of the main cluster and of the NW sub-cluster are marked in Fig.~\ref{fig:opt}. \cite{Joffre2000} obtained weak-lensing maps of Abell 3667, identifying, albeit with low significance, the presence of mass excesses to the NW and SE of the main cluster. \cite{Owers2009} finally provided a redshift estimation of $z=0.0553\pm0.0002$, which we adopt thought this paper.

% ICM/cold front
Being nearby and massive, Abell 3667 is one of the strongest X-ray sources in the southern sky. It is luminous \citep[$L_{0.4-2.4~\rm keV} = 5.1 \times 10^{44}$~\ergs;][]{Ebeling1996} and hot \citep[$kT = 6.3^{+0.5}_{-0.6}$~keV;][]{Knopp1996}. ROSAT observations show that the cluster is elongated in the NW-SE direction. Significant X-ray emission is associated with the NW sub-cluster, suggesting that Abell 3667 is not in dynamical equilibrium \citep{Knopp1996}. Observations taken with the Advanced Satellite for Cosmology and Astrophysics (ASCA) show the presence of a surface brightness discontinuity to the SE of the cluster centre, which was initially interpreted as a merger shock \citep{Markevitch1999}. Deeper observations at higher angular and spectral resolution from both \textit{Chandra} and \textit{XMM-Newton} show that the discontinuity was a cold front, tracing cold and dense gas moving at near-sonic speed through the hotter and more diluted ICM \citep{Vikhlinin2001, Briel2004, Ichinohe2017} and producing large-scale hydrodynamical instabilities in the cluster atmosphere \citep{Mazzotta2002}. The cold front in Abell 3667 was one of the first discovered and the first named as such. A detailed analysis of the cluster's metallicity showed an inhomogeneous distribution, with the highest metallicity located to the SE of the X-ray peak \citep{Lovisari2009}. 

% shocks/relics
Abell 3667 is one of the few clusters that contain a pair of radio relics, which indicates the presence of merging shock waves. The brightest source, the NW relic, was studied in \cite{Goss1982}, but the authors could not come to a conclusive interpretation due to the limited resolution and sensitivity of their observations. The first in-depth study of the NW radio relic was performed with the Australia Telescope Compact Array (ATCA) by \cite{Rottgering1997}, while the southern one was studied by \cite{Johnston-Hollitt2003}, again with ATCA. The NW radio relic is the brightest cluster radio shock known, with a flux density at 1.4 GHz of $3.7\pm0.3$~Jy, while the southern relic is fainter, with a flux density of $0.30\pm0.02$~Jy \citep{Johnston-Hollitt2003}. More recently, \cite{Hindson2014} explored the low-frequency emission ($105-241$~MHz) from Abell 3667 with the Murchison Widefield Array (MWA). They measured the integrated spectral index between 120 and 1400 MHz of both relics to be $\alpha = -0.9\pm0.1$ but also discuss inconsistencies that could lead to steeper values. Finally, \textit{XMM-Newton} and \textit{Suzaku} observations suggest the presence of a shock front at the north-west edge of the NW relic with Mach number, derived from the \textit{XMM-Newton} average temperature and density jump, of $\mach = 2.54^{+0.80}_{-0.43}$ \citep{Finoguenov2010, Akamatsu2012, Sarazin2016}. This is one of the strongest shocks found in a cluster. \textit{XMM-Newton} data also show the presence of an X-ray `mushroom', possibly related to the disrupted cool core of sub-cluster B, whose edge also shows a surface discontinuity \citep{Sarazin2016}. The observations can be explained as the result of an offset binary merger with the cluster $\sim 1$~Gyr after core passage.

% halo/bridge
The presence of a radio halo in the system has been investigated by a number of authors, resulting in no detection \citep{Hindson2014, Riseley2015, Johnston-Hollitt2017}. Using the Parkes radio telescope, \cite{Carretti2013} detected an unpolarised radio `bridge'\footnote{Not to be confused with radio bridges in pre-merger clusters \citep{Govoni2019, Botteon2020b}.} running from the NW radio relic to the centre of the cluster. They conclude that the synchrotron emission from this source is linked to the post-shock turbulence trailing the relic shock. Given the central position and correlation with the X-ray emission, in the rest of the paper we call this feature a radio halo.

% simulations
The merger scenario of Abell 3667 has been simulated by
\cite{Roettiger1999}. Their conclusion is that Abell 3667 is an offset binary merger observed $\sim1$ Gyr after its first core passage. The mass ratio used for modelling the merger was 5:1, with the smaller cluster merging from the SE. Their model fits some of the most prominent features of the X-ray and radio data, for instance generating two radio relics, with the NW one being larger, more curved, and slightly off-axis. More recently, \cite{Datta2014} presented a more detailed magnetohydrodynamic (MHD) simulation with initial conditions drawn from a cosmological simulation. Their work shows that the combined shock statistics from the X-ray and radio data are in agreement with those derived from the simulated cluster.

\begin{figure}
\includegraphics[width=.5\textwidth]{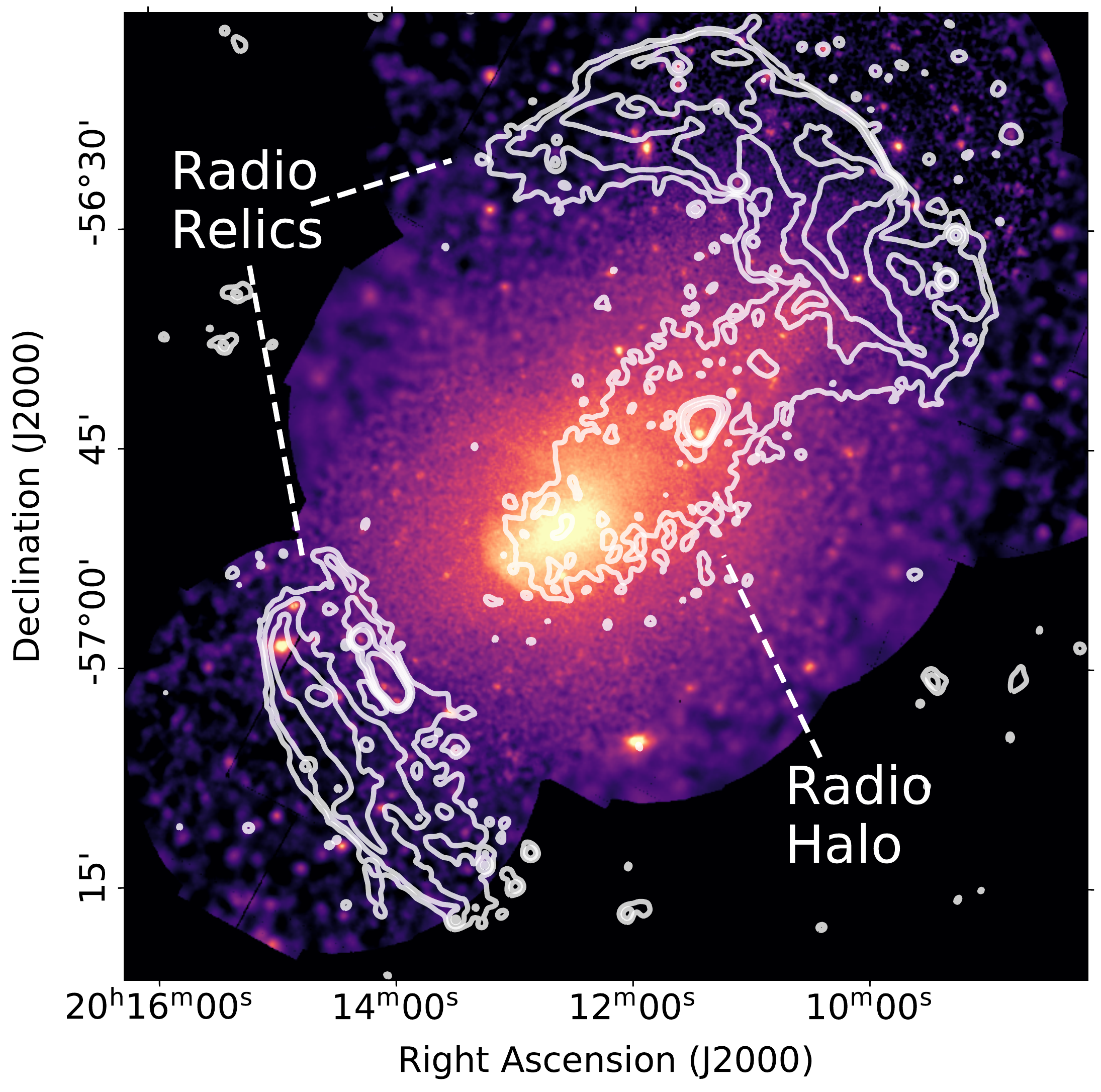}
\caption{Adaptively smoothed \textit{XMM-Newton} background-subtracted and exposure-corrected 0.5-2.0 keV image. Contours are taken from the flat-noise map and trace the source-subtracted low-resolution MeerKAT data at $(5,20,80,320)\times \sigma$ with $\sigma=17$~\mujybeam{} and beam \beam{35}{35}.}\label{fig:xmm}
\end{figure}

% this paper
This paper is organised as follows. We describe the MeerKAT observations as well as the data reduction procedure and image processing in Sect.~\ref{sec:radio}. The re-analysis of \textit{XMM-Newton} data is described in Sect.~\ref{sec:xray}. In Sect.~\ref{sec:results} we present our results, including polarisation and spectral analysis of the sources in Abell 3667. The discussion and conclusions are given in Sects.~\ref{sec:discussion} and \ref{sec:conclusions}. Throughout this paper, we adopt a fiducial $\Lambda$ cold dark matter cosmology with $H_0 = 70\rm\ km\ s^{-1}\ Mpc^{-1}$, $\Omega_m = 0.3,$ and $\Omega_\Lambda = 0.7$. At the redshift of Abell 3667 ($z\approx0.05525$), we have 1\arcsec = 1.073447 kpc. Unless otherwise specified, uncertainties are at $1\sigma$. The spectral index, $\alpha$, is defined as: $S_{\nu} \propto \nu^\alpha$, where $S_\nu$ is the flux density at frequency $\nu$.

\begin{figure*}[tb!]
\centering
\includegraphics[width=\textwidth]{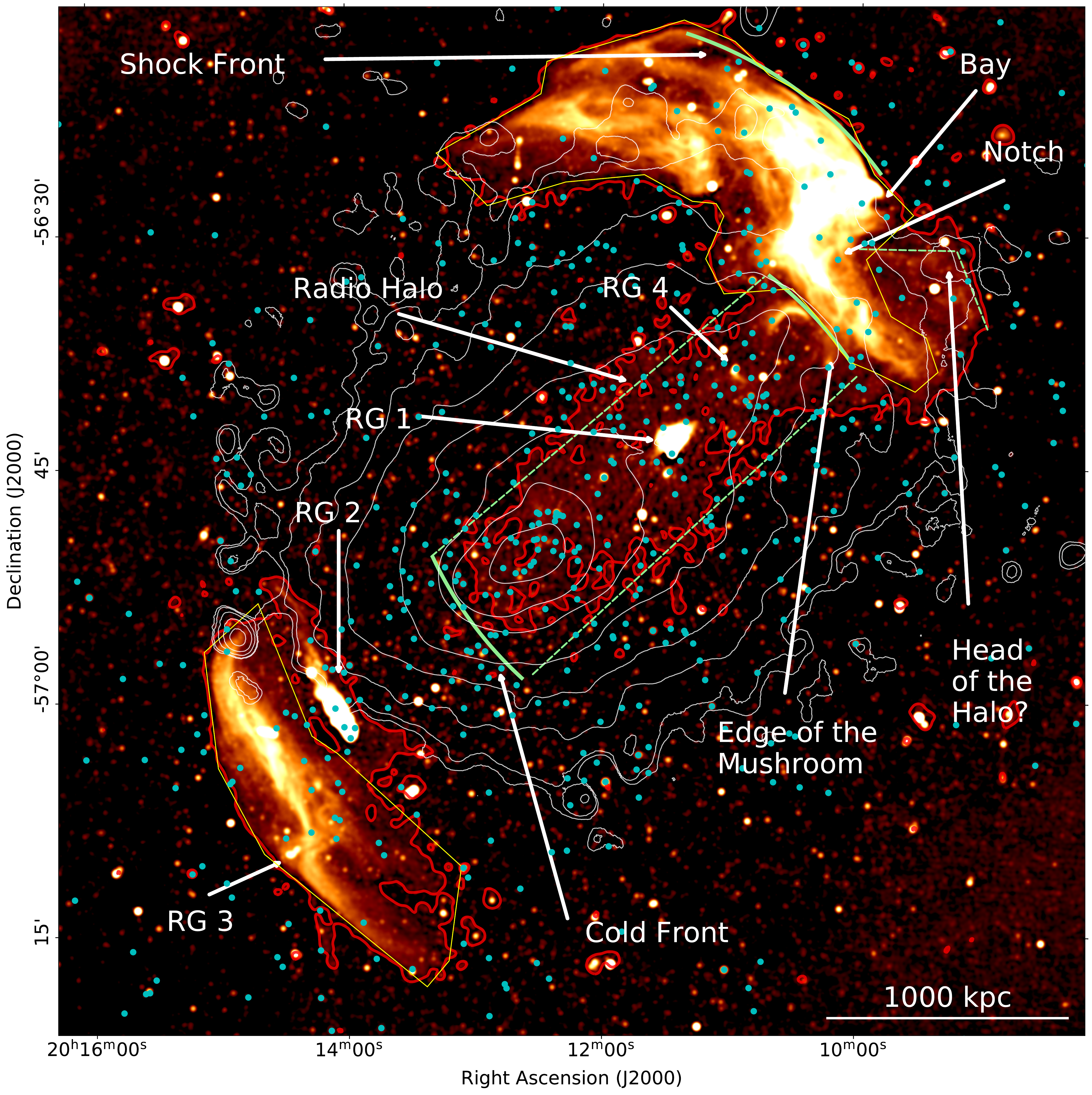}
\caption{Flat-noise image of the cluster region with some features and sources labelled. Solid green lines trace the location of known X-ray discontinuities. Dashed green lines show the region occupied by the radio halo and its possible starting point. The red contour traces the $5\sigma$ level in the low-resolution map ($\sigma=17$~\mujybeam{}, with beam: \beam{35}{35}). White contours trace the X-ray emission from \textit{XMM-Newton}. Light blue dots mark the positions of spectroscopically identified cluster members from \cite{Owers2009}. The yellow lines show the region used to extract the integrated flux density of the radio relics.}\label{fig:wide}
\end{figure*}

\section{Observations}
\label{sec:observations}

In this section we describe the MeerKAT data acquisition, reduction and the techniques used for the data analysis. We also present the reprocessing of the \textit{XMM-Newton} observations.

\subsection{Radio observations}
\label{sec:radio}

MeerKAT \citep{Jonas2016} is a radio interferometer that consists of 64 offset Gregorian parabolic dish antennas, each with an effective diameter of 13.5 m. The instrument is located in the Northern Karoo desert of South Africa. Most of the MeerKAT antennas (75\%) are located within the inner 1 km region, providing the telescope with a good sensitivity to extended emission. At the central frequency of 1283 MHz, the largest source the interferometer can observe without resolving it out is about 27.5\arcmin, whilst the longest baseline of 7698 m provides a nominal resolution of about 6\arcsec.

% MGCLS
The radio data used in this publication were taken in the framework of the MeerKAT Galaxy Cluster Legacy Survey \citep[MGCLS;][]{Knowles2021}. This is a programme of long-track MeerKAT L-band ($900-1670$ MHz) observations that covered 115 galaxy clusters for $6-10$ hours each in full polarisation. The first legacy product data release includes the visibilities (used for this project), image cubes, as well as spectral and polarisation image cubes.

The dataset used for this work was taken on 27 October 2018. The most important parameters of the observation are listed in Table.~\ref{tab:obs}. The main target of the observation was the centre of the galaxy cluster Abell 3667, with interleaved observations of the phase calibrator ($1934-638$). The phase calibrator was observed for 40 seconds every 10 minutes spent on the target.

\begin{table}
\centering
\begin{threeparttable}
\begin{tabular}{ll}
Number of antennas & 64\tnote{a} \\
-- min baseline length & 29 m\\
-- max baseline length & 7698 m\\
Target's phase centre & \\
-- RA & \hms{20}{12}{33.68}     \\
-- Dec & \dms{-56}{50}{26.3}  \\
Frequency range & 886 MHz -- 1682 MHz\\
-- Bandwidth & 797 MHz \\
Frequency resolution & \\
-- initial & 208 kHz\\
-- final & 836 kHz \\
Total integration time & 8 h \\
Time resolution & 8 s\\
\end{tabular}
\begin{tablenotes}    
\item[a] 61 operating at the time of this observation.
\end{tablenotes}
\end{threeparttable}
\caption{Observation parameters.}\label{tab:obs}
\end{table}

The calibration of the dataset is described in the MGCLS survey paper \citep{Knowles2021}. The flux scale is set to follow \cite{Reynolds1994}, based on the spectrum of PKS~B1934-638. In order to fine tune the imaging parameters, we did not use the images released as part of the MGCLS DR1 but we produced a new set of images starting from the calibrated visibilities\footnote{Only the raw visibilities are provided as part of the MGCLS DR1.}. For calibration purposes, the visibilities are divided into eight frequency blocks, which we refer to as spectral windows (SPWs). Here we explain the details the imaging and post-processing of the images.

\subsubsection{Imaging}
\label{sec:imaging}

We produced a set of images with different parameters. The software used for the inversion and deconvolution was \texttt{WSclean} \citep{Offringa2014}. Unless otherwise specified, we used a Briggs weighting of $-0.5$ and we divided the data into seven channels -- one per SPW, excluding SPW 3 that was completely flagged mostly due to radio frequency interference from satellites. To speed up the process, we used the baseline-based averaging, parallel gridding, and parallel deconvolution options of \texttt{WSclean}. The beam was forced to be circular and the pixel size was set to 2 arcsec. Multi-scale cleaning was enabled, and we set a 1 $\sigma$ threshold to halt the deconvolution. Finally, the spectral profile of the clean components was forced to follow a second order polynomial.

Images were corrected for primary beam attenuation assuming a cosine beam shape \cite[eq. 3.95;][]{Condon2016}, which should provide a good approximation in the central degree$^2$. Primary beam correction has been applied in all cases where we estimated physical quantities, such as luminosity or flux density. In some figures, where specified, contours are taken from flat-noise (primary-beam-uncorrected) images to derive where the detection is significant. Those contours should not be used to evaluate physical quantities.

The uncertainties on the flux densities were estimated with
\begin{equation}\label{eq:errors}
    \delta S_\nu = \sqrt{(f \cdot S_\nu)^2 + N_{\rm beams} (\sigma_{\rm rms})^2},
\end{equation}
where $S_\nu$ is the flux density, $f$ is the fractional error on the flux density, and $\sigma_{\rm rms}$ is the local rms noise estimated from the maps. $N_{\rm beams}$ is the number of beams covering an extended source. We assumed a 5\% fractional error on the absolute flux density that includes the flux scale uncertainty, the primary beam uncertainty, and possible loss of signal due to missing short baselines.

% radio halo
We re-imaged the dataset with increasing Gaussian tapering to obtain lower-resolution maps with the aim to better describe the extended emission (see e.g. Fig.~\ref{fig:wide}). To image the radio halo we produced an image at high resolution (\beam{3}{3}) using uniform weighting. This image is used to detect compact sources and subtract them from the $uv$ data. We excluded from this process the two large radio galaxies and the region of the radio relics since their filamentary structure was not fully resolved out in the high-resolution map. Any detected source in the halo region, beside the two large radio galaxies, has been subtracted. Since at normal resolution the halo is completely resolved out, this did not affect its properties. Finally, we re-imaged the dataset using a Gaussian taper to bring the angular resolution down to 35 arcsec. A summary of all the images produced and their properties is given in Table~\ref{tab:images}.

\begin{table}
\centering
\begin{threeparttable}
\begin{tabular}{lcccc}
\hline
\hline
Weighting & Resolution & Taper & rms Noise\tnote{a} & Figures\\
\hline
Uniform & \beam{3}{3} & -- & 13~\mujybeam & ~\ref{fig:wat} \\
Briggs $-0.5$ & \beam{5}{5} & -- & 5.3~\mujybeam & \ref{fig:relicN}, \ref{fig:relicS}\\
Briggs $-0.5$ & \beam{10}{10} & 10\arcsec & 6.3~\mujybeam & -- \\
Briggs $-0.5$ & \beam{15}{15} & 15\arcsec & 10~\mujybeam & \ref{fig:loop}, \ref{fig:boomerang} \\
Briggs $-0.3$ & \beam{35}{35} & 35\arcsec & 17~\mujybeam & \ref{fig:wide}\\
\hline
\end{tabular}
\begin{tablenotes}
    \item[a] Calculated on the flat-noise map.
\end{tablenotes}
\end{threeparttable}
\caption{Imaging parameters.}\label{tab:images}
\end{table}

\subsubsection{Spectral index maps}
\label{sec:spidx}

The fractional bandwidth of MeerKAT is large enough (62\%) so that we could carry out a spectral index analysis by using the in-band spectra. For the spectral index maps we produced two sets of 7 images each, again one image per usable SPW. The first set had the lowest frequency channel imaged at 12\arcsec and the second one at 26\arcsec resolution. Both resolutions were achieved applying a Gaussian taper. Both image sets were produced using uniform weighting and applying an inner $uv$ cut of $164 \lambda$ so to harmonise the $uv$ coverage across the different SPWs. Images of each set were finally convolved to the resolution of the lowest frequency channel and corrected for the primary beam shape at the image frequency. We then produced pixel-by-pixel spectral index maps using all pixels that had a surface brightness above $3\sigma$ in all images, with $\sigma$ estimated using the rms noise of each individual image. We then carried out the linear regression to estimate the spectral index using a bootstrap Monte-Carlo method obtaining 1000 estimations per pixel. The errors on the values used in the regressions were calculated as in Eq.~\ref{eq:errors}. Although a flux scale error would be irrelevant in the derivation of the in-band spectral index, we maintain the 5\% absolute flux density uncertainty due to the beam errors and different weighting of the visibilities. We used the mean of the distribution as the estimator of the spectral index and the standard deviation as the estimator of the error.

\subsubsection{Spectral tomography}
\label{sec:tomography}

Spectral tomography is a technique introduced by \cite{KatzStone1997}. The main aim of this analysis is to disentangle overlapping components of a radio source characterised by a different spectral index. The goal is achieved by scaling an image at a certain frequency ($\nu_2$) with a set of spectral indexes $\alpha_t$. The scaled image is then subtracted from a reference image obtained at another frequency, $\nu_1 < \nu_2$, obtaining a series of maps with features having spectral index $\alpha_t$ subtracted:

\begin{equation}
    I_\nu (\alpha_t) = I_{\nu_1} - \left(\frac{\nu_1}{\nu_2}\right)^{\alpha_t} I_{\nu_2}.
\end{equation}Here we used $\nu_1=1.035$~GHz, $\nu_2=1.480$~GHz, and $\alpha_t$ ranging from $-0.5$ to $-1.5$. 

\subsubsection{Polarisation and rotation measure}
\label{sec:polarisation}

\begin{figure}
\centering
\includegraphics[width=.49\columnwidth]{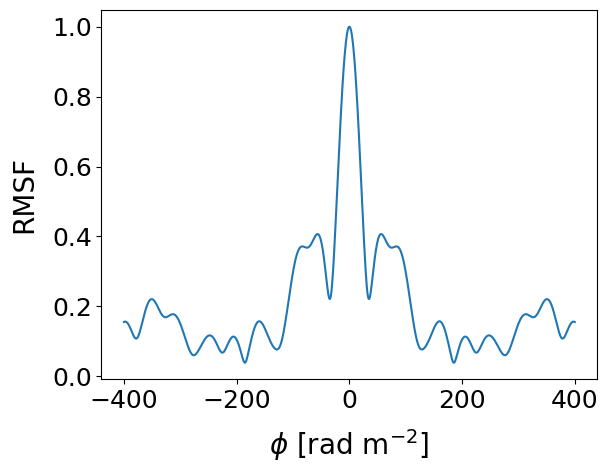}
\includegraphics[width=.49\columnwidth]{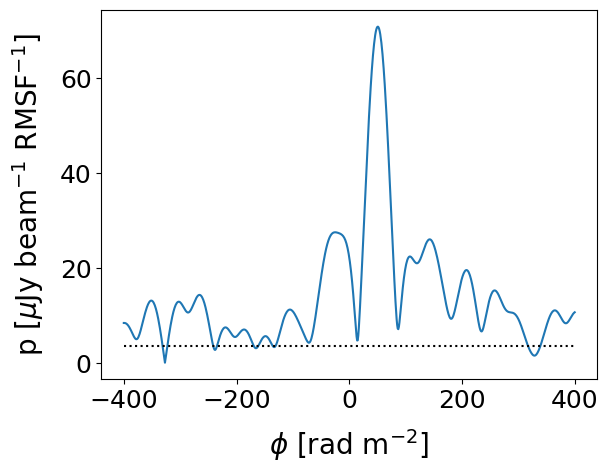}
\caption{Rotation measure analysis. \textit{Left:} RM spread function (FWHM: 42~\radmm). \textit{Right:} Example of a Faraday spectrum taken from one pixel in the northern relic. The dotted line is the estimated noise level.}\label{fig:rmspectra}
\end{figure}

The Q- and U-Stokes were imaged separately from the I-Stokes and using the \textit{squared-channel-joining} option of \texttt{WSclean} so that the peak finding is performed in the sum of squares of the Q and U images, without losing signal due to Faraday rotation. We divided the bandwidth into 119 channels. Of these, 25 were removed because they covered flagged regions of the band. The remaining 94 channels were convolved to the same resolution (\beam{10}{10}) and processed with \texttt{RMsynth}\footnote{\url{https://github.com/mrbell/pyrmsynth}} using a sampling in Faraday space $\delta\phi=2$~\radmm{} with Faraday depth ($\phi$) ranging $\pm 400$~\radmm{}. The maximum theoretical resolution -- the full width at half maximum (FWHM) -- of the rotation measure spread function (RMSF) is 42~\radmm{}. No correction for spectral index has been applied. The RMSF and one example of an extracted Faraday spectra are presented in Fig.~\ref{fig:rmspectra}.

The rotation measure (RM) is derived searching for the maximum of the Faraday spectrum in each pixel. When needed (and specified in the text) the RM has been corrected for Galactic Faraday rotation subtracting $\phi=39.5$~rad m$^{-2}$ for the NW relic region and $\phi=55.0$~rad m$^{-2}$ for the SE relic region. These values have been derived from the Galactic Faraday map presented in \cite{Hutschenreuter2020}. We derived the fractional polarisation for each pixel with a polarised signal above $3\sigma$ and the polarisation angle. For the latter we rotated the polarisation vector by 90 degrees to show the local projected magnetic field orientation and the vector angle has been de-rotated to its intrinsic orientation at zero wavelength using the observed Faraday rotation. In both cases we considered only pixels with signal to noise larger than three, both in the total intensity and polarisation maps. Fractional polarisation maps were corrected for Ricean bias.

In order to examine the possibilities of using a deconvolved Faraday synthesis, the data were also imaged in Obit task MFImage using a maximum fractional bandwidth of 0.3\% giving 223 channels across the bandpass, numerous of these were blanked due to radio interference. The deconvolution used the band average channel polarised intensity ($\sqrt{Q^2+U^2}$) to drive the CLEAN and reduce the sensitivity to depolarisation due to Faraday rotation \citep{Condon2021}. A frequency-dependent taper was used to obtain a nearly constant spatial resolution with frequency. The Briggs' robust factor of $-1.5$ (AIPS/Obit usage) resulted in a spatial resolution (FWHM) of \beam{7.4}{7.3} at position angle $-87\deg$.

Faraday synthesis used Obit task RMSyn (Cotton \& Rudnick in prep.) to first compute the complex Faraday depth function followed by a complex Hogbom CLEAN deconvolution on a pixel by pixel basis. The residual Faraday depth function was restored with a complex Gaussian whose width $\sigma$ was that fitted to the central real lobe of the Faraday function point spread function (`Beam') and using the phase of the Faraday function point spread function. The final Faraday synthesis cube is the modulus of the restored Faraday function. The range of RMs covered was $\pm$300 rad m$^{-2}$ in steps of 2 rad m$^{-2}$; the FWHM of the restoring function in Faraday space of 15.5 rad m$^{-2}$.

\subsection{X-ray data}
\label{sec:xray}

%%% General information
Abell 3667 was extensively observed by \textit{XMM-Newton} covering up to the virial radius \citep[2.26 Mpc, 34\arcmin;][]{Akamatsu2012}. European Photo Imaging Camera (EPIC) MOS \citep{mos} and pn-CCD camera~\citep{pn} instruments were used for these observations. To perform data reduction of these data, we used SAS version 17.0.0 and the latest Current Calibration File (CCF) version as a calibration database. HEASOFT v6.28 was also used for the data reductions. 
%% Data reduction flea clipping
We used the data analysis pipeline ESAS (Extended Source Analysis Software) presented by \citet{snowden08}. 
The data were screened for periods of high particle background by rejecting events outside $\pm2\sigma$ in the count-rate distribution. The details of the resulting screened exposure time and the background-subtracted and exposure-corrected 0.5-2.0 keV image are given in Table~\ref{tab:xmm_data} and Fig~\ref{fig:xmm}, respectively.

\begin{table}[ht]
    \caption{Observation log and exposure time after data screening of the EPIC MOS1 instrument.}
    \label{tab:xmm_data}
    \centering
    \begin{tabular}{ccccccccccccccccccc}
    \hline \hline
    OBSID   &   (RA, Dec)   & Start date    & Exp (ks)\\
    \hline 
0105260201 & (302.83, -56.66) & 2000-10-13 & 13.8  \\
0105260401 & (303.10, -56.63) & 2000-10-03 & 16.2  \\
0105260601 & (303.02, -56.93) & 2000-10-02 & 21.5  \\
0105260101 & (303.31, -56.90) & 2000-09-21 & 4.7   \\
0105260301 & (303.33, -56.73) & 2000-10-03 & 16.4  \\
0105260501 & (302.78, -56.82) & 2000-10-04 & 12.5  \\
0206850101 & (303.24, -56.88) & 2004-05-03 & 54.5  \\
0553180101 & (302.59, -56.39) & 2008-10-12 & 47.3  \\
0653050301 & (302.30, -56.65) & 2010-10-03 & 23.3  \\
0653050501 & (302.66, -56.45) & 2010-09-25 & 73.0  \\
0653050801 & (302.61, -56.43) & 2010-11-02 & 61.1  \\
0653050201 & (303.18, -56.37) & 2010-09-21 & 27.5  \\
0653050401 & (302.57, -56.41) & 2010-09-22 & 78.0  \\
0653050601 & (302.63, -56.42) & 2010-09-27 & 21.3  \\
0761210101 & (303.67, -57.10) & 2015-10-12 & 60.3  \\
    \hline
    \end{tabular}
\end{table}

\section{Results}
\label{sec:results}

The radio environment of Abell 3667 is complex and includes a variety of types of radio sources. The dominant radio source is the radio relic (cluster radio shock) located towards the NW. A counter radio relic is located on the opposite side of the cluster at the distance of just about 1\deg{} (3.86 Mpc), measured from the sources' external edges. The radio relic properties are discussed in Sect.~\ref{sec:relics}. Another cluster-size but fainter extended radio source is the radio halo that extends for more than 2 Mpc from the NW relic along the merging axis. While this source can probably be related with the well known giant radio halos, it has some peculiarities. We discuss the linear radio halo in Sect.~\ref{sec:halo}. The brightest radio galaxies in the cluster are discussed in Appendix~\ref{app:rg}.

\subsection{Radio relics}
\label{sec:relics}

\begin{figure*}
\centering
\includegraphics[width=\textwidth]{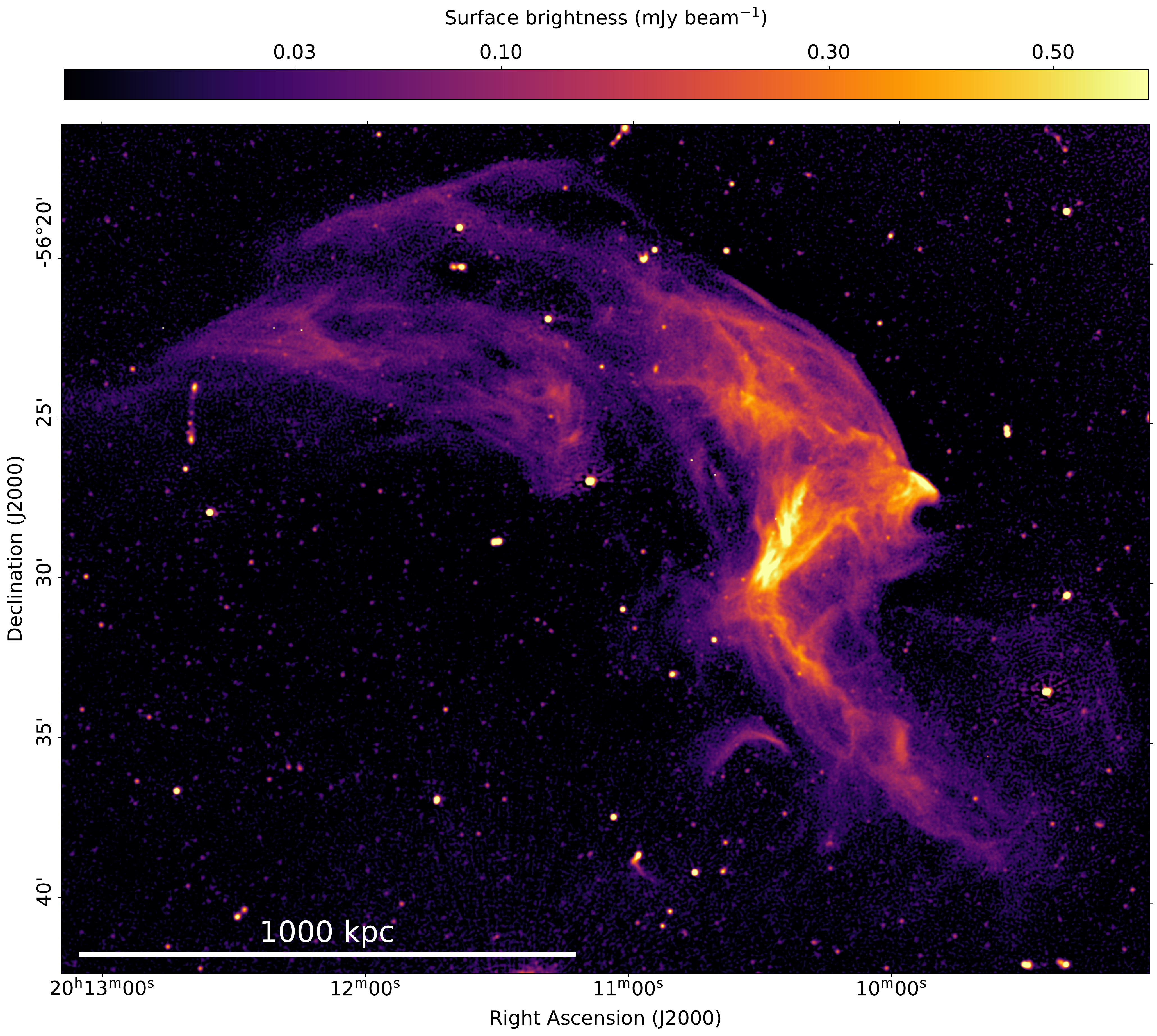}
\caption{Radio emission from the northern radio relic. The image has a local rms noise of 10~\mujybeam{} and a beam of \beam{5}{5}.}\label{fig:relicN}
\end{figure*}

\begin{figure}
\includegraphics[width=.5\textwidth]{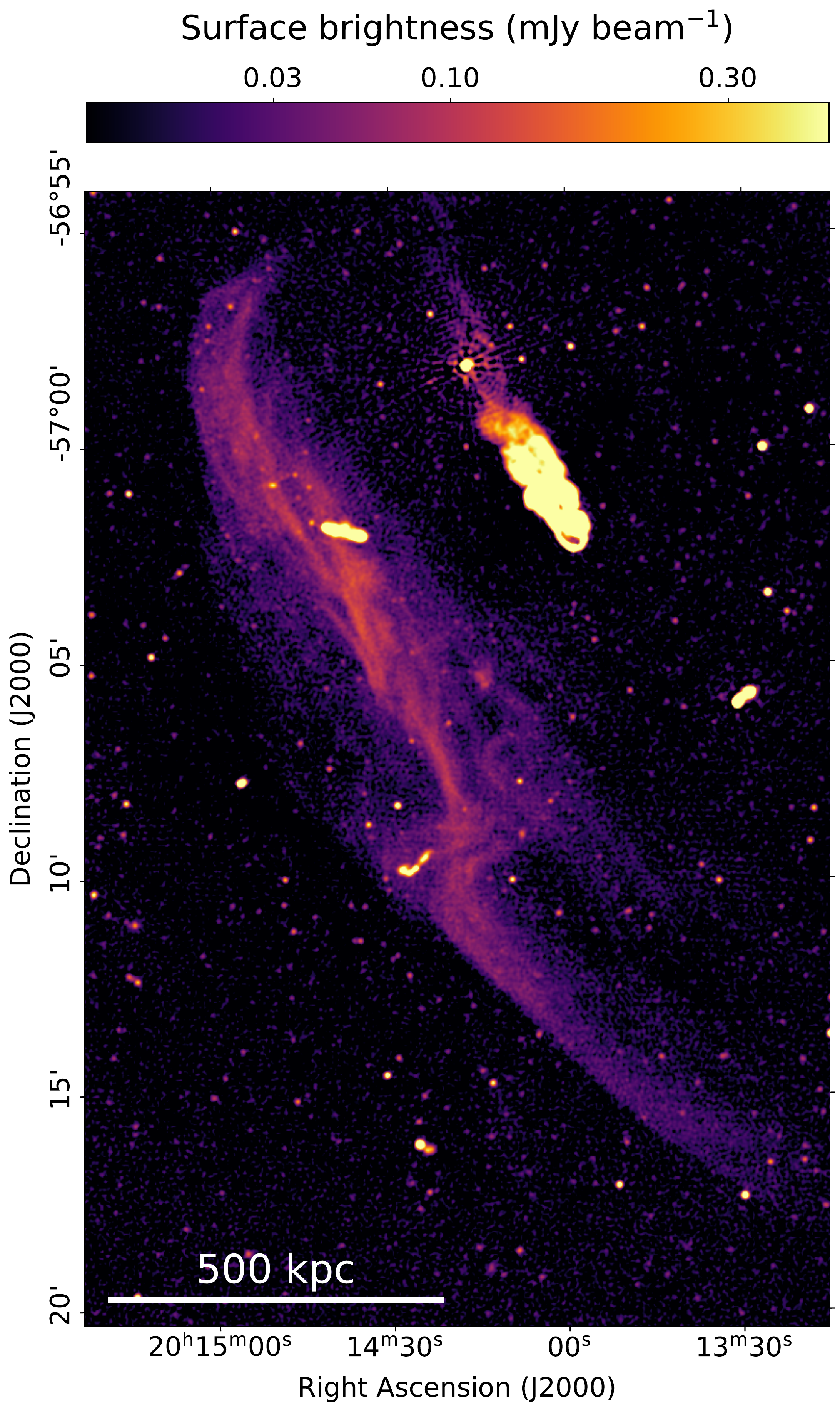}
\caption{Radio emission from the southern radio relic. The image has a local rms noise of 8~\mujybeam{} and a beam of \beam{5}{5}.}\label{fig:relicS}
\end{figure}

We present the radio image of the NW radio relic in Fig.~\ref{fig:relicN}. The radio relic is the largest known in terms of apparent size with an angular size of 0.6\deg. The large angular size is due to the cluster being rather close as well as the NW relic being one of the largest in terms of physical (projected) size, with an extension of 2.3 Mpc. This large angular size translates into an unprecedented level of details in the internal structure of the radio relic. The vast majority of the flux density coming from the radio source originates from filaments of various sizes, orientations, and with non-uniform surface brightness. Filamentary structures in radio relics have been seen in other nearby cluster radio shocks with high-resolution data such as those in Abell 2256 \citep{Owen2014, Rajpurohit2021a}, CIZA J2242.8+5301 \citep[Sausage cluster;][]{diGennaro2018}, 1RXS J0603.3+4214 \citep[Toothbrush cluster;][]{Rajpurohit2018}, MACS J0717.5+3745 \citep{vanWeeren2009d}, and Abell 3376 \citep{Bagchi2006}.

The SE relic is shown in Fig. \ref{fig:relicS}. It is smaller than the companion, with an extension of 0.42\deg (1.6 Mpc). It also shows the presence of filaments although they appear globally ordered along the relic extension and less bright.

\begin{figure}
\includegraphics[width=.5\textwidth]{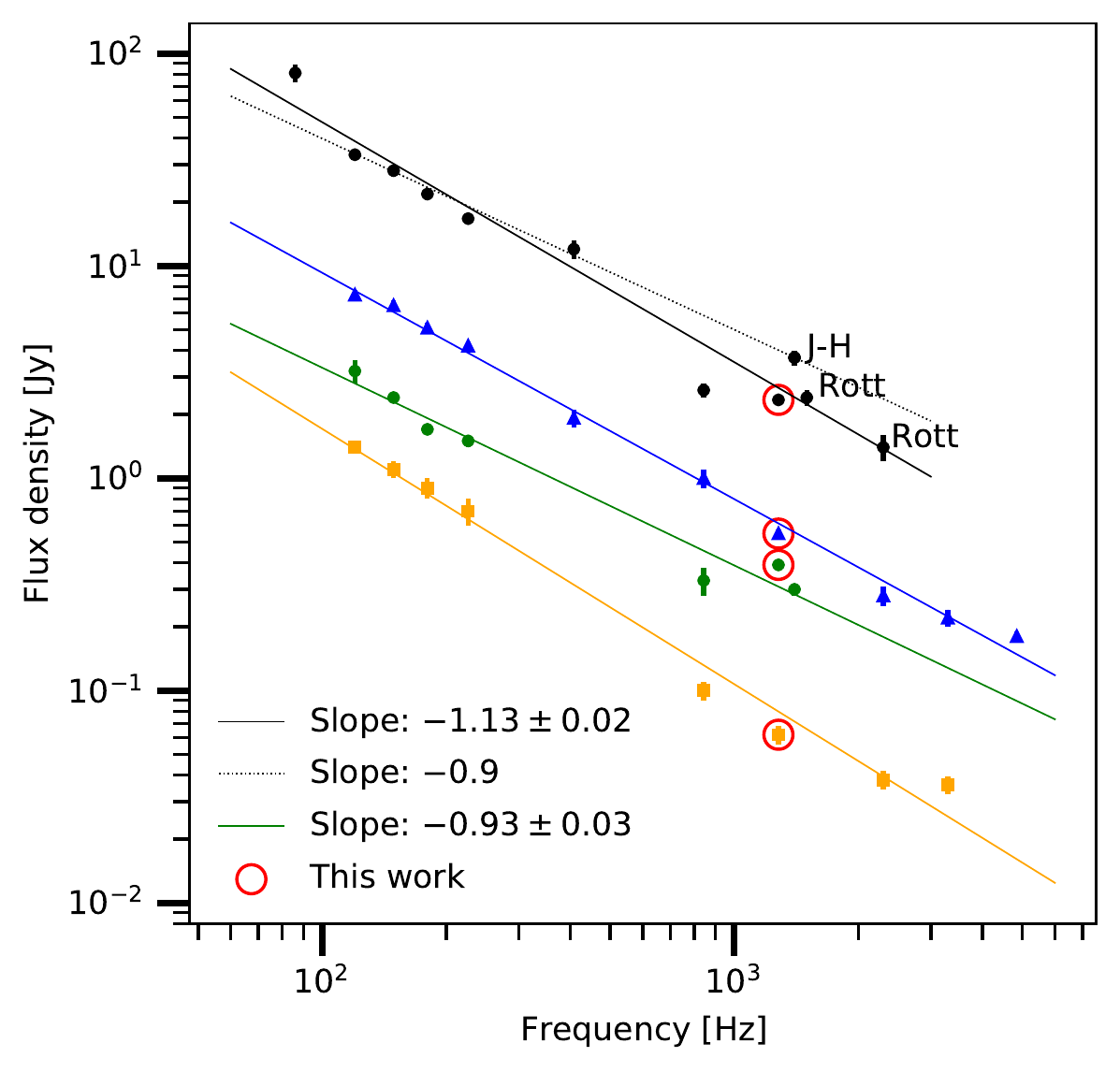}
\caption{Spectral energy distribution of four sources in the field. From top to bottom: the NW radio relic (black), RG1 (blue), the SE radio relic (green), and `Source C' from \citet{Hindson2014}, located close to the Notch (yellow). Flux densities from lower to higher frequencies are taken from \citet{Mills1961}, \citet{Bolton1964}, \citet{Large1991}, \citet{Hindson2014}, \citet{Johnston-Hollitt2003}, \citet{Rottgering1997}, \citet{Carretti2013}, and \citet{Wright1994}. Solid lines are linear regressions.}\label{fig:integrated_spectrum}
\end{figure}

% luminosity
Measuring the luminosity of the relics in Abell 3667 is not trivial because of the large number of foreground or background radio galaxies superimposed to the extended emission. The attempt of isolating their emission making a high-resolution map that resolved out the relic failed because of the many small-scale structures embedded in the relic itself that are detected even in the longest baselines. Therefore, we masked all compact sources we could identify by eye and extracted the flux density of the relic using the non-contaminated pixels. To compensate for the missing pixels, we re-scaled up the relic flux density by the fraction of the pixels that were masked. The correction is $<2\%$ for both relics. To estimate the integrated flux densities of the relics we used the map at 15\arcsec{} resolution to minimise the loss of flux on the largest scales, while for radio galaxies we used the map at 5\arcsec{} to better disentangle their emission from other sources. In any case, the values derived for the NW relic should be treated as lower limits because the source might be partially resolved out due to its large extension. We note that our integrated flux density estimations are in line with those derived by \cite{Rottgering1997}, while being 44\% lower than those derived by \cite{Johnston-Hollitt2003}, once re-scaled at 1.4 GHz. The region used to extract the values are visible in Fig.~\ref{fig:wide}, the estimated flux densities are luminosities are listed in Table~\ref{tab:fluxlum}. For a few bright sources with archival data, we present an integrated spectral energy distribution in Fig.~\ref{fig:integrated_spectrum}. The good agreement for the radio galaxies shows that our flux density scale is not biased. The large scatter in the archival data due to the complexity of the extended northern relics, makes it hard to estimate if or how much flux loss we have. Our flux density estimation for the SE relic is about 15\% higher than the expectations based on archival data; however, we recovered emission from a larger area.

\subsubsection{Spectral properties}

\begin{figure*}[htb]
\centering
\includegraphics[width=\textwidth]{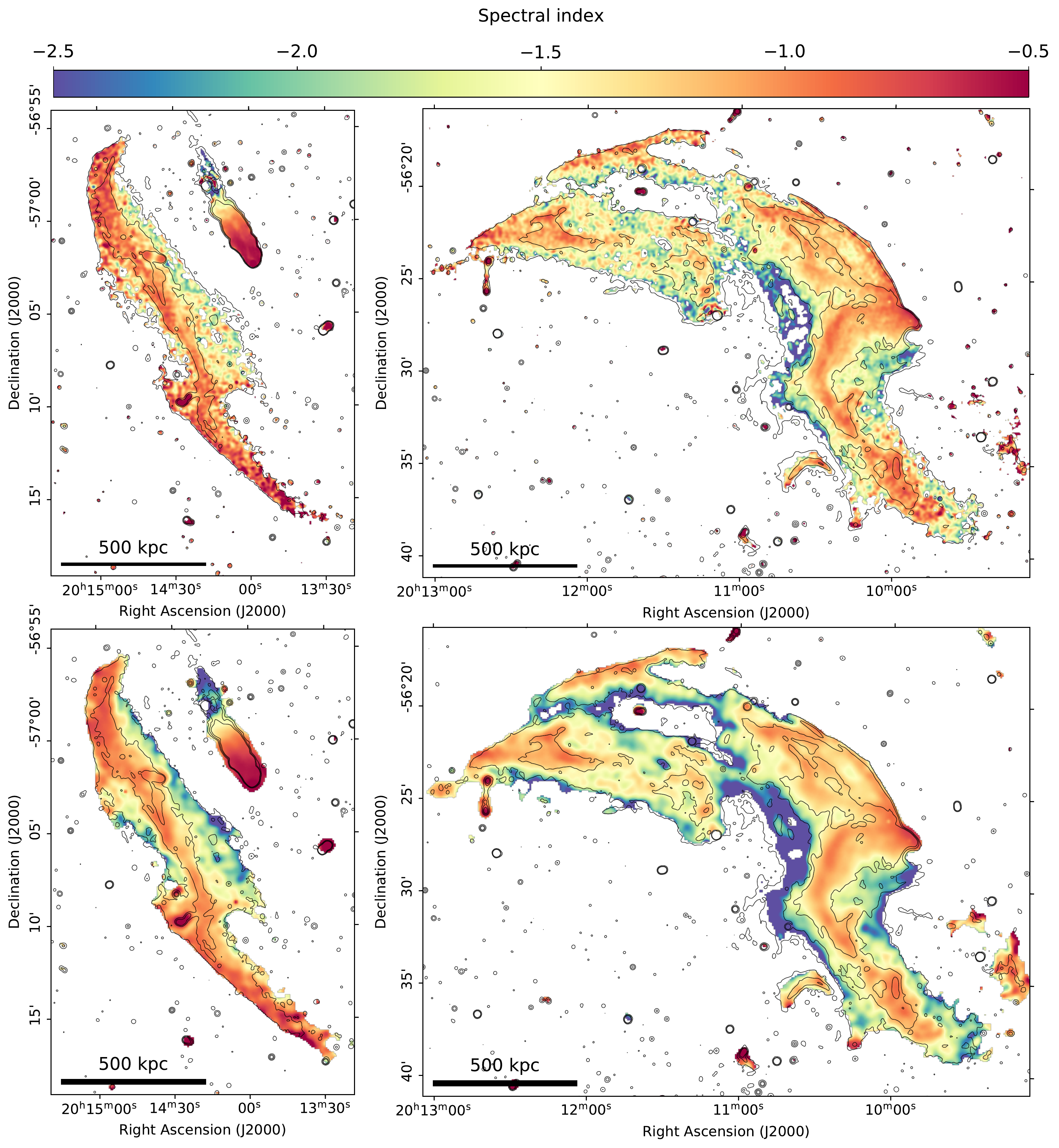}
\caption{Spectral index maps of the radio relic regions. \textit{Top panels:} Spectral index maps derived from the high-resolution image set (beam: \beam{12}{12}). \textit{Bottom panels:} Spectral index maps derived from the low-resolution image set (beam: \beam{26}{26}). Corresponding error maps are shown in Fig.~\ref{fig:spidx-err}. In all panels, contours are taken from the flat-noise map at $(7,25,60)\times\sigma$ with $\sigma=6.3$~\mujybeam{} and beam \beam{10}{10}.}\label{fig:spidx}
\end{figure*}

% integrated spectral index
We estimated the integrated spectral index of the two radio relics using the seven images (one per usable SPW) at 12\arcsec{} resolution used to produce the spectral index maps. The frequency range covered is $936-1632$~MHz. We used the same regions for the integrated flux densities shown in Fig.~\ref{fig:wide}. For this estimation we also masked all pixels belonging to background point sources. We found $\alpha = -1.4 \pm 0.1$ for the NW relic and $\alpha=-1.2 \pm 0.1$ for the SE relic. To verify if our estimation might be biased due to flux loss at higher frequencies, we repeated the experiment using low and high-resolution maps, obtaining consistent results and therefore excluding that bias. In a similar frequency range, \cite{Johnston-Hollitt2003} found a similar value for the SE relic ($\alpha = -1.2\pm0.2$). Their value for the NW relic ($\alpha = -0.9\pm0.2$) is flatter than ours. At the same time, \cite{Hindson2014} found compatible but different values, with $\alpha=-1.2\pm0.1$ for the NW relic (revised to $\alpha=-0.9\pm0.1$ after removing some data points) and $\alpha=-0.9\pm0.1$ for the SE relic. To better understand the origin of the discrepancy we did a linear regression using archival data of the integrated of flux density of both relics (see Fig.~\ref{fig:integrated_spectrum}). The integrated flux density for these sources are complex to measure and a large scatter is present. For the NW relic we obtained a spectral index $\alpha=-1.13\pm0.02$, flatter than our in-band estimation but steeper than what reported in the literature. The southern relic integrated spectral index ($\alpha=-0.93\pm0.03$) is also flatter than our in-band estimation but compatible with the literature values. The values measured here, with $\alpha \lesssim -1$ are generally in line with other integrated spectral index estimations for radio relics \citep[e.g.][and references therein]{Bonafede2012a, Feretti2012, deGasperin2014c}.

\begin{table}
\centering
\begin{threeparttable}
\begin{tabular}{lccl}
\hline
\hline
Source & Flux density & Luminosity & Notes\\
       & (mJy)                  & ($10^{31}$ erg s$^{-1}$ Hz$^{-1}$) & \\
\hline
Relic N & $2430 \pm 120$\tnote{a} & $17.7 \pm 0.9$\tnote{a} & \\
Relic N & $2300 \pm 110$\tnote{a} & $16.7 \pm 0.8$\tnote{a} & no point sources \\
Relic N & $2340 \pm 120$\tnote{a} & $17.0 \pm 0.9$\tnote{a} & corr.\tnote{b} \\
Relic N & $2070 \pm 100$\tnote{a} & $15.1 \pm 0.8$\tnote{a} & 1.4 GHz\tnote{c} - corr.\tnote{b}\\

Relic S & $446 \pm 22$ & $3.3 \pm 0.2$ & \\
Relic S & $375 \pm 19$ & $2.7 \pm 0.1$ & no point sources \\
Relic S & $390 \pm 20$ & $2.8 \pm 0.1$ & corr.\tnote{b}\\
Relic S & $350 \pm 18$ & $2.6 \pm 0.1$ & 1.4 GHz\tnote{c} - corr.\tnote{b}\\

Halo & $99 \pm 6$ & $0.72 \pm 0.04$ & \\
Halo & $104 \pm 6$ & $0.75 \pm 0.04$ & corr.\tnote{b}\\

RG1 & $549 \pm 27$ & $4.0 \pm 0.2$ & Fig.~\ref{fig:wat} (left)\\
RG2 & $378 \pm 19$ & $2.8 \pm 0.1$ & Fig.~\ref{fig:wat} (centre)\\
RG3 & $7.3 \pm 0.4$ & $0.053 \pm 0.003$ & Fig.~\ref{fig:wat} (right) \\
RG4 & $5.7 \pm 0.3$ & $0.041 \pm 0.002$ & BCG "B" \\
%Boomerang & $9.9 \pm 0.5$ & $0.071 \pm 0.004$ & \\
\hline
\end{tabular}
\begin{tablenotes}
\item[a] Due to its large extension, part of the source might be resolved out and the flux density underestimated. This value should be treated as a lower limit.
\item[b] Corrected to compensate for the blanked pixels contaminated by background point sources.
\item[c] Re-scaled at 1.4 GHz assuming $\alpha=-1.4$ for the NW relic and $\alpha=-1.2$ for the SE relic.
\end{tablenotes}
\end{threeparttable}
\caption{Radio flux densities and luminosities at 1.28 GHz.}\label{tab:fluxlum}
\end{table}

% spectral index maps
Spectral index maps of the two radio relics are presented in Fig.~\ref{fig:spidx} at two angular resolutions (12\arcsec\ and 26\arcsec). The relative error maps are presented in Appendix~\ref{app:spidx_err}. The high-resolution map of the NW relic shows a flat spectral index along the entire outer edge of the radio relic, with several regions with spectral index $\alpha<-1$.

The shock associated with the NW radio relic (see also Fig.~\ref{fig:wide}) has a Mach number derived using X-ray observations from the density jump of $\mach = 2.05^{+0.73}_{-0.38}$ and from temperature jump of $\mach = 3.34^{+0.91}_{-0.50}$. The average of these values provide an estimation of the Mach number of $\mach= 2.54^{+0.80}_{-0.43}$ \citep{Sarazin2016}. According to diffuse shock acceleration, this translates into an injection spectral index of 

\begin{equation}
    \alpha=\frac{3+\mach^2}{2-2\mach^2}=-0.87^{+0.27}_{-0.15}.
\end{equation}
This value is consistent with the average spectral index of the flat filament at the relic's edge (i.e. $\alpha = -1.0 \pm 0.1)$), which is compatible with what found by \citep{Johnston-Hollitt2003}. The injection index derived from the X-ray-inferred Mach number is also consistent with the integrated spectral index if we assume continuous injection, where electrons are accelerated at the shock front and then age following a Jaffe-Perola model \citep{Jaffe1973a}, in a constant magnetic field producing an integrated spectra $\alpha_{\rm int} = \alpha_{\rm inj}-0.5$ \citep{Kardashev1962}. However, this expectation is simplistic as it assumes that the shock properties do not evolve with time \citep[see also][]{Bruggen2020a}.

Behind the relic's edge, most of the relic emission is characterised by a steep spectrum with $\alpha$ ranging between $-1$ and $-2$, with the exception being the bright filaments where the spectrum is closer to $\alpha=-1$. Another region where the NW relic has a particularly flatter spectrum is in the northern (bright) part of the radio bay, further discussed in Sect.~\ref{sec:bay}. The northern edge of the relic also has some filamentary, although fainter, structures, probably associated with other regions of shock acceleration. In that region, the outer edge has a rather flat spectrum, with a mean value of $\alpha=-1$. In many regions where the emission becomes diffuse and faint, the spectrum also steepens to values $\alpha<-2$. Interestingly, as noted by \cite{Johnston-Hollitt2003}, only a moderate spectral gradient across the source is visible. There is clearly a faint, diffuse, steep spectrum region in most of the inner part of the NW relic, but a similar region is also visible below the bay and on the outer edge in the southern part of the relic. The diffuse and more uniform region to the NE shows only a mild steepening gradient from E to W. In most cases, the flattening (or steepening) of the spectral index seems more correlated with the presence (or absence) of a filament than with the distance from the cluster centre as it would be expected from simple particle ageing behind a well defined region of acceleration (the shock front). The SE relic also shows a flatter spectrum, $\alpha \sim -1$, close to the outer edge and along the brighter filament. Diffuse emission, mostly located towards the cluster centre but also surrounding the main filament at the relic's centre, has a steeper spectrum ($\alpha < -1.5$).

\subsubsection{Polarisation and rotation measure}

\begin{figure*}
\centering
\includegraphics[width=\textwidth]{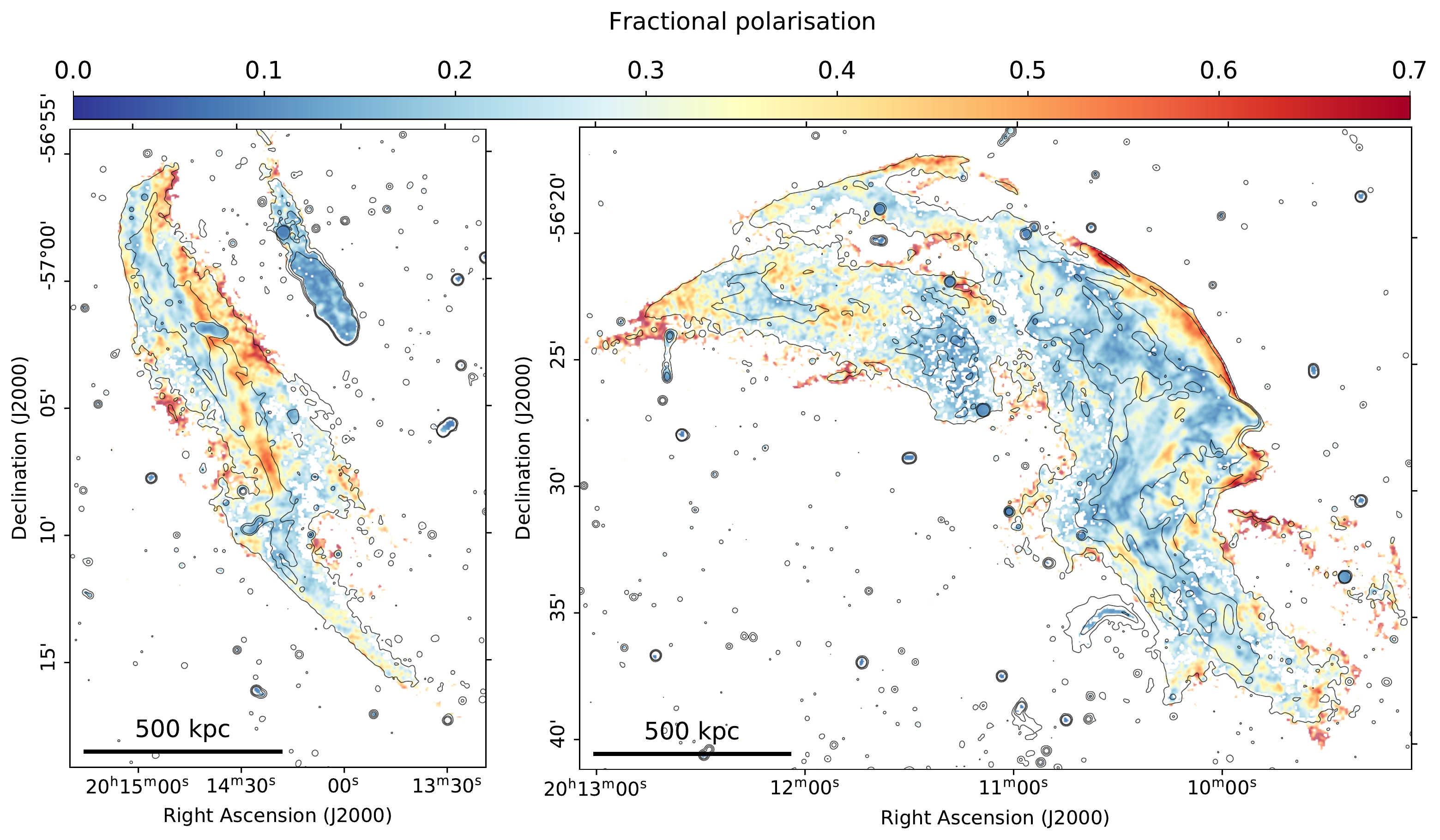}
\caption{Fractional polarisation maps after Faraday correction at the nominal frequency of 1.2 GHz. Resolution is \beam{10}{10}. Contours are the same as in Fig.~\ref{fig:spidx}.}\label{fig:fpol}
\end{figure*}

%FRACTIONAL POLARIZATION
We were able to detect the polarised flux and determine the polarisation angle from almost the entire area covered by the radio relics, as seen in total intensity. The fractional polarisation map of the radio relics is presented in Fig.~\ref{fig:fpol}. The maximum expected fractional polarisations depend on the spectral index. At the shock front, the electron energy distribution is a power-law $dN(E)/dE \propto E^{-\delta}$ with slope $\delta = 1 - 2\alpha$. A region with a homogeneous magnetic field will have an intrinsic polarisation equal to \citep{Rybicki1986}
\begin{equation}
    p_0 = \frac{3\delta+3}{3\delta+7}.
\end{equation}
In the case of the NW relic, a maximum polarisation of 74\% is expected. The NW relic has maximal fractional polarisation close to the outer edge, where the emission is almost fully polarised ($\sim 70\%$), implying magnetic fields ordered on scale lengths larger than the beam size ($\approx 10$~kpc). Other regions with high fractional polarisation are the northern edge, reaching $40-50\%$, and the southern part of the bay region. Most of the NW radio relic has a fractional polarisation of $10-30\%$. Beside from the edge filament tracing the X-ray detected shock front, all other filaments in the NW relic show a decrease in fractional polarisation compared to the surrounding emission (see Fig.~\ref{fig:fpol}), with a fractional polarisation between 5\% and 15\%. A noticeable difference are the filaments oriented perpendicular to the shock front in the region that connects the main body of the relic to the southern arm. Those filaments appear more polarised than the surrounding, with polarisation fractions up to 40\%. The SE relic shows mild enhancement of polarisation fraction along the shock front, with the highest fraction ($\sim 50\%$) along the main filament and in the northern part of the downstream region. Any regions dominated by background emission from radio galaxies present a strong contrast in the fractional polarisation map, with background emission being on average less polarised compared to the relics' radiation. The flux-weighted average fractional polarisation of the NW relic is $\sim20\%$, while for the SE relic is $\sim 25\%$. In ideal conditions, this would imply a relatively small viewing angle $\delta \sim 50\deg$, with $\delta = 90\deg$ being the shock wave propagating the plane of the sky \citep{Ensslin1998}. However, the complex morphology and filaments superimposition of the relics is far from the ideal case, likely reducing substantially the estimated viewing angle.

\begin{figure*}
\centering
\includegraphics[width=\textwidth]{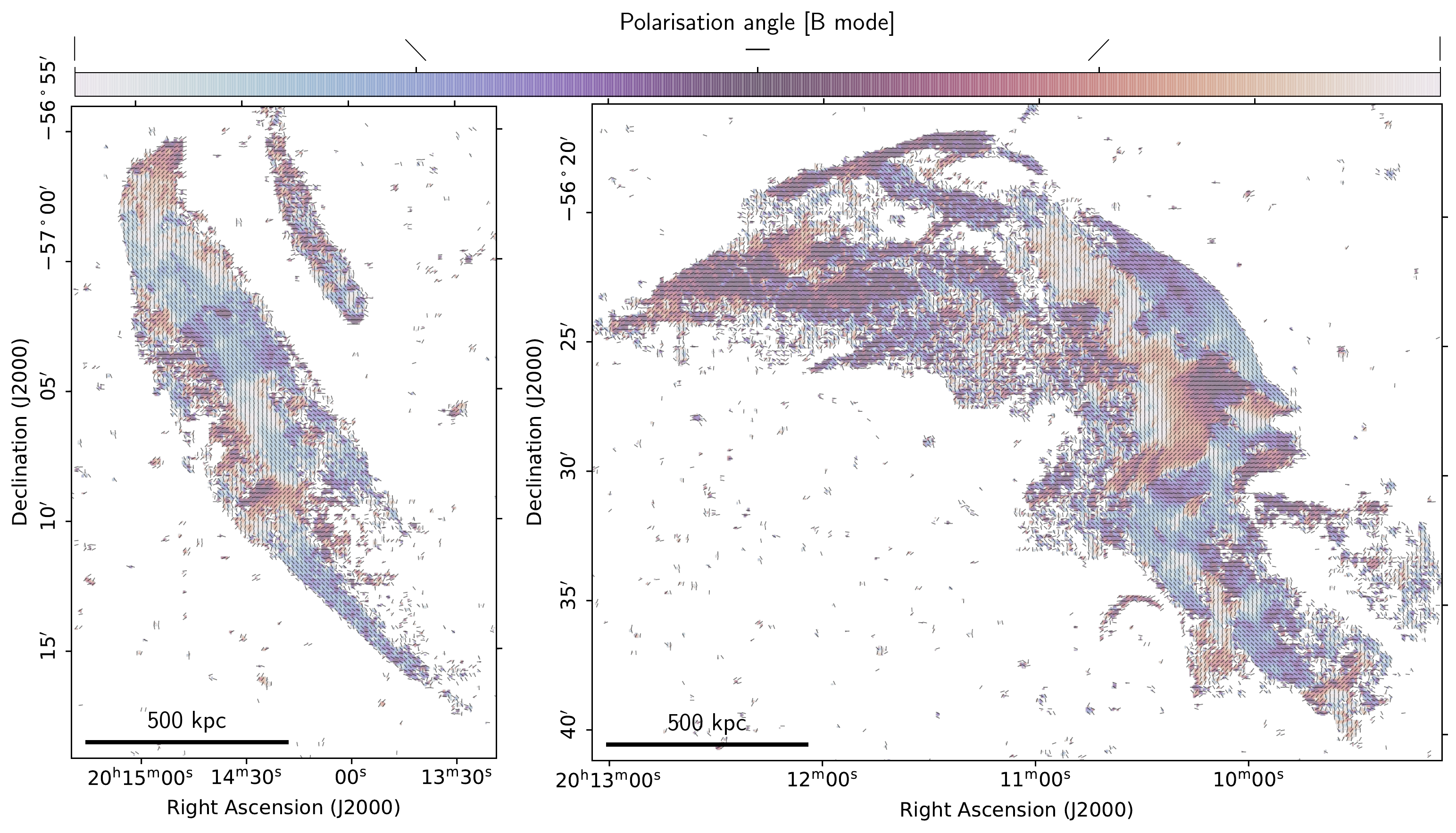}
\caption{Polarisation $B$-vector map at the resolution of \beam{10}{10}, de-rotated to its intrinsic orientation at zero wavelength using the observed Faraday rotation. The orientation of the magnetic field is also colour-coded. An alternative view of the magnetic field vector orientation is presented in Fig.~\ref{fig:pola2}.}\label{fig:pola}
\end{figure*}

%POLARIZATION ANGLE
The orientation of the polarisation vectors, rotated by 90\deg{} to show the projected magnetic field orientation, is presented in Fig.~\ref{fig:pola} (see also Fig.~\ref{fig:pola2}). In both cases we plotted the B vectors of the polarised radiation, which correspond, for synchrotron emission, to the local orientation of the magnetic field projected on the plane of the sky. The main driver deciding the orientation of the magnetic field seems to be the local orientation of the dominant filament, with the magnetic field oriented along its extension. In regions where the filamentary structure is less prominent, as in the faint and steep spectrum regions, the magnetic field orientation is more chaotic. Regions of interface between different filaments or filaments-diffuse regions show a decrease in the polarisation fraction, pointing towards beam depolarisation. In the region behind the shock front, the magnetic field is ordered and perpendicular to the shock normal. It is interesting to note that, while projection effects might play a role in altering X-ray derived Mach numbers, \cite{Sarazin2016} suggested the presence of such a field configuration as the driver for a larger-than-isothermal effective adiabatic index ($\gamma_{\rm eff} \sim 2$) that in turn would explain the inconsistency between the Mach numbers derived from the density ($\mathcal{M} = 2.05^{+0.73}_{-0.38}$) and the temperature ($\mathcal{M} = 3.34^{+0.91}_{-0.50}$) jumps across the shock front. A significant energy density in relativistic particles or a turbulent magnetic field would have favoured $\gamma_{\rm eff} \sim 4/3$, in friction with X-ray observations.

\begin{figure*}[ht!]
\centering
\includegraphics[width=\textwidth]{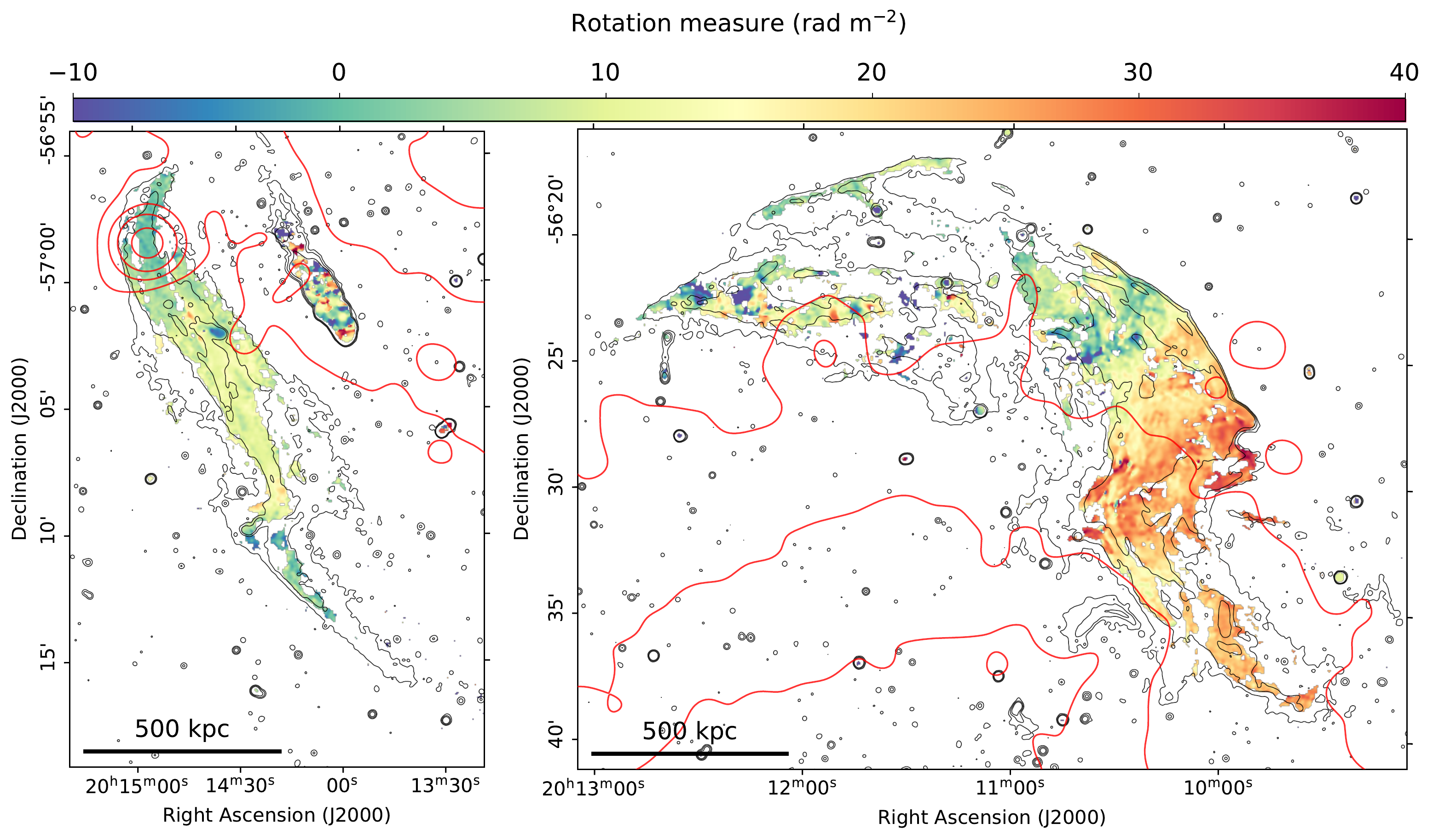}
\caption{Rotation measure map at the resolution of \beam{10}{10}. Local estimates for the Galactic Faraday rotation ($\phi=39.5$~rad m$^{-2}$ for the NW relic and $\phi=55.0$~rad m$^{-2}$ for the SE relic) have been subtracted. Regions with a polarisation signal lower than $7\sigma_P$ ($\sigma_P=3.6$~\mujybeam{}) are blanked. Red contours show the X-ray emission from the ICM. Black contours are the same as in Fig.~\ref{fig:spidx}.} \label{fig:rm}
\end{figure*}

%ROTATION MEASURES
The RM maps of the radio relics, corrected for average local galactic contribution (see Sect.~\ref{sec:polarisation}) are presented in Fig.~\ref{fig:rm}. The values of the RM appear to change smoothly across the radio relics. For the SE relic the RM stays between 0 and 10 rad m$^{-2}$. The northern portion of the SE relic shows a change in RM in a region that corresponds to an X-ray excess (see Fig.~\ref{fig:rm}). While a portion of the X-ray excess may be associated with the ICM, we note that the compact X-ray region seen in Fig.~\ref{fig:rm} is spatially co-incident with a double star system (HIP~\,99803). This system is a member of the Tucanae Association \citep{zuckerman2000}, a co-moving group of young stars within 50 pc of Earth. One of the characteristic signatures of stars in young stellar associations is the presence of X-ray emission \citep{gaidos98}. The association of X-ray emission with the foreground double star system is presented in \citet{stelzer2000}. Analysis of the compact X-ray excess in this region reveals the emission is thermal. Detailed analysis of the RM in this region would require consideration of the contribution of the stellar X-ray emission to the observed excess in Fig.~\ref{fig:rm}. For the NW relic, the RM varies from about 0 to 40 rad m$^{-2}$ moving from north to south. The region with the larger residual RM is closer to the cluster centre, where significant X-ray emission from the ICM is detected. A possibility is that the ICM in front of the NW radio relic rotates the polarised signal. The large apparent size of the radio relics (0.6\deg and 0.4\deg) and the galactic latitude of $-33\deg$ suggests that there could be some small contributions from galactic Faraday rotation gradients across their extension. A similar gradient between some bright point sources located behind the outer edge of the NW relic supports this possibility.   A more detailed look at the RMs of the brightest NW relic filaments is discussed below.

% head tail
In contrast with the SE relic, the head-tail source parallel to it and closer to the cluster centre, shows strong Faraday effects. Focusing on the first 225\arcsec{}, where the signal-to-noise ratio is high, the overall RM is within a few rad m$^{-2}$ of the relic's, but the rms variations are $\sim 13$ rad m$^{-2}$, compared to the $\sim 2$ rad m$^{-2}$ for similar scales on the relic. The RM fluctuations are mostly unresolved by a $\sim$10\arcsec{} beam, and the fractional polarisation is $\sim$8\%. The complex RM is probably linked to the strong interaction of head-tail plasma with the ICM generating instabilities in the tail (see the right panel of Fig.~\ref{fig:wat}) and a locally disturbed magnetic field.

\subsection{Linear radio halo}
\label{sec:halo}

The MeerKAT observations were able to detect very clearly an elongated radio halo already noticed by \cite{Carretti2013} with single dish Parkes observations. The emission starts at the cold front, covers the remnant of the cool core of the main sub-cluster (A), and extends until the NW radio relic, for a total length of about 1.8 Mpc (see Fig.~\ref{fig:wide}). The halo has an average width of 800--900 kpc, which makes it much more elongated than standard radio halos that usually have a round shape. Interestingly, a patch of emission of surface brightness similar to the one of the halo is also present on the other side of the NW relic's southern arm, suggesting that the halo might not end at the radio relic, reaching a possible total length of about 2.5 Mpc.

The flux density and the luminosity of the radio halo, covering the region from the cold front up to the radio relic, are listed in Table.~\ref{tab:fluxlum}. The extension of the source combined with its low surface brightness makes it barely detectable in the highest frequency channels. We therefore do not feel confident in extracting an in-band spectral index value for the radio halo.

%%%%%%%%%%%%%%%%%%%%%%%%%%%%%%%%%%%%%%%%%%%%%%%%%%%%%%%%%%%%%%%%%%%
\section{Discussion}
\label{sec:discussion}

In this section we discuss the main results of this paper. In Sect.~\ref{sec:filaments} we focus on the radio relics discussing the properties and origin of the filaments. In Sect.~\ref{sec:mushroom} we focus on the mushroom region combining X-ray and radio data to infer local magnetic properties. In Sect.~\ref{sec:origin_halo} we discuss the origin of the linear radio halo. Finally, in Sect.~\ref{sec:simul} we compare our findings with simulations.

\subsection{Post shock region and filaments in the radio relics}
\label{sec:filaments}

In this section we focus on the radio relics, examining their morphology, spectral and polarisation properties. When radio relics are observed at high resolution, they appear to consist of a number of filaments, whose origin is still debated \citep[e.g.][]{Rajpurohit2018, diGennaro2018, vanWeeren2017b, Owen2014}.

\begin{figure*}[ht!]
\includegraphics[width=.85\textwidth,left]{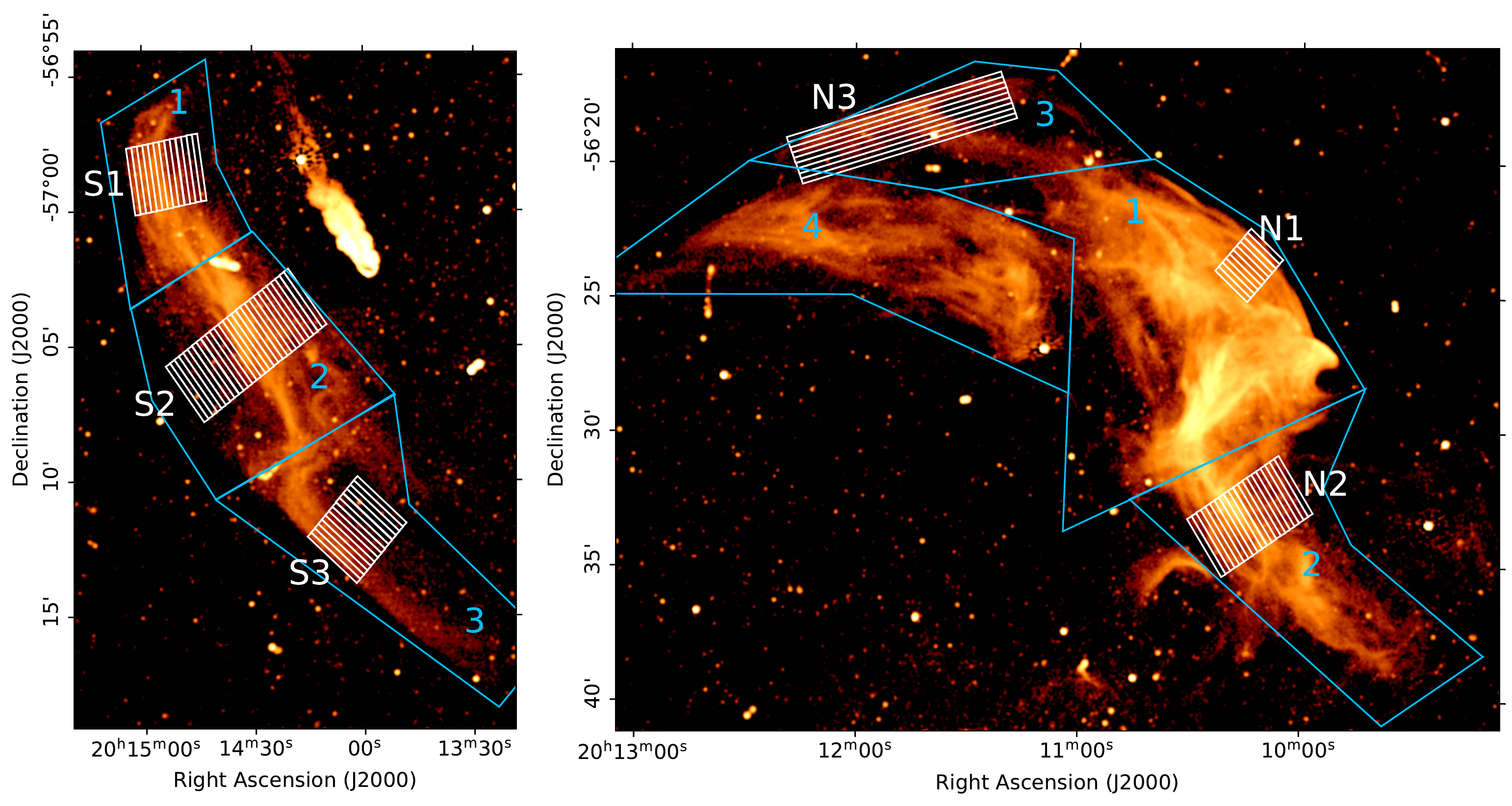}\\
\includegraphics[width=.48\textwidth]{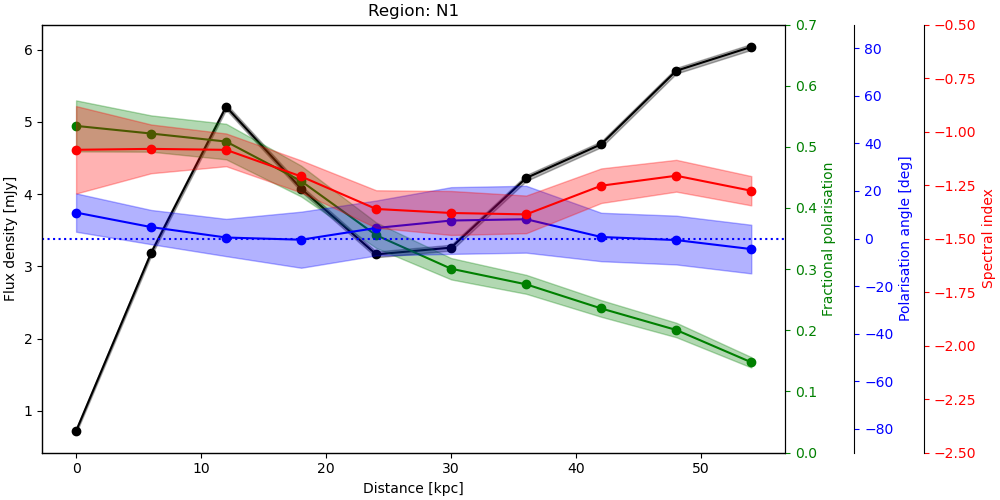}
\includegraphics[width=.49\textwidth]{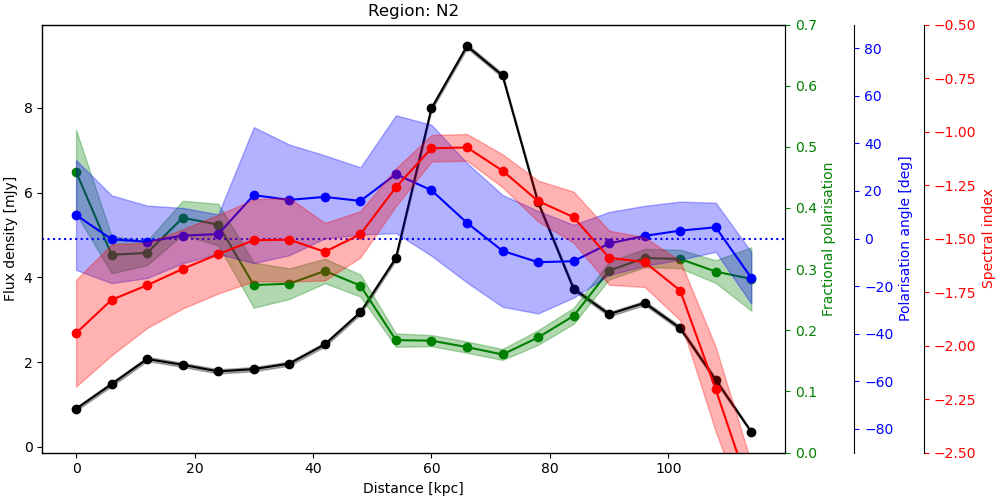}\\
\includegraphics[width=.49\textwidth]{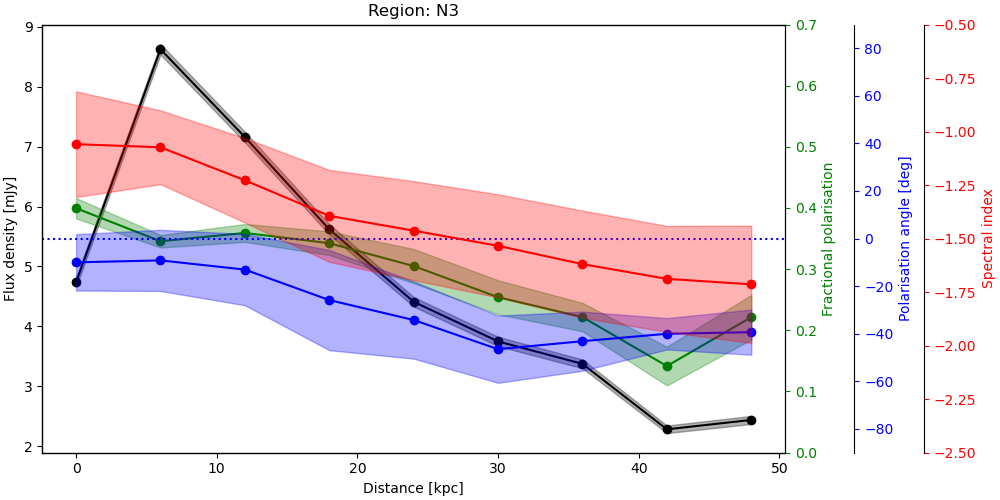}
\includegraphics[width=.49\textwidth]{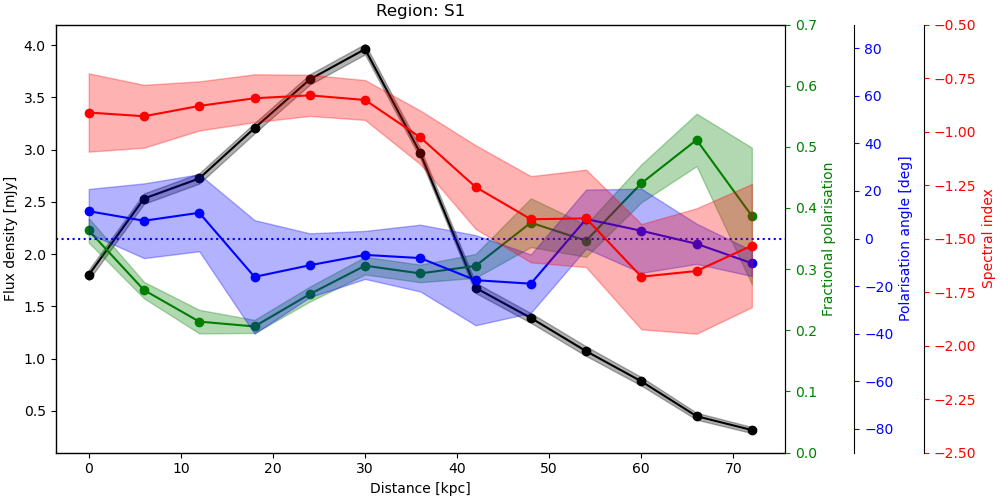}\\
\includegraphics[width=.49\textwidth]{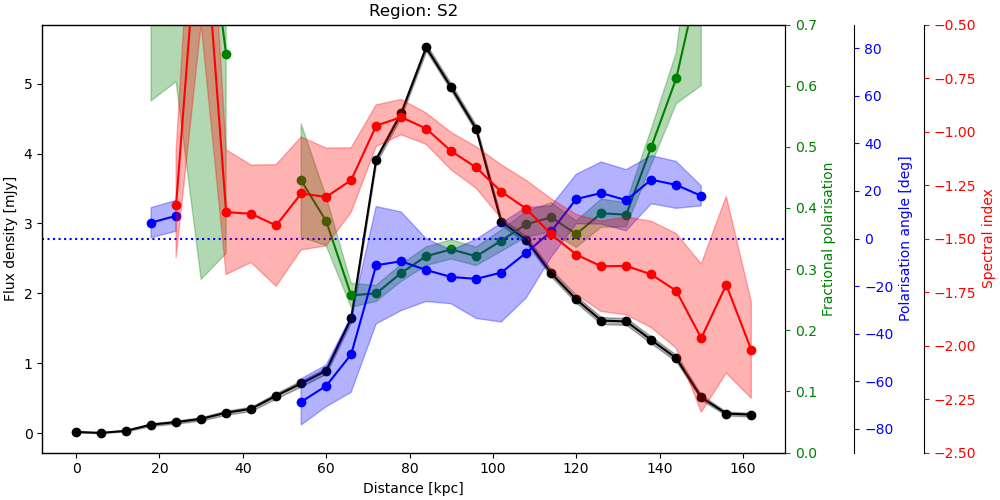}
\includegraphics[width=.49\textwidth]{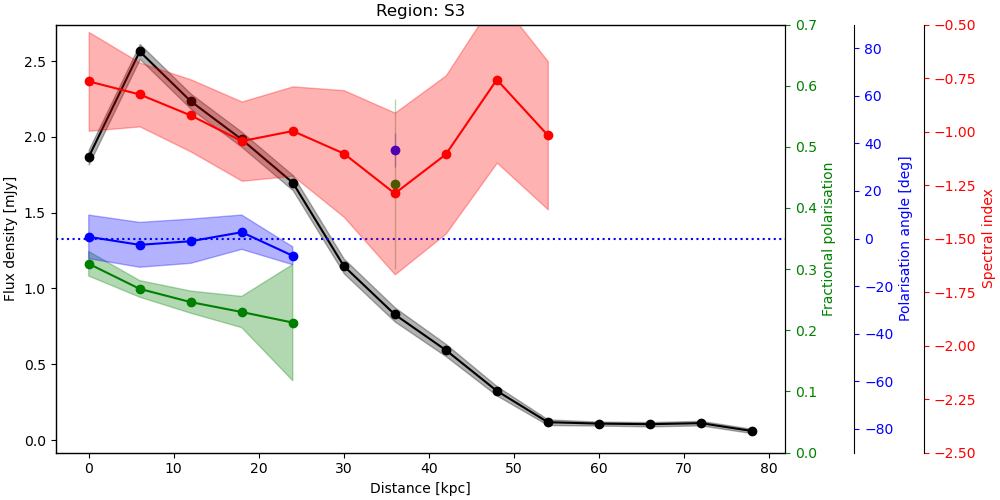}
\caption{Radio relic profiles. \textit{Top panel:} Definition of the regions used for profile extractions and in the discussion. \textit{Other panels:} Flux density, spectral index, fractional polarisation, and polarisation angle profiles along the regions specified in the top panel. The polarisation angle is relative to the orientation of the region, so  0\deg (dotted line) means a B-mode vector oriented parallel to the extension of the slice.}\label{fig:profile_relics}
\end{figure*}

Both radio relics in Abell 3667 present a complex structure, made by dozens of filaments, some only visible inspecting the polarisation maps or through spectral tomography (see below). Some filaments are thin, below our maximum resolution of 3\arcsec (3.2 kpc). Some are marginally resolved, on scales from a few to tens of kiloparsecs. In order to describe the relic features, we follow the labels used in Fig.~\ref{fig:profile_relics}. We selected six representative regions were we extract profiles of flux density, spectral index, fractional polarisation, and polarisation angle. Each region is used to extract the average values of the aforementioned quantities in slices of width of 12\arcsec, which is larger or equal to the resolution of the maps (10\arcsec{} for polarisation and total intensity maps, 12\arcsec{} for spectral maps). For the polarisation angle the average value is computed using a circular mean after rotating the coordinate system so that a magnetic field angle of 0\deg{} corresponds to the field being parallel to the slice extension. The error-bars for the polarisation angle represent the circular standard deviation of the values across the slice and gives an information on the uniformity of the magnetic field in each slice.

% profiles north
The NW relic is divided into four zones.\ Zone 1 is the main body of the relic and is edged by the X-ray shock front. Apart from the first 20 kpc, the region does not show a clear gradient of decreasing brightness and spectral index moving downstream from the shock front. As noted by \cite{Finoguenov2010} and \cite{Sarazin2016} the expected thickness of the radio relic can be estimated using

\begin{equation}
    d_{rad} = 70 \left(\frac{\nu_b}{1.4\ \rm GHz}\right)^{-1/2} \left(\frac{B}{3\ \rm \mu G}\right)^{-3/2}
    \left[\left(\frac{3.6\ \rm \mu G}{B}\right)^{2} + 1\right]^{-1} \rm kpc,
\end{equation}with $\nu_b$ the observing frequency and $B$ the local magnetic field. The predicted thickness of the relic is much smaller than what we measure ($\sim 0.5$ Mpc) and the relic's body has several flat-spectrum filaments crossing it, incompatible with a steady ageing of the electrons in the post shock flow. The outermost region, however, follows the expected behaviour of acceleration$\rightarrow$ageing behind the shock front, until other filaments start dominating the emission. This is visible in Fig.~\ref{fig:profile_relics} (region: N1), where the relic intensity peak at the shock front and decrease for a few kiloparsecs behind it. The decrease in the intensity is accompanied by a decrease in spectral index, likely due to the ageing of the accelerated particles, and a decline in fractional polarisation that might be due to a less ordered magnetic field, for instance due to an increasingly turbulent medium, coupled with beam depolarisation or projection effects. The magnetic field orientation is generally perpendicular to the shock normal, and it becomes mildly less uniform across the regions while moving away from the shock front.

The NW relic region 2 is dominated by a bright central filaments that is located more than 300 kpc behind the measured shock front location from X-ray data. Observing the profiles, the relic intensity appears to decrease gradually on both sides of the filament, the same is true for the spectral index, with the spectra getting steeper both moving away and towards the cluster centre. This is in contrast with the filament tracing a outward moving shock front. The magnetic field orientation is parallel to the filament extension and gradually deviate from this orientation when the distance from the filament increases. Finally, the fractional polarisation seems lowest on the filaments, and appears to increase at larger distances. This trend might be caused by a superimposition of emitting regions with different magnetic properties. Region 3 of the NW relic has a fainter average intensity and its edge seems to follow the continuation of the edge of region 1. The region is characterised by a bright filament, with a well defined edge and the magnetic field oriented along its extension. The intensity decreases gradually towards the cluster centre together with a steepening of the spectrum and a decrease in polarisation fraction. As discussed above, a decrease in polarisation fraction in a downstream direction can be related to a turbulent ICM. Nevertheless, the amount of depolarisation in the downstream region is mild, suggesting an upstream magnetic field with small fluctuations \citep{Dominguez-Fernandez2021, Dominguez-Fernandez2021a}. In general, all diagnostics point towards another region of shock acceleration. Finally, region 4 appears different from the others, with less bright filaments and a large component of diffuse emission. The spectral index is generally flat ($\alpha \sim -1$), mostly in the eastern part and show only a mild spectral gradient. This region may trace a part of the shock front seen face-on.

% profiles south
The SE relic is divided into three zones, the northern and southern zones have a sharp edge, are dominated by a filament, and show a slow decline of steep emission towards the cluster centre. Their profiles also show the typical characteristics of a shock front followed by a region characterised by ageing plasma. To note in region S1 is the slow rise of the radio intensity in the 30 kpc region in front of the bright filament. Region S2, on the other hand, behaves similarly to region N2, where the bright, flat-spectrum filament is surrounded on both sides by fainter, steeper spectrum emission.

% properties of filaments
The nature of filaments in radio relics is still debated \citep[e.g.][]{Rajpurohit2020}. One model suggests that a shock front with substructures with complex shapes can create thin filaments when seen in projection. The effect should be maximal when the shock front propagates in the plane of the sky. In this case the bright filaments would trace the underlying distribution of Mach numbers and consequently a range of electron acceleration efficiencies. This model is supported by recent simulations \citep[see e.g.][]{Dominguez-Fernandez2021a, Wittor2021}.

% filaments as high B
An alternative hypothesis is that filaments trace the strength and topological structure of the magnetic field within the relic region. The simplest assumption is that cosmic rays are accelerated only at the shock front and the filaments are due to local enhancements of the magnetic field in the downstream region. A high magnetic field $B$ results in a radio emission at a certain frequency $\nu_c$ being generated by electrons at lower energies $E$, with $\nu_c \propto E^2 B_\perp$. Since low-energy electrons are less affected by losses due to ageing, their spectra is still close to the injection spectra. Therefore, in this scenario and for a given age, the radio emission in regions with high magnetic field is dominated by electrons with a flatter energy distribution producing a flatter radio spectrum than those emitting in environments with lower magnetic fields. We simulated this scenario for the northern relic, assuming particle acceleration at the relic edge with an injection spectral index of $\alpha_{inj} = -1$ and an ageing caused only by unavoidable inverse Compton losses with an age $t=d/v$, where $d$ is the distance from the shock front and $v=530$~\kms{} is the downstream velocity. We then used the measured spectral index in each pixel to obtain a lower limit on the magnetic field in that place. We found that the filaments located downstream of the shock front need to have an extremely strong magnetic field ($B>100~\mu$G) to justify the measured spectral index without localised energisation of particles. Any realistic scenario that includes also synchrotron losses or a longer lifetime of particles due to projection effects, would further increase the minimum local magnetic field in the sampled filaments. This findings favour the interpretation of filaments as regions of particle acceleration seen in projection.

\subsubsection{Spectral analysis}

\begin{figure*}[ht]
\centering
\includegraphics[width=\textwidth]{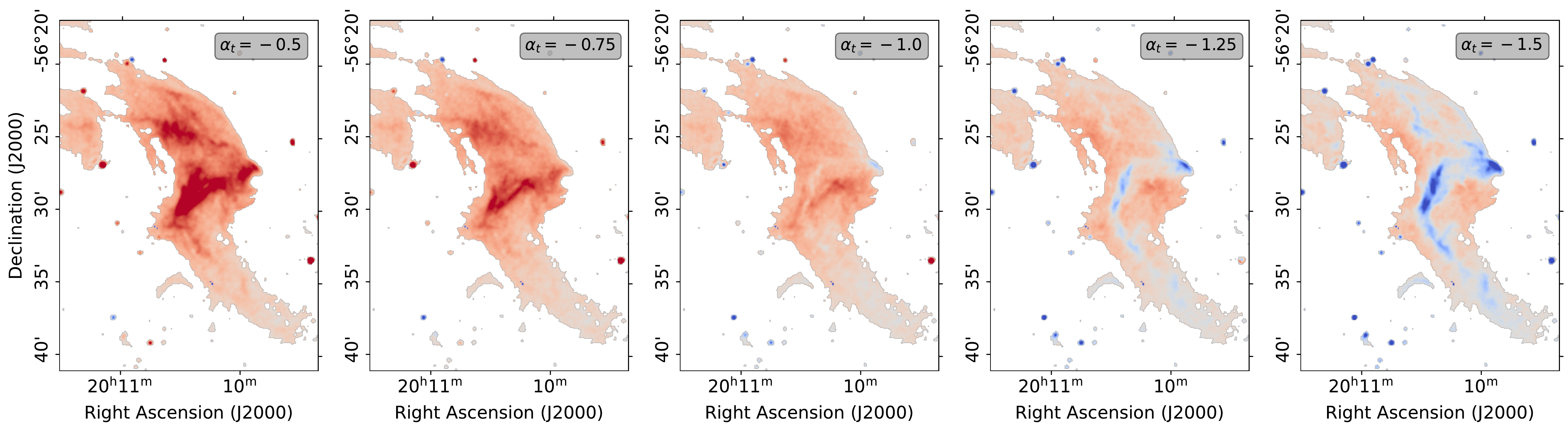}
\caption{Radio spectral tomography maps of the NW radio relic obtained using the $\alpha_t$ specified in the box. Red regions have a spectra steeper than $\alpha_t$, and blue regions have a spectra flatter than $\alpha_t$. The central panel shows evidence for filaments both steeper and flatter than $\alpha=-1$.}\label{fig:tomo}
\end{figure*}

\begin{figure}[tb]
\centering
\includegraphics[width=\columnwidth]{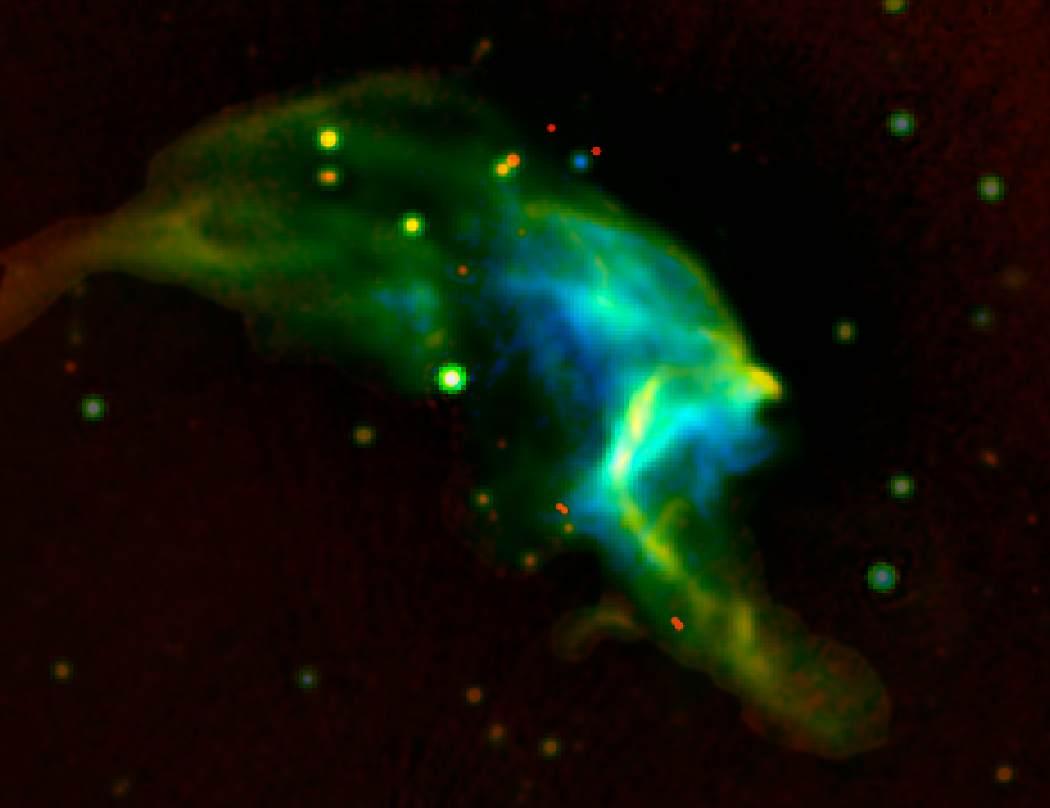}
\caption{Colour image at \beam{15.8}{15.8} resolution, created from a combination of tomography and total intensity images, qualitatively illustrating spectral variations among the filaments.  There is not a 1:1 correspondence between colour and spectral indices, but yellow and red regions are typically flatter than $\sim-1$, while blue regions are typically steeper than $\sim-1.5$.}\label{fig:tomo_multicolor}
\end{figure}

% spectral index + depol
Given the importance of the filament spectral indices, we visualised them in two new ways, using the spectral tomography images.  To first order, the spectral tomography map in Fig.~\ref{fig:tomo} shows that most of the bright filaments have spectral indices $\sim$-1, since they largely disappear against the background when $\alpha_t=-1$.
When the emission of the filament is removed in the tomography maps with $\alpha_t \lesssim -1$, the relic appears filled with a rather uniform, steep-spectrum emission. Going a step further and subtracting all flux with spectral index $\alpha>\alpha_t=-1.5$ reveals the region of the relic where the very steep spectrum emission is concentrated, which is in the downstream part of regions 1 and 3, in a large fraction of region 4, and, interestingly, on both sides of the main filament in the region of the notch. However, since it is difficult to picture what is ``missing'' from an image, we made an initial attempt at combining two different tomography maps with a total intensity map in a red-green-blue image, so that their differences would show up as features with different colours.  The results of this visualisation are shown in Fig. \ref{fig:tomo_multicolor} and show that different filaments indeed have different spectral indices.

\begin{figure}[ht!]
\centering
\includegraphics[width=.8\columnwidth]{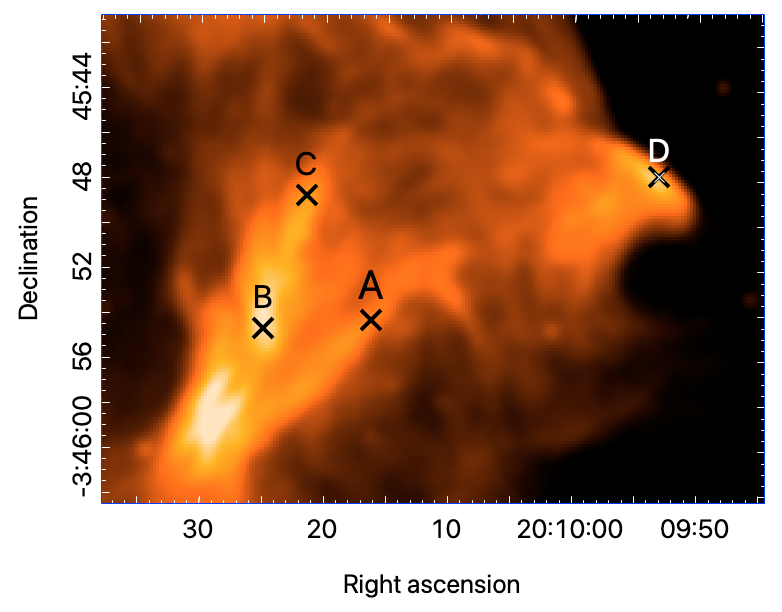}\\
\includegraphics[width=.8\columnwidth]{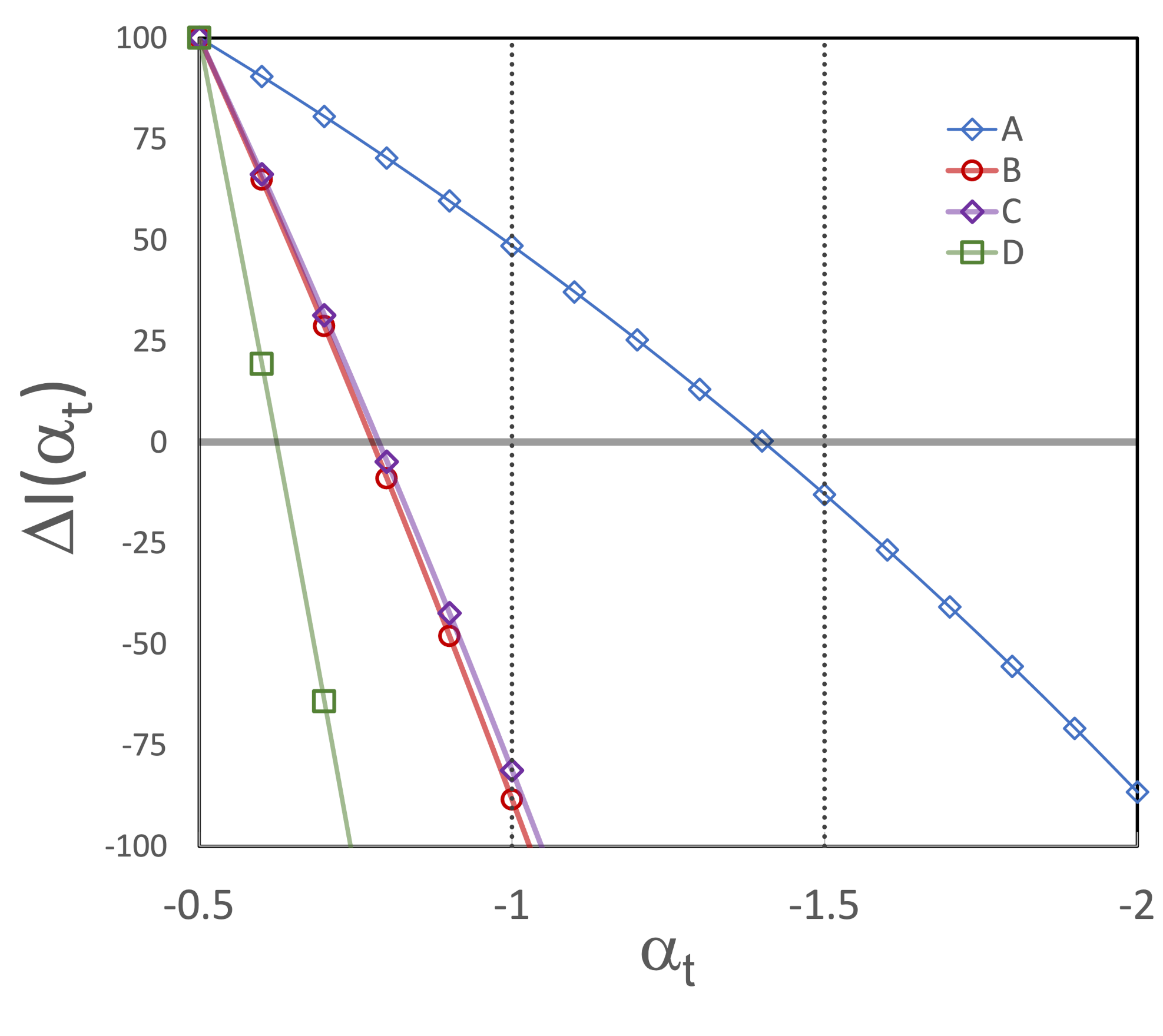}
\caption{Spectral index of the filaments. \textit{Top panel:} Positions for isolating filament spectra.  \textit{Bottom panel:} Background-subtracted flux for filaments as a function of tomographic spectral index. The brightnesses are all normalised to 100 at $\alpha_t$=-0.5 . The zero crossing points indicate the respective filament spectral indices when they disappear against the local background.}\label{fig:filaments}
\end{figure}

The results of a more quantitative, though limited examination of filament spectra are shown in Fig. \ref{fig:filaments}.  We have applied the tomography analysis in a somewhat new way here, similar to that illustrated in Fig. 7 of \citet{1999ApJ...516..716K}. We determine at what value of $\alpha_t$ the feature of interest goes through zero with respect to the local background.  We accomplished this by measuring, in each tomography image, the difference ($\Delta~I(\alpha_t)$) between the flux density of the filament and the average of the local background. We then plot $\Delta~I$  as a function of $\alpha_t$, as shown in Fig.~\ref{fig:filaments}.  Several filaments (two are shown here) were found to have spectra of $\alpha \sim -0.8$; one filament  has a significantly steeper spectrum, $\alpha =-1.4$, while the bright patch/filament located close to the bay has the flattest spectrum, with $\alpha = -0.6$ (see Sect.~\ref{sec:bay}). The range of indices could reflect local particle acceleration differences as predicted by simulations \citep[e.g.][]{Skillman2013}, or changes due to magnetic field strength with a curved underlying particle distribution, and a more detailed analysis would require data at other frequencies.

\subsubsection{Rotation measure analysis}

\begin{figure*}[!ht]
    \begin{minipage}{0.45\textwidth}
        \centering
        \includegraphics[width=\textwidth]{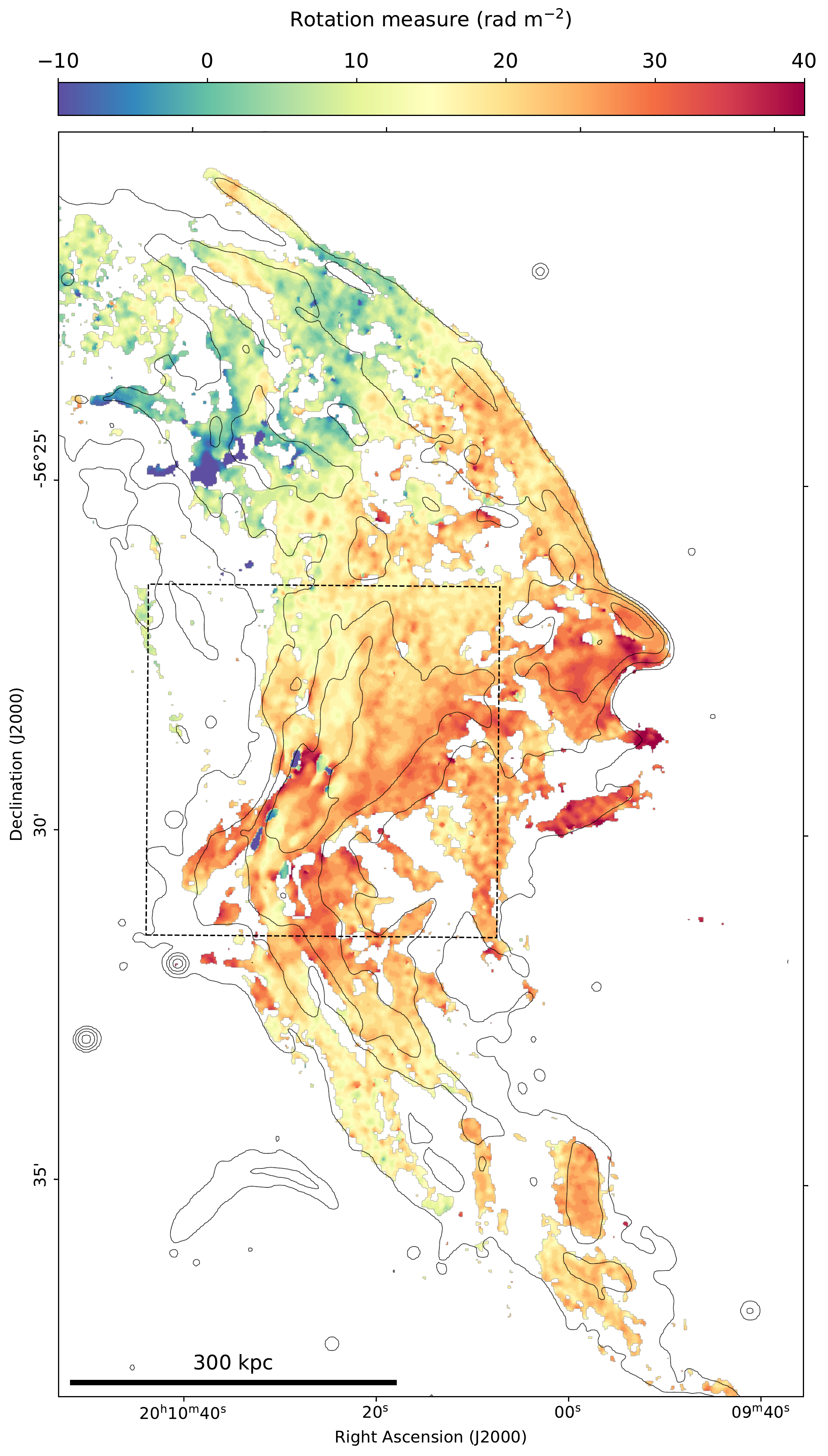}
    \end{minipage}\hfill
    \begin{minipage}{0.55\textwidth}
        \centering
        \includegraphics[width=\textwidth]{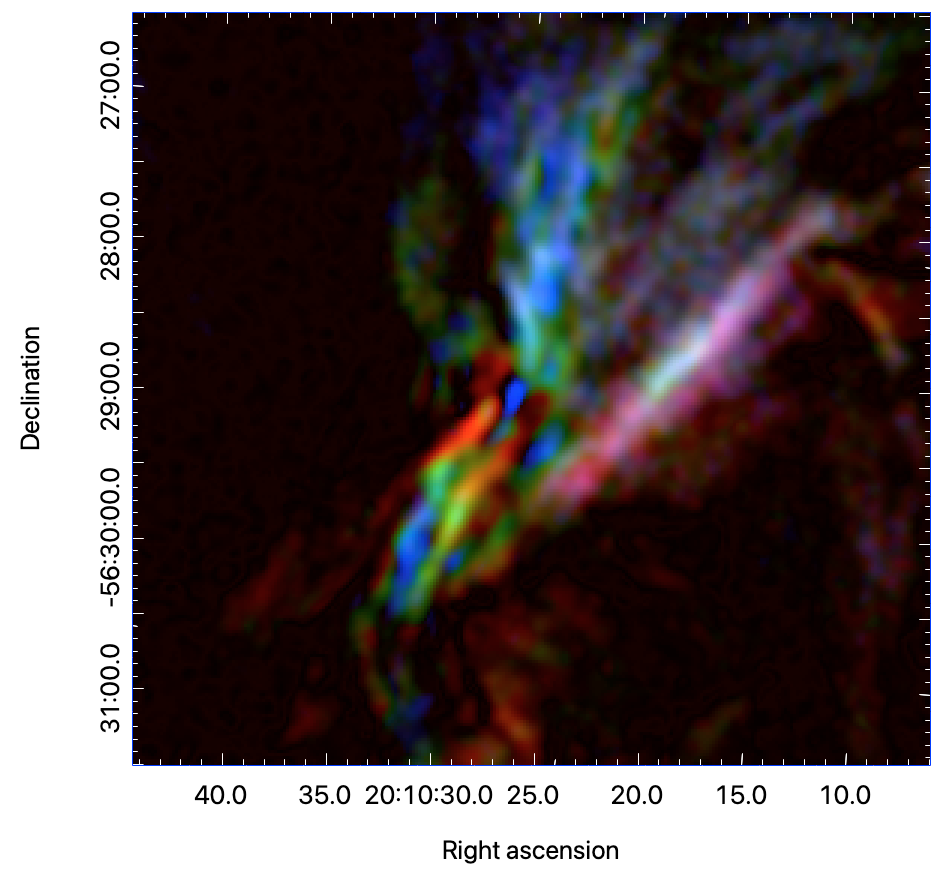}
        \caption{Rotation measure of the filaments. \textit{Left:} High spatial resolution (\beam{7.4}{7.3}) RM map. The local Galactic Faraday rotation has been subtracted ($\phi = 39.5$~\radmm). Black contours are the same as in Fig.~\ref{fig:spidx} but at levels $(20,50,100,200)\times\sigma$. \textit{Top:} More detailed look at the Faraday depth structure within the box of the left panel, using three planes from the deconvolved Faraday depth cube. After subtracting a Galactic contribution of 39.5 ~rad m$^{-2}$, the respective Faraday depths were red=8.5, green=20.5, and blue=32.5~\radmm. There is sometimes emission at more than one Faraday depth at the same location, producing other colours, but they do not correspond to a single Faraday depth (or RM).}
        \label{fig:rm2}
    \end{minipage}
\end{figure*}

%ROTATION MEASURE ZOOM
Figure~\ref{fig:rm2} shows a zoom-in of the Faraday rotation of the main filaments in the central part of the NW radio relic. These filaments are the brightest in the NW relic and show  large fractional polarisations.  The filaments have RMs that are coherent over scales much larger than their widths but which differ from one another. This is not what is expected from foreground Faraday rotating patches, which would impose random-looking patterns on the filament structure. These coherent RM structures, and gradients in them, likely reflect the intertwining of relativistic and magnetised thermal plasmas within the relic.  Similar behaviour is seen in the simulations discussed in Sect. \ref{sec:simul1}. 

The filament in the west of this figure appears enshrouded in higher Faraday depth material, and is the only one with a clear steep spectrum ($\alpha \sim -1.4$; see Sect.~\ref{sec:filaments}). It can be traced for $>$170~kpc in length; its width in total intensity is $\sim$13~kpc in the north, rising to $\sim$22~kpc in the south, where it is surrounded by faint material.  The large-scale coherent RMs and the differences between filaments in both RM and spectra suggest that the physical conditions  are different from place to place along the line of sight, although the filaments appear close in projection.  

We can  roughly estimate the magnetic field strengths in the thermal plasma of the relic, by looking at the differences in RM between the different filaments. Using the definition of Faraday depth \citep[e.g.][]{Brentjens2005b}, the electron densities ($n_e$) inferred from the X-ray observations, and RM variations $\sim$20~\radmm, we derive characteristic values of the magnetic field strength,

\begin{equation}
  B=0.15 \left(\frac{n_e}{1.5 \cdot 10^{-3}\ {\rm cm}^{-3}}\right)^{-1} \left(\frac{L}{100\ {\rm kpc}}\right)^{-1}\ \mu\rm{G},
\end{equation}where L is the assumed depth of the Faraday screen. In this situation, since we are looking at the differences between filaments, L would represent a characteristic separation between filaments along the line of sight, which we have scaled to 100~kpc.

We find a magnetic field strength $B\sim0.15\ \mu$G, which is consistent with those of typical clusters modelled by \citet{2019MNRAS.490.4841L} at $\sim$1~Mpc from their cores (see their Fig. 2).  To reproduce the behaviours seen in Fig. \ref{fig:rm2}, variations in magnetic field strength and orientation in the thermal plasma must be on similar scales to those of the bright filaments, $\gtrsim$100~kpc. We underline that the magnetic field strength {within} the filaments themselves can be significantly higher and in line with upper limits given by, for example, \citealt{Finoguenov2010}, who found $B>\ \mu$G. The overall picture is that radio relics are regions where most of the volume is permeated by magnetic fields with strengths not too different from surrounding unperturbed regions. Within this volume, filamentary structures, possibly coincident with regions of acceleration and therefore magnetic compression, have substantially higher magnetic fields. We look more closely at the Faraday structures expected in these intertwined plasmas in the simulations described in Sect. \ref{sec:simul1}.

\begin{figure}
\centering
\includegraphics[width=.95\columnwidth]{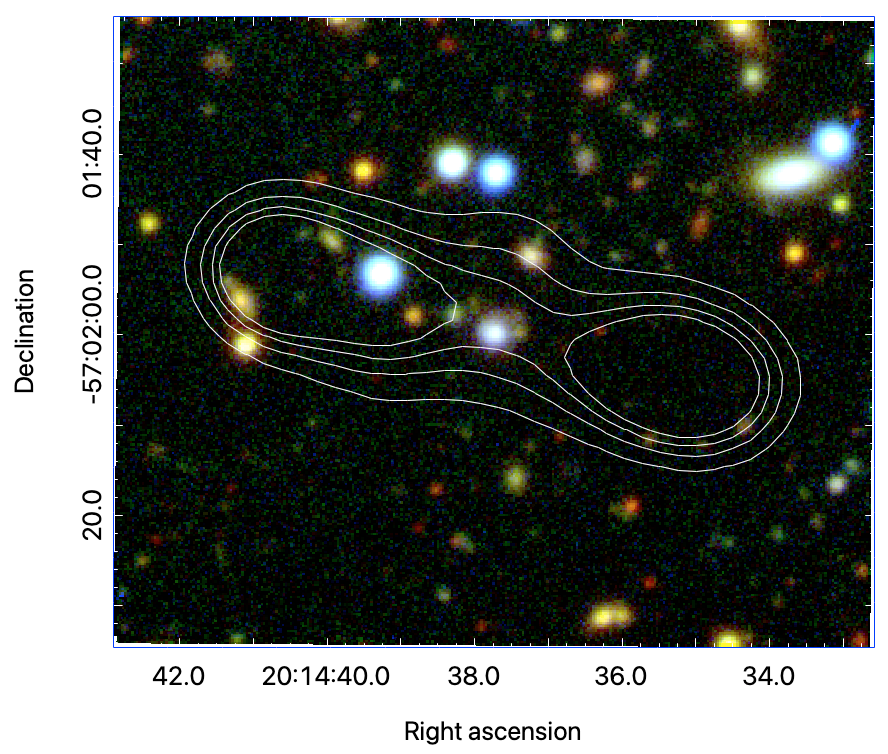}\\
\includegraphics[width=\columnwidth]{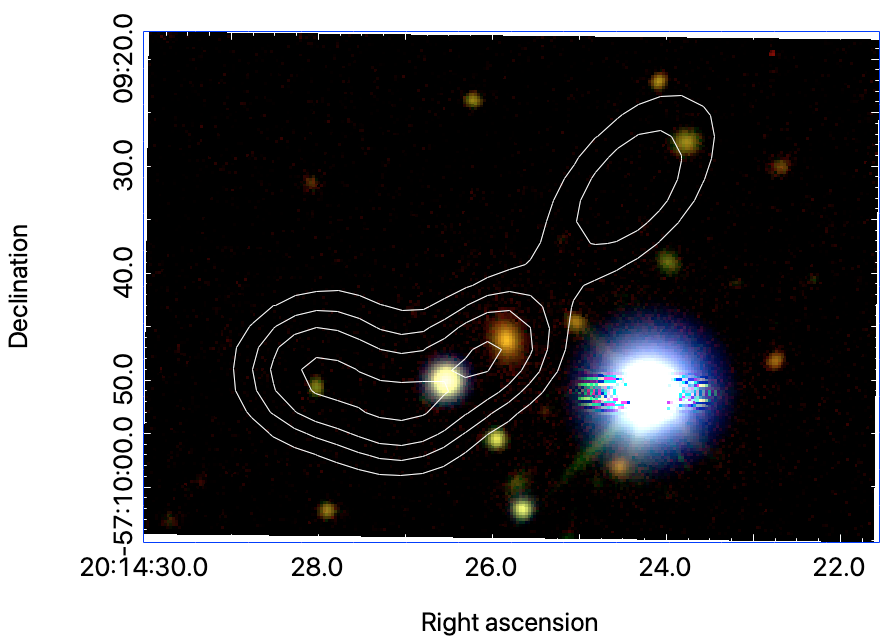}\\
\caption{Optical images from the Dark Energy Survey of two radio galaxies embedded into the emission of the SE radio relic. The contour levels in the top panel are $(0.4, 0.8, 1.2, 1.6)\,\times$~\mjybeam{} with beam \beam{10}{10}, and in the bottom panel they are $(0.2, 0.3, 0.4, 0.525)\,\times$~\mjybeam{} with beam \beam{10}{10}.} \label{fig:GalSErelic}
\end{figure}

\begin{figure}
\centering
\includegraphics[width=\columnwidth]{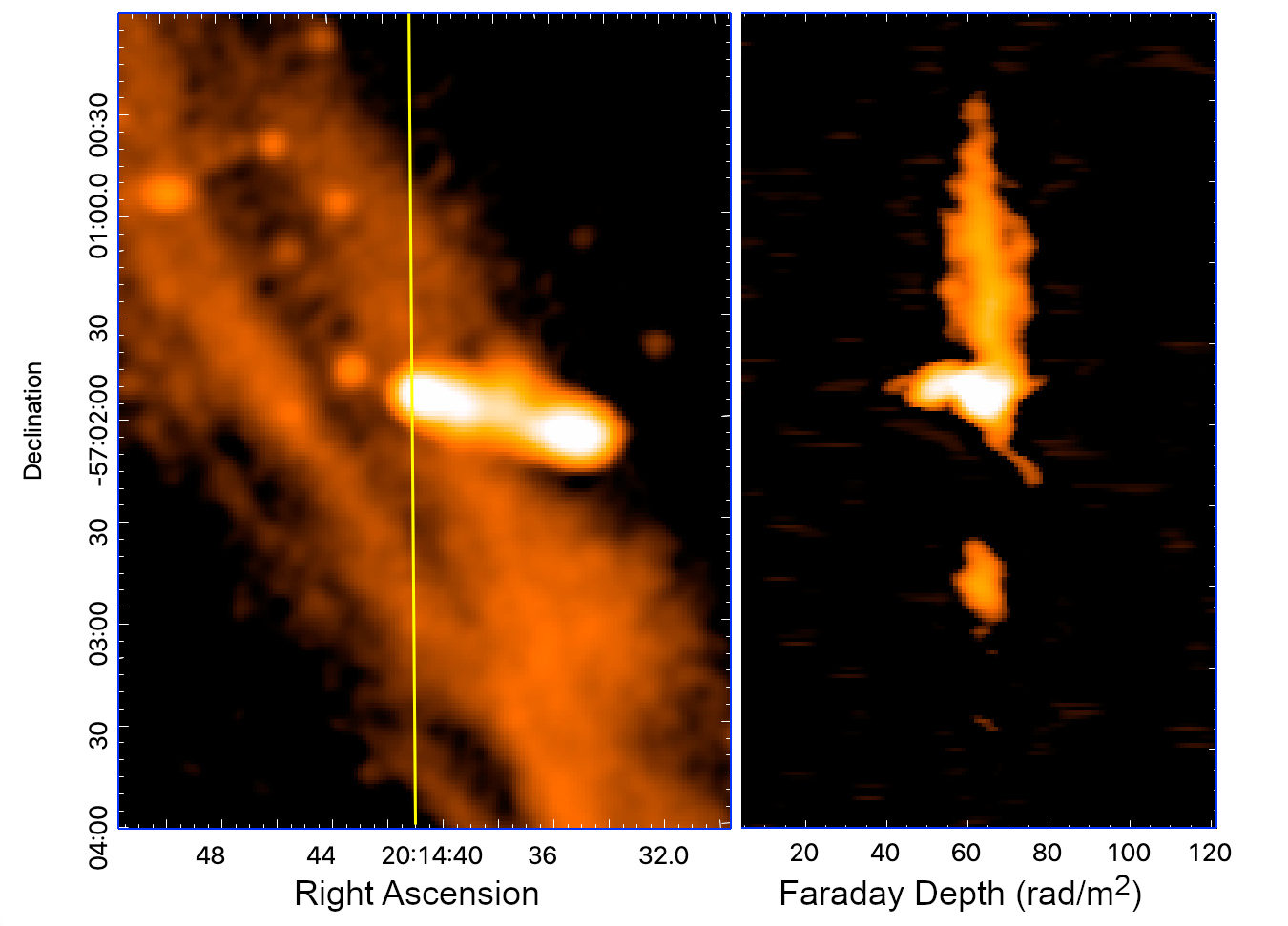}
\includegraphics[width=\columnwidth]{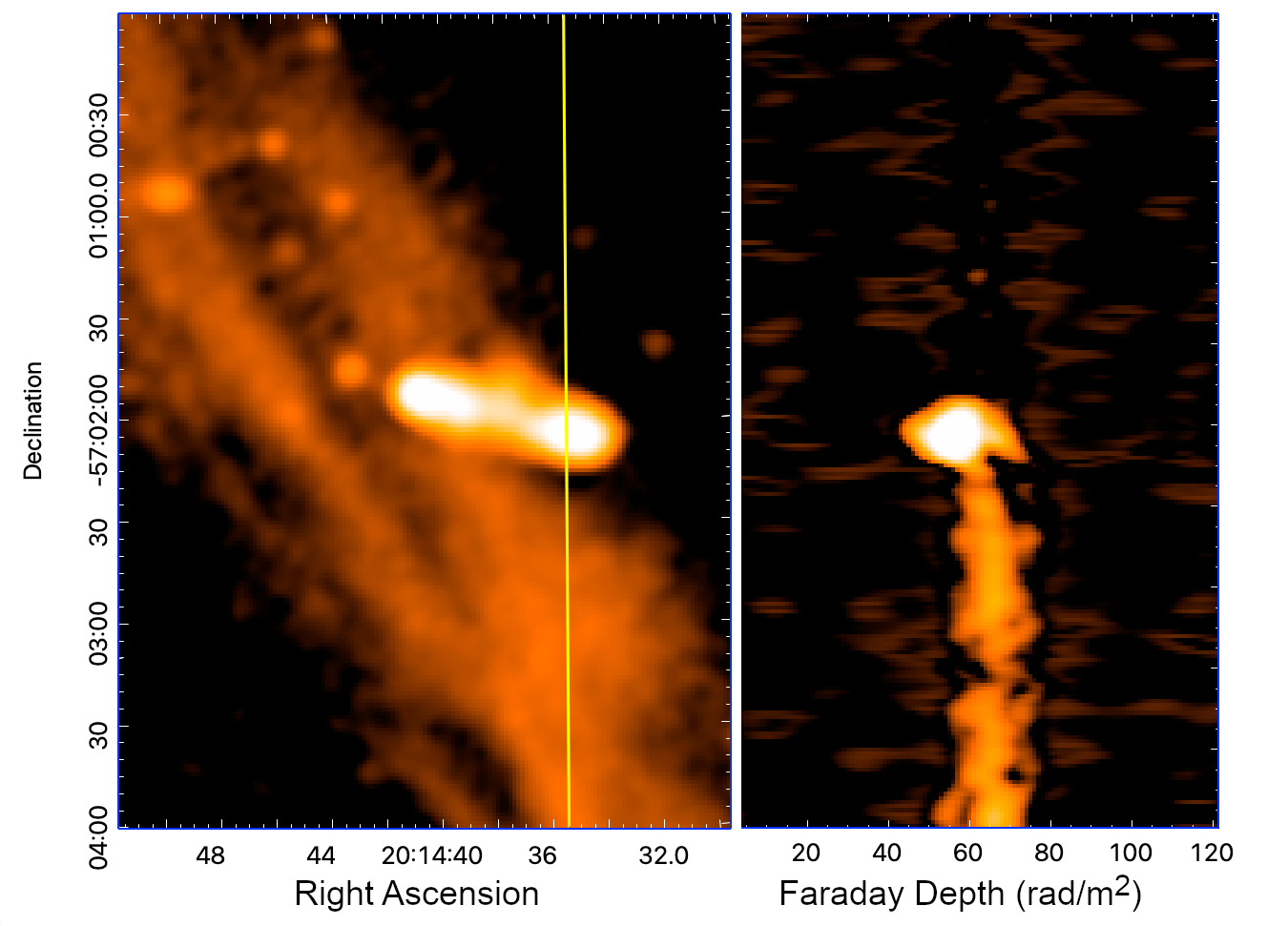}
\caption{Total intensity image, with the yellow line indicating a plane of interest in the RA, Dec, RM cube at fixed RA, shown to the left. The corresponding image on the right is that Dec versus RM plane (not corrected for Galactic rotation). The radio relic filament is the vertical stripe at approximately constant RM, and the bright spot is the double radio galaxy, showing complex RM structure (horizontal extension). } \label{fig:rmdouble}
\end{figure}

% other radio sources
We can also use polarised background radio galaxies to probe the Faraday structure of the relic.  There are two bright radio galaxies seen in superposition with the SE relic and visible in Fig.~\ref{fig:relicS}. The northern one (RA = \hms{20}{14}{37.8}, Dec = \dms{57}{02}{02.5}) is associated with WINGS J201433.69-570142 \citep{Moretti2017} at a redshift of $z=0.05895$ and is therefore a cluster member (see Fig.~\ref{fig:GalSErelic} for an overlay). It also is associated with the X-ray source 4XMM J201437.8-570158 \citep{Flesch2021}. In Fig.~\ref{fig:rmdouble} we take a closer look at the entire Faraday depth structure in this region. This is a new technique, enabled by the extended nature of the relic emission and the information from Faraday synthesis.  In this figure, we show two 2D cuts in declination versus Faraday depth from the Faraday spectrum, at fixed right ascensions on the eastern and western lobes. On the right panels, the vertical strip shows the Faraday depth of the relic, at a fairly constant value of 65~\radmm{} (25.5~\radmm{} after a galactic foreground subtraction). The eastern lobe of the double shows complex Faraday structure, implying that it has more than one RM (Faraday depth) component. If its associated Faraday medium were in the foreground of the relic, then we would expect to see the relic Faraday depth change as well, at that location.  Although the emission from the relic and the double are somewhat overlapping, it appears that the relic depth stays approximately constant. Therefore, the double source is likely behind the relic, and the variations in Faraday structure across the double could be due to the relic. Using the same calculation as above, with a change in RM of $\sim$20~\radmm, we derive magnetic fields of less than $1\ \mu$G. If the depth through the relic is much larger than 100~kpc, the inferred magnetic fields would be even smaller. The western lobe shows very little difference, so the post-relic fields may be $\lesssim 0.1\ \mu$G.

The second, bent, radio galaxy is RG3, seen projected against the middle of the SE relic (RA = \hms{20}{14}{25.847}, Dec = \dms{-57}{09}{46.86}). There are two potential hosts, the galaxy WISE J201425.90-570946.7 and the more compact 2MASS J201426.52-5709497 (see Fig.~\ref{fig:GalSErelic}). Neither of these has an available redshift. Although the polarisation drops dramatically over part of the structure of this galaxy, no change in RM from the local relic value is detectable at the level of $\sim 5$ rad m$^{-2}$, again implying very small magnetic fields.

We note that much higher values for the magnetic field in the NW relic, $3-5\ \mu$G have been claimed by \cite{Johnston-Hollitt2004} for two polarised background sources behind the diffuse relic emission. The data come from \cite{Johnston-Hollitt2003}. Source A3667\_{17} (at RA=\hms{20}{09}{05.368}, Dec=\dms{-56}{33}{26.76}) is reported to have an RM of $-108$~\radmm{} and source A3667\_A (at RA = \hms{20}{11}{09.272}, Dec = \dms{-56}{26}{59.59}) a value of $-175$~\radmm.  However, Fig.~3.4 in \cite{Johnston-Hollitt2003} show that the fits are quite poor, and we measure very different RM values of 50~\radmm{} and 43~\radmm{} for the two sources, respectively.  Our values are within the range of those observed for the relic emission, and would thus lead to magnetic field values at least an order of magnitude smaller.

\subsubsection{The bay}
\label{sec:bay}

\begin{figure}[ht!]
\centering
\includegraphics[width=\columnwidth]{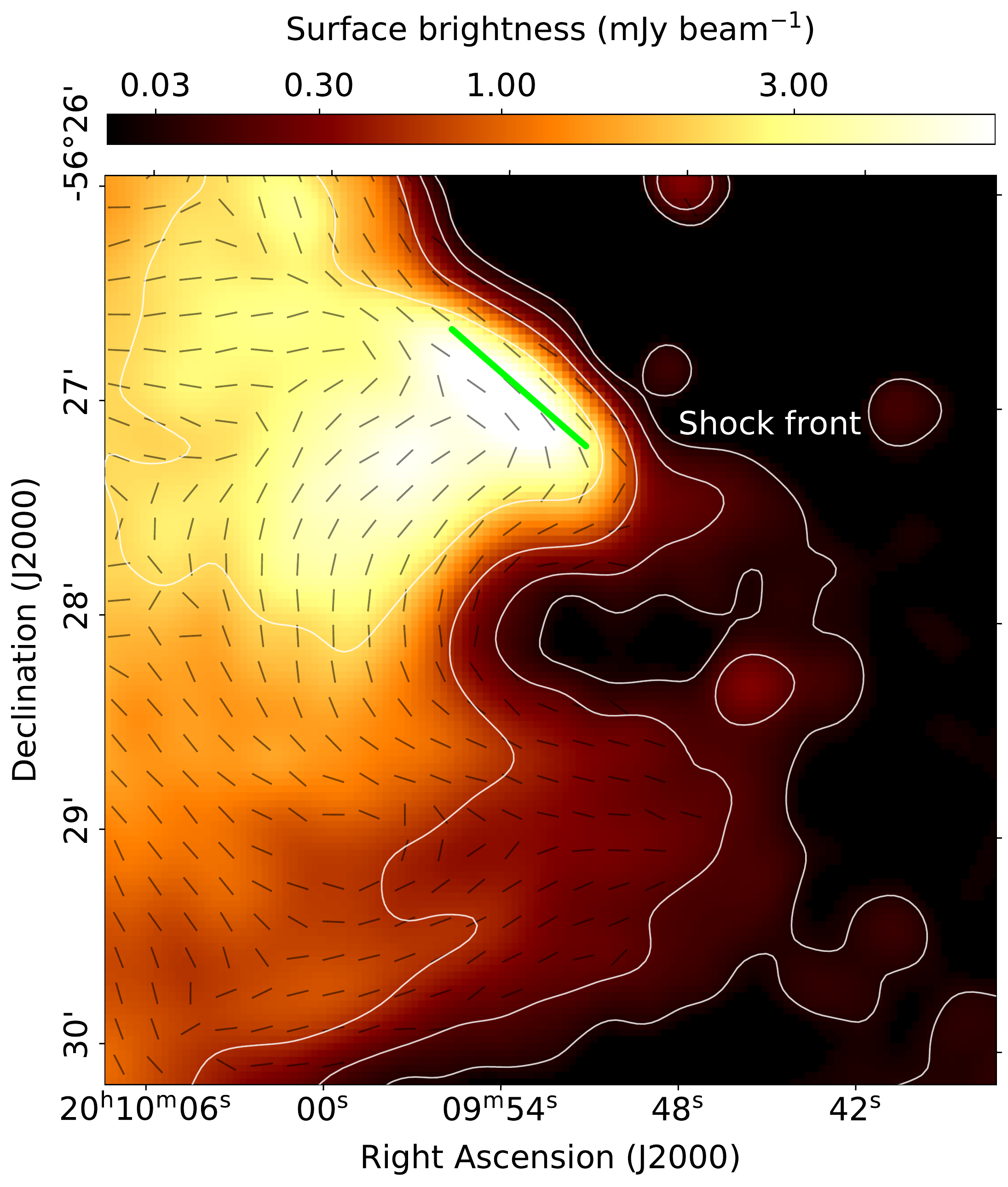}
\caption{Zoomed-in view on the bay region. Intrinsic polarisation vectors have been rotated by 90 degrees to show the local orientation of the magnetic field. Contours are at $(3,12,48,192) \times \sigma$ with local noise $\sigma=12$~\mujybeam and beam \beam{15}{15}. The green line shows the location of the shock front.}\label{fig:loop}
\end{figure}

We note the presence of a peculiar structure on the western side of region 1 of the NW relic (see Fig.~\ref{fig:loop}). We named the structure `Bay' following a similar example in the Toothbrush cluster \citep{Rajpurohit2020}. Similarly to this case, the bay in Abell 3667 is a bridge of faint emission that appear to extend out of the relic's external edge. However, in this case, the half of the loop that is on the relic side, seems to deeply affect its morphological and polarisation properties. This creates a hole with no detected emission, surrounded by the bay to the west and by a very clean semi-circular cut in the relic emission to the east, most likely driven by magnetic confinement. The bay region is too faint to be detected in polarisation, but on the relic side the magnetic field line follow a very precise semi-circular pattern.

The radio intensity and the spectral properties of the relic in the bay region appears to follow the normal behaviour associated with shock acceleration, with the shock being located north of the hole (see Fig.~\ref{fig:loop}) and a gradual decrease in both intensity and spectral index moving south. The emission at the shock front appear to have a flatter spectral index than the rest of the relic's edge, with $\alpha_{inj} = -0.7 \pm 0.1$. The unusual aspect is that the flow of aged plasma appears to carefully avoid the hole.

\subsection{The X-ray mushroom}
\label{sec:mushroom}

\subsubsection{Connection with the notch}

There are three regions of interest that are aligned along the radio halo major axis: the `edge of the mushroom', the bright filament of the relic to the south of the notch, and the `head of the halo' (see Fig.~\ref{fig:wide}). They are relatively close to one another, with the separation between the first two being 130\arcsec{} (139 kpc), and about 400\arcsec{} (428 kpc) between the others two.

\textit{XMM-Newton} observations show a mushroom consisting of a tail, with a flattened region at the top and possible vortexes at the two sides. The top edge of the mushroom is marked in Fig.~\ref{fig:wide}. Due to the low entropy and high X-ray emissivity, \cite{Sarazin2016} excluded that this structure is caused by buoyant gas. Most likely, the cool gas within the mushroom is the remnant of the cool-core of sub-cluster B moving through the atmosphere of the main sub-cluster. About 130\arcsec{} above the mushroom head the temperature profile shows a decrease that could indicate the presence of a shock at that location. That would be co-located with the bright filament of radio emission in the southern part of the NW relic. However, the X-ray surface brightness profile does not show any signature of a shock at that location.

In general, the nature of the Notch is still unclear: the radio properties of the bright filament are dissimilar to those typical shock edges (see Sect.~\ref{sec:filaments}) and no consistent X-ray discontinuities have been found. However, its location right on top of the mushroom and consequently at the end of the elongated radio halo is unlikely to be casual. A possibility is that the shock that generated the main body of the NW relic has encountered a different environment during its development: a region towards the north where it could expand freely, and one aligned with the merging axis where possibly more material is infalling from a filament oriented as the one that originally linked the two main sub-clusters of Abell 3667. 

\subsubsection{Magnetic draping}
\label{sec:draping}

\begin{figure}[t]
\centering
\includegraphics[width=\columnwidth]{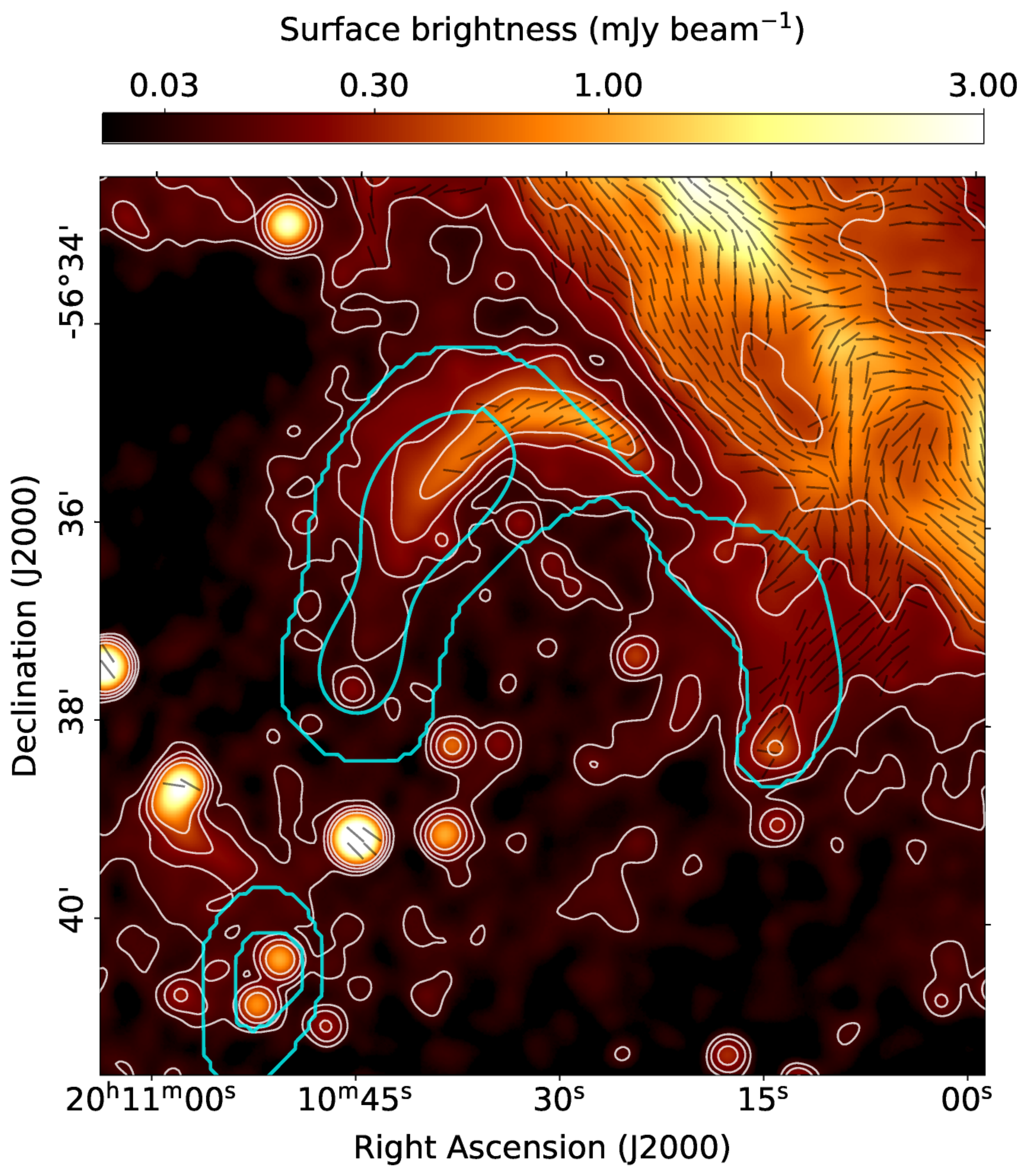}
\caption{Thin layer of enhanced polarised radio emission surrounding the head of the mushroom, possibly caused by magnetic draping. The X-ray mushroom is outlined by the cyan lines obtained by applying wavelet filtering to a background- and point-source-subtracted \textit{XMM-Newton} image in the 0.5--2 keV energy band, combining MOS and pn data. The intrinsic polarisation vectors are rotated by 90 degrees to show the local orientation of the magnetic field. White contours are at $(5,10,20,40) \times \sigma$ with local noise $\sigma=12$~\mujybeam and beam \beam{15}{15}.}\label{fig:boomerang}
\end{figure}

A zoomed-in view of the mushroom region is presented in Fig.~\ref{fig:boomerang}, where the edge of the X-ray mushroom is outlined by the cyan lines. These contours trace the X-ray emission from a background-subtracted and point-source-removed map \citep{Finoguenov2010}. The point source and background-subtracted image has been wavelet filtered keeping the scales 32\arcsec and 64\arcsec. A discontinuity in the X-ray profile at the mushroom edge suggests the presence of the contact interface between the cool core gas of the merging sub-cluster B and the surrounding hot gas from the main sub-cluster. While no shock is detected from the temperature profile, the speed of the mushroom cold front relative to the gas ahead is measured to be $1340^{+250}_{-140}$~\kms, close to the sound speed \citep{Sarazin2016}.

% magnetic draping
As the cool core of the moving sub-cluster B moves like a projectile through the ambient ICM, it produces several effects. To understand these effects, we first measured the density in the region preceding the cold front finding $n_e = 1.5 \times 10^{-4}$~cm$^{-3}$. Assuming a local magnetic field of $3~\mu$G, the Alfvén velocity is
\begin{equation}
v_A \approx \left(\, 2.18 \times 10^{6} \,\mbox{km s}^{-1}\right)\,\left(\frac{n}{\, {\rm cm}^{-3} \,}\right)^{-\frac{1}{2}} \left(\frac{B}{\, \rm G \,}\right)=  534~\mbox{km s}^{-1}.     
\end{equation}
The motion of the projectile is therefore super-Alfvénic. A  solid object moving super-Alfvénically in a magnetised medium can sweep up a significant magnetic layer that is compressed and bent (`draped') over the projectile \citep[e.g.][]{Bernikov1979}. The same process has been proposed to stabilise the boundaries of over-dense regions, such as cold fronts in a galaxy cluster \citep{Vikhlinin2001, Asai2004}, as well as boundaries of under-dense regions, such as bubbles generated by active galactic nuclei \citep{Lyutikov2006, Dursi2007} over dynamically long times. The process of `magnetic draping' has been discussed with analytical and numerical predictions \citep{Lyutikov2006, Dursi2007, Dursi2008} and its effect has been linked with observable features in the Faraday rotation of active galactic nucleus lobes \citep{Guidetti2011}. However, the direct observation of the magnetic draping in clusters is not trivial due to the thinness of the magnetic layer that cannot be resolved at the typical cluster distance with current X-ray instruments. An observation of the phenomenon has been claimed in the case of the bending in the jets of a radio galaxy due to the presence of strong magnetic fields around a cool core remnant in Abell 3376 \citep{Chibueze2021}.

% draping effect and radio emission
In the case of Abell 3667, the motion of the cold front in the magnetised plasma of the post-shock flow will compress and bend the magnetic field lines around the `cap' of the mushroom, thus amplifying the ambient magnetic field at the interface. Moreover, the cold front is moving through the downstream plasma of the preceding shock. This plasma contains the shock-accelerated cosmic ray electrons that have lost only part of their energy though radiative processes. Since the edge of the mushroom is a contact interface between gases of different origins, relativistic electrons would not leak into the mushroom unless there was significant diffusion or mixing at its upper surface. This is inhibited by the short diffusion lengths \citep{Finoguenov2010}. Moreover, the clean X-ray edge suggests that the interface is stable and no mixing occurs on the scales that we can resolve. As the electrons cannot penetrate into the mushroom, they accumulate at the front and are advected along its edges.

The combination of a compressed magnetic field with the presence of the relativistic electrons should produce a radio source around the mushroom cap. This source should follow the cap's edge and is expected to brighten where the projection of the shell increases the column density of emitting electrons, as in the case of shell-type supernova remnants. Furthermore, the emission is expected to be polarised along the magnetic field lines that follow the mushroom edges. Finally, if the thickness of the magnetic layer is not much lower than our resolution element, a gradient in intensity and spectral index might be visible when moving away from the interface as the magnetic field intensity decreases to the background level of the post shock plasma.

In Fig.~\ref{fig:boomerang}, we see that the top edge of the mushroom is limited by radio emission, in the region were the magnetic draping predicts most of the accumulation of magnetic field lines and the consequent strengthening of the magnetic field. The source is bent backward wrapping around only about half of the high-density region traced by X-ray observations, meaning that some projection effect might hide the real, likely complex, shape of the projectile. The radio source does not have an optical counterpart. The emission is stronger on the NE side, and mirrored by two possible counterparts on the SW side of the mushroom cap, one following the X-ray profile, one fainter and more internal. The radio source appears polarised, although not simply along its extension as the draping model would predict. However, the fractional polarisation of the region is low ($<20\%$) and the error on the polarisation angle can be significant. The NE side of the cap is the brightest, and the emission declines rapidly on the concave part, while it decreases gradually on the convex part, excluding the acceleration from a shock front moving outwards. The region at lower surface brightness appear to have a steeper spectra (see Fig.~\ref{fig:spidx}). \cite{Dursi2008} shows that the thickness of the magnetic layer $l$ is related to the Alfvénic Mach number through

\begin{equation}
l =  \frac{1}{6 \mach_A^2} R ,
\end{equation}where $R$ is the projectile radius, $\mach_A = u/v_A$ is the Alfvénic Mach number of the core and $v_A^2 = B_0^2/4\pi\rho_0$ is the ambient Alfv\`en speed with $B_0$ and $\rho_0$ the ambient magnetic field strength and density. If the thickness of the layer is close to our maximum resolution of $3.3\arcsec=3.5$~kpc, as suggested by its gradual decrease in intensity and steepening of the spectra, we can invert the above equation to estimate the ambient magnetic field to be $B_0\lesssim 5~\mu$G. Finally, in order for the mechanism of magnetic draping to work, the correlation length of the magnetic field needs to be larger than the size of the sub-cluster, which is about 300 kpc \citep[e.g.][]{Ruszkowski2007}. If the origin of this radio source is be confirmed, this is the first case where we can detect magnetic draping in a system of merging clusters.

\subsection{Origin of the linear radio halo}
\label{sec:origin_halo}

% morphology
\cite{Carretti2013} first discovered the linear radio halo in Abell 3667, suggesting that it was a case of a radio bridge \citep[similar to the one present in the Coma cluster;][]{Bonafede2021} whose origin was related to the mechanism creating the relics. In a few other clusters the radio emission from the halo appears bounded by merger shock fronts \citep{Markevitch2005, Brown2011, Markevitch2012, Vacca2014, Shimwell2014}. A famous example is the Toothbrush cluster \citep{vanWeeren2016, Rajpurohit2018, deGasperin2020a}. It has been suggested that the shock might pre-accelerate the electrons that are subsequently re-accelerated by the merger turbulence. However, in some cases the shock front appears unpolarised \citep{Shimwell2014} and in the Toothbrush cluster images at ultra-low frequencies (50 MHz) showed that the halo and the relic emission are likely superimposed in projection rather than blended together \citep{deGasperin2020a}.

In Abell 3667, the linear halo does not seem to link both radio relics. It is only connected to the western part of the NW relic, in the region of the Notch. The radio halo seems to be correlated with the X-ray emission (see Fig.~\ref{fig:wide}). Its SE edge coincides with the cold front associated with the main sub-cluster core \citep{Vikhlinin2001}. The radio halo then follows the trail of X-ray-emitting gas up to the edge of the mushroom described by \cite{Sarazin2016}, which is likely the remnant of the core of sub-cluster B (see also Sect.~\ref{sec:draping}).

\begin{figure*}[ht]
\centering
\includegraphics[width=\textwidth]{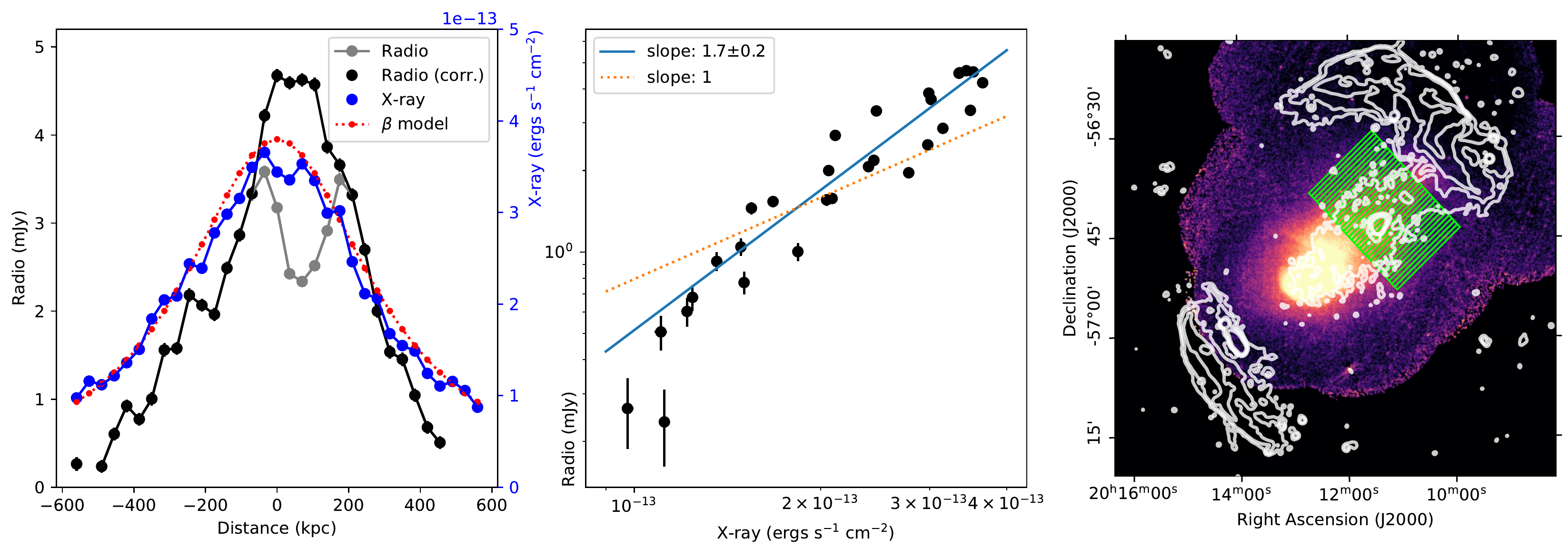}
\caption{Analysis of the linear radio halo. \textit{Left panel:} Radial profile of the X-ray intensity (blue) and the $\beta$ model described in the text. The radio profile with the pixels belonging to the bright radio galaxy (RG1) blanked is shown in grey. We corrected for the missing pixels assuming their value to be equal to the mean of the other pixels in the same region (shown in black). \textit{Central panel:} X-ray--radio intensity relation with an orthogonal distance regression linear fit (solid line) that includes error on both variables. \textit{Right panel:} In green, the regions used to evaluate the profile (background and contours are the same as in Fig.~\ref{fig:xmm}).}\label{fig:profile_trail}
\end{figure*}

% head of the halo
In the region of the Notch, we find a patch of extended, smooth radio emission. The average surface brightness is similar to that of the radio halo and the location is fairly aligned with the radio halo extension, as if this region were a continuation of the radio halo on the other side of the radio relic's southern arm. The diffuse emission in this region seems to end abruptly at a radial distance similar to the one of the detected shock front generating the main body of the radio relic. Furthermore, the emission in the NW edge of the region (called `head of the halo' in Fig.~\ref{fig:wide}) is marginally enhanced, it has a curved shape and a spectral index of $\alpha \sim -1$ (see Fig.~\ref{fig:spidx}). Also, it is polarised with the magnetic field vectors aligned with the edge (see Fig.~\ref{fig:pola2}). While these features indicate shock acceleration, we note that the signal in this region has a low significance.

% acceleration mechanisms, excluding hadronic model
The absence of a surface brightness gradient along the halo extension suggests the presence of \textit{in situ} acceleration of the cosmic ray electrons.
%Historically, two theories have been proposed for the origin of the synchrotron radiation of radio halos: the turbulence reacceleration and the hadronic models. In the former scenario, CR electrons are energised by Fermi II type acceleration in the turbulence injected in the ICM during merger events \citep{Brunetti2001, Petrosian2001}. According to hadronic models instead, the radio emitting electrons are generated as secondary products of inelastic collisions between CR protons and thermal protons in the ICM \citep{Dennison1980, Blasi1999a}. In both cases, the net result is the production of CR electrons (and positrons in the hadronic model) that produce synchrotron radiation. The hadronic model predicts an emissivity that increases with the density of the thermal protons \citep[see e.g.][]{Brunetti2017}. 
In Abell 3667, the ICM density varies by a factor of four along the halo while the radio surface brightness remains nearly constant. Assuming a cylindrical symmetry for the source, this also implies a constant emissivity, which favours a turbulent re-acceleration origin \citep{Brunetti2001, Petrosian2001} over a hadronic origin for the radio halo.

% turbulence model
Elongated radio halos have been found in some clusters of the MGCLS \citep{Knowles2021} and in a few famous clusters, for instance, in `El Gordo' \citep{Lindner2014}, MACS J1752.0+4440 \citep{Bonafede2012a}, MACS J1149.5+2223 \citep{Bruno2021}, SPT-CL J2023-5535 \citep{Hyeonghan2020}, and Abell 2142 \citep{Venturi2017}. In two of these systems, MACS J1149.5+2223 \citep{Bruno2021} and Abell 2142 \citep{Markevitch2000}, as well as in Abell 3667, cold fronts have been detected and the cool core of the main cluster has not been disrupted. Mergers with low mass ratios (1:5 or less) and low impact parameters can produce turbulence in the wake of a bullet-like sub-cluster. Such conditions are met in Abell 3667, where sub-cluster B merged from the south-east, without disrupting the main cluster core. In the process, it leaves a trail of X-ray-emitting gas, likely the remnant of the merging sub-cluster core. The trail is coincident with the linear radio halo. The X-ray and the radio emission are well correlated along the minor axis of the trail (see Fig.~\ref{fig:profile_trail}). The X-ray transverse profile follows a $\beta$-model \citep{Cavaliere1978} with $\beta = 2/3$ and an effective radius of 319 kpc. The radio profile has a steeper slope so that the X-ray -- radio brightness correlation is super-linear, with slope $1.7\pm0.2$. Super-linear relations between X-ray and radio brightness were also found for a sample of mini halos \citep{Ignesti2020}, while giant radio halos typically show a linear or sub-linear relation \citep{Govoni2001, Feretti2001, Giacintucci2005, Vacca2010, Hoang2019, Botteon2020a}.

If the linear halo traces the turbulence in the wake of the merging sub-cluster, its narrow size implies a high Reynolds number flow \citep[see e.g.][]{Ormieres1999}. Flow past spheres has been studied extensively using experiments and hydrodynamical simulations. It is generally understood that at Reynolds numbers $> 800$ the wake flow becomes turbulent \citep{Sakamoto1990}. With increasing Reynolds number the vortex structure in the wake becomes increasingly irregular until it becomes fully turbulent. Assuming that the radio emission in the linear halo is powered by turbulence produced by wake flow, this suggests Reynolds numbers of the ICM of $> 800$, in agreement with arguments given by, for example, \cite{Miniati2015}. For Abell 3667, the study of the cold front property along its azimuthal angle also suggests high Reynolds number \citep{Ichinohe2017}.

% comparison with other halos
Finally, we compared the properties of the radio halo in Abell 3667 with the known population of radio halos. Assuming a cylindrical geometry, we used the flux density extracted from the radio halo region\footnote{We excluded all the pixels contaminated by the bright radio galaxy (RG1) and re-scaled the final flux to compensate for the missing pixels.} to compute its average emissivity. Typical giant radio halos have emissivity at 1.4 GHz $J_\nu \sim 10^{-42}$~ergs s$^{-1}$ Hz$^{-1}$ cm$^{-3}$ \citep{Cuciti2021a}. We re-scaled the radio halo flux density measurement to 1.4 GHz assuming a spectral index $\alpha=-1.3$, obtaining an emissivity $J_\nu = (2.9 \pm 0.2) 10^{-43}$~ergs s$^{-1}$ Hz$^{-1}$ cm$^{-3}$. This is a factor of three to four lower than the average halo emissivity and lower than any emissivity measured in the sample of \cite{Cuciti2021a}. The luminosity of the radio halo, re-scaled to 1.4 GHz, is $L_r = (6.8 \pm 0.4) \times 10^{30}$~ergs s$^{-1}$ Hz$^{-1}$ --- about a factor of two lower than what predicted by the power -- mass relation for radio halos, putting the halo of Abell 3667 in a region typically populated by ultra-steep spectrum radio halos \citep{Cuciti2021a}. Unfortunately, due to the large extent of the source, we could not derive a reliable measure of the spectral index.

\subsection{Comparison with simulations}
\label{sec:simul}

We have compared the morphology, polarisation fraction, and spectral properties of the main body of the NW relic (region 1) with simulations. The main objective was to try to reproduce some of the relic's properties, such as its filamentary structure, and infer their physical origins. When studying the filamentary structures of radio relics, various physical scales have to be considered. Achieving a large range of scales in a single simulation is still technically challenging. Hence, we use two type of simulations for the comparison. The first one is a large-scale cosmological MHD simulation (see Sect. \ref{sec:simul1}). This simulation allows the effects of the cluster dynamics to be properly modelled. However, this simulations only resolves the physics of the ICM on scales of $\sim 12$ kpc. Hence, the second comparison we make is to a hybrid MHD-Lagrangian simulation (see Sect. \ref{sec:simul2}). This simulation resolves the ICM on a kiloparsec scale and, hence, captures the relevant MHD processes while neglecting the global cluster dynamics.

\subsubsection{Cosmological MHD simulations}
\label{sec:simul1}

\begin{figure*}[tb]
    \centering
    \includegraphics[width=\textwidth]{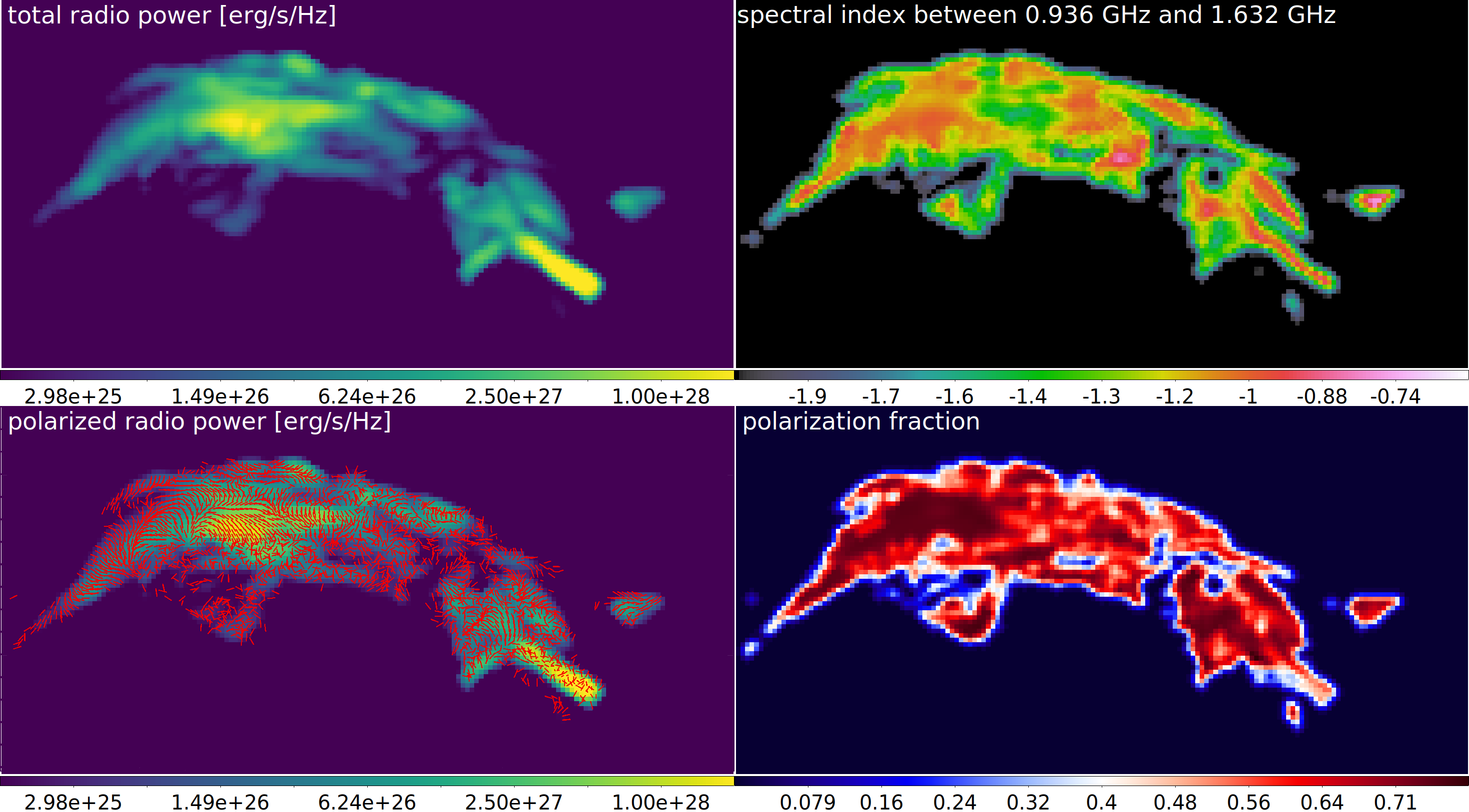}
    \caption{Maps of the simulated relic. \textit{Top left:} Radio power at 1.4 GHz. \textit{Top right:} Spectral index measured between 4.85 GHz and 140 MHz. \textit{Bottom left:} Polarised radio power at 1.4 GHz overplotted with the magnetic field angle (B vectors). \textit{Bottom right:} Polarisation fraction at 1.4 GHz. In each map, the resolution is $(12.6292\ \mathrm{kpc})^2$ per pixel, which corresponds to 13.5\arcsec{} at the distance of Abell 3667. Each map was smoothed with a Gaussian of size $\sigma = 27"$.}
    \label{fig:simulation}
\end{figure*} 

\begin{figure}[tb]
    \centering
    \includegraphics[width=0.5\textwidth]{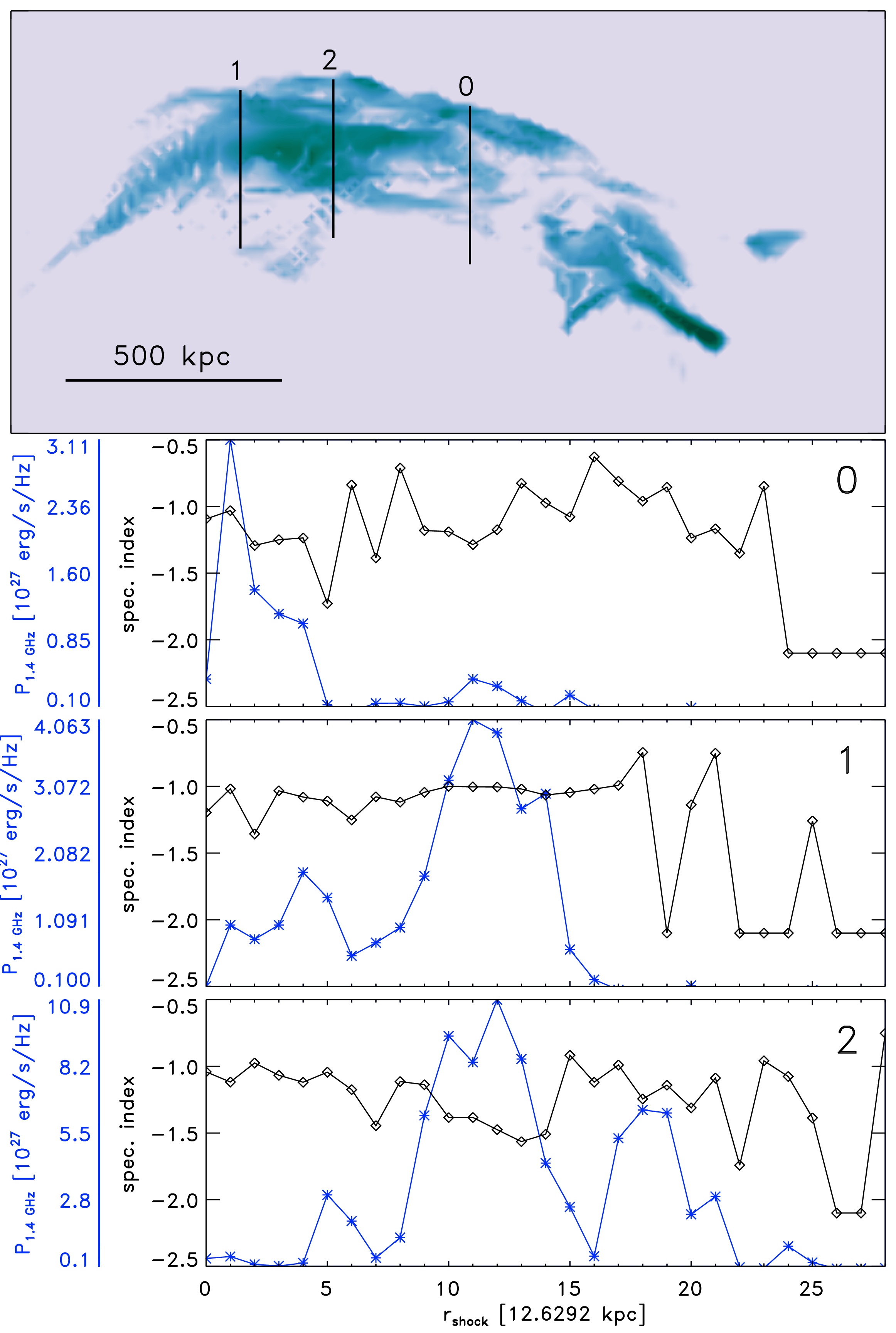}
    \caption{Profiles of the radio power (blue asterisks) and the spectral index (black diamonds) extracted from three different regions of the simulated radio relics. The top panel shows a map of the relic and the locations of the three regions where the profiles have been extracted.}
    \label{fig:simulation_prof}
\end{figure} 

\begin{figure}[tb]
    \centering
    \includegraphics[width=0.5\textwidth]{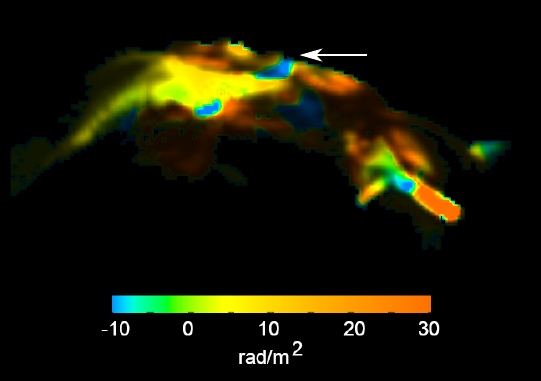}
    \caption{Total intensity of the simulated relic in Fig. \ref{fig:simulation}, colour-coded by an approximation to the observed RM, as described in the text. The white arrow marks the height from where Fig. \ref{fig:simulation_sideRM} was extracted.}
    \label{fig:simulation_sliceRM}
\end{figure}

\begin{figure}[tb]
    \centering
    \includegraphics[width=0.5\textwidth]{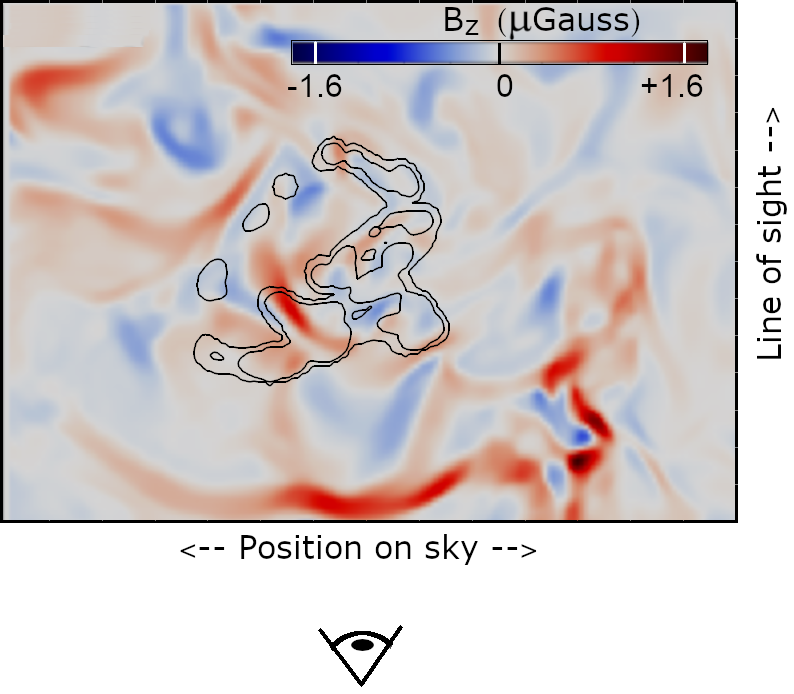}
    \caption{2D slice from the cube shown in projection in Fig. \ref{fig:simulation}, where the horizontal axes are the same, but here the vertical axis is along the line of sight.  The contours show the location of radio-emitting material, at various depths along the line of sight, and the coloured map shows the line-of-sight (z) component of the magnetic field.}
    \label{fig:simulation_sideRM}
\end{figure}

The simulation we use in this section is presented in \citet{Wittor2021} and is part of the \textit{SanPedro}-cluster catalogue, a set of simulations targeted to model radio relics. It was initialised with \textit{MUSIC} \citep{Hahn2011} and it has been carried out with the \textit{ENZO} code \citep{Bryan2014}. Here, we summarise the main specifications of the simulation. For more details, especially on the numerical schemes, we refer the reader to \citet{Wittor2021}. 

The root grid of the simulation covers a volume of the size $(140 \ \mathrm{Mpc}/h)^3$ and it is sampled with $256^3$ grid cells. Using nested grid and adaptive mesh refinement, the central $(4.04 \ \mathrm{Mpc}^3)$ around the cluster have been refined $2^5$ times for a uniform resolution of $12.6292 \ \mathrm{kpc}$. Shock waves are detected using a velocity-jump based shock finder \citep{Vazza2009}.

One simulated relic shows similar spectral properties and filamentary structures as the NW relic of Abell 3667 (region 1). The corresponding radio emission is computed using the method described in \citep{Wittor2019}. This model spatially resolves the downstream cooling behind the shock front assuming that the ICM is stationary and that the shocks runs across it. The aged spectrum at a distance $x_i$ from the shock front is computed based on the local gas properties and the time since the last shock passage. We only considered synchrotron and inverse Compton losses. A recent simulation by \cite{Wittor2021a} showed that the shocked gas undergoes only mild compression in the downstream. Hence, any modifications of the particle spectrum due to adiabatic processes are negligible. 

In Fig.~\ref{fig:simulation}, we show the total intensity, the spectral index, the polarisation fraction, and the polarised intensity of the simulated relic. The simulated relic has a morphology similar to the one in Abell 3667. The relic shows a few bright parts and filamentary structures that are also identified in the spectral index map as regions with flatter spectra. As in Abell 3667, we find spectral indices flatter than $-1$ in the post-shock region. That is because the shock front is seen at an oblique angle to the observer and flat indices are projected into the downstream region. 

Across the simulated relic, the polarisation fraction is fairly high and dominated by a few bright pixels along the line of sight. Hence, the polarisation fraction mirrors their intrinsic degree of polarisation. Furthermore, in the central region and the lower right part of the simulated relic, the polarisation fraction increases in the downstream. This is in contrast to its left part, where the polarisation fraction decreases in the downstream. These features are also found in the NW relic, for example a decreasing polarisation fraction in N1, and the SE relic, for example an increasing polarisation fraction in S1, of Abell 3667. 

Similar to what we have done in Sect.~\ref{sec:filaments} and Fig.~\ref{fig:profile_relics}, we extracted profiles of the radio power at 1.4 GHz and the spectral index in three different regions of the simulated relics (see Fig.~\ref{fig:simulation_prof}). The radio profiles measured in the three regions all have different shapes. In region 0, the radio profile shows the expected shape: the radio emission peaks at the shock front and decreases into the downstream. This is similar to the profile observed in regions N3 and S3 of the observed relics. In region 1 of the simulated relic, the radio power does not peak at the shock front but at an intermediate distance in the downstream. This is comparable to regions N2, S1 and S2 of the observed relic. Region 2 shows the most complex radio power profile. Here, the radio power peaks twice. This is similar to the profile observed in region N1 of the observed relic. Here, the radio emission first peaks, then decreases before it increases again. A closer look at the profiles measured in region 0 and region 1 reveals that these profiles also contain smaller peaks in the radio power. The different peaks, observed in the radio profiles, mirror the various filamentary structures observed in the maps of the radio power. 

As for the radio power, we also find regional variation in the spectral index profiles. Again, region 0 shows the expected spectral index profile: it is flat at the shock front and it steepens in the downstream, before it flattens again. Here, the second flattening correlates with an increase in the radio power. The spectral index in region 1 is mostly above $-1.0$ and it decreases behind the peak of the radio flux. Finally, the spectral index in region 2 gradually decreases in the downstream. However, it shows local variations that appear to correlate with the increase or decrease in the radio power.

We can also use the simulations to visualise the RMs that would be observed along each line of sight (see Fig. \ref{fig:simulation_sliceRM}). At this stage, we make a very rough characterisation of the RM by integrating the Faraday depth along each line of sight weighted by the local radio brightness --  $RM = \sum(RM \cdot I_{1.236 \ \mathrm{GHz}}) / \sum(I_{1.236 \ \mathrm{GHz}})$ -- where the sum is taken across all radio-emitting grid cells along the line of sight. A full calculation would integrate the Stokes parameters along each line of sight at many different wavelengths, and perform Faraday synthesis, \citep{Brentjens2005b}, which reveals the emission at each Faraday depth.   Our simplified calculation already shows that different filaments show different values of RM along with gradients along the filaments, similar to what is seen for A3667 in Fig. \ref{fig:rm2}. This is largely due to the fact that different filaments originate at different depths along the line of sight, and therefore experience different amounts of Faraday rotation.  

To see a little more clearly how each radio-emitting region traverses different magnetic field regions along the line of sight,  we show a 2D slice of horizontal position versus line-of-sight distance through the full cube at a fixed vertical position. The white arrow in Fig. \ref{fig:simulation_sliceRM} marks the vertical position of the slice. Figure \ref{fig:simulation_sideRM} shows the 2D location of the radio emission and magnetic field component parallel to the line of sight along this line. At any given position, there can be radio emission from more than one depth along the line of sight, and thus each filament will experience different amounts of Faraday rotation.  We note that in this figure there is little correspondence between the filaments and the magnetic field structure.  This is because the variations in radio emission are dominated by the local population of relativistic particles generated in the shocks.

The simulation only includes particle injection at shocks and subsequent cooling. Hence, the radio bright regions must correspond to a strong shock wave that lies along the line of sight. Indeed, numerical simulations have shown that shock fronts in the ICM are neither uniform nor planar structures \citep[e.g.][]{Dominguez-Fernandez2021, Dominguez-Fernandez2021a, Wittor2019}. In fact, shock fronts in the ICM are 3D structures and the shock strength varies across the shock surface. The bright filaments are then caused by the projection of the shock surface onto the plane of the sky.

\subsubsection{Hybrid MHD-Lagrangian simulations}
\label{sec:simul2}

\begin{figure}[ht!]
    \centering
    \includegraphics[width=.9\columnwidth]{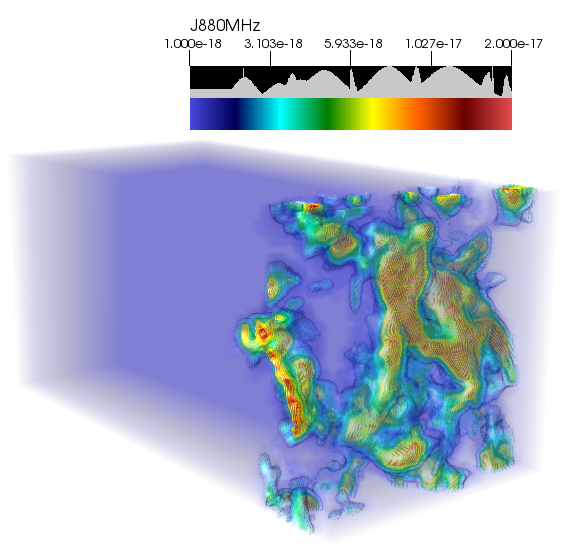}\\
    \includegraphics[width=.9\columnwidth]{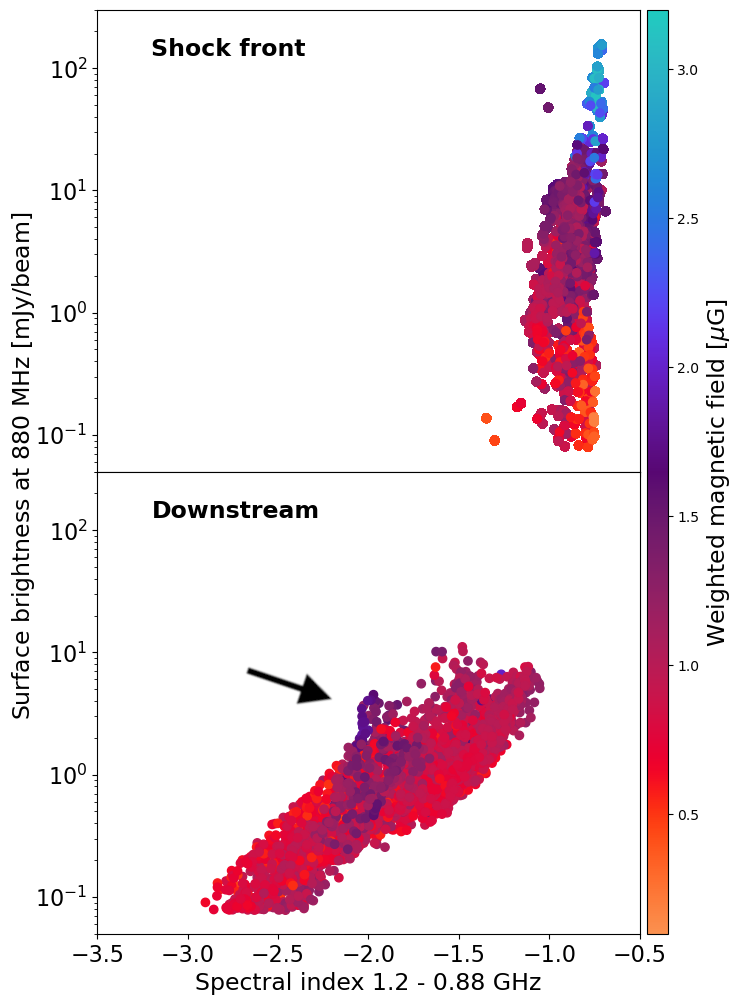}
    \caption{Results from Hybrid MHD-Lagrangian simulations. \textit{Top panel:} Visualisation of synchrotron emissivity at 880 MHz (in code units). \textit{Bottom panels:} Phase plot of the spectral index computed between 880 MHz and 1.2 GHz versus surface brightness at 880 MHz. We show the values at selected pixels in the surface brightness maps. In the upper (lower) panel we show the values corresponding to the shock front (downstream) region. We obtain magnetic field maps by projecting the 3D magnetic field using the 880 MHz emissivity as a weight, and finally, we colour-code each point in the plot with its corresponding weighted magnetic field value. The arrow shows a group of brighter particles associated with stronger magnetic fields. See Fig.~\ref{fig:simulation_comparison} for a comparison with real data.}
    \label{fig:simulation2}
\end{figure} 

\begin{figure}[ht!]
    \centering
    \includegraphics[width=.9\columnwidth]{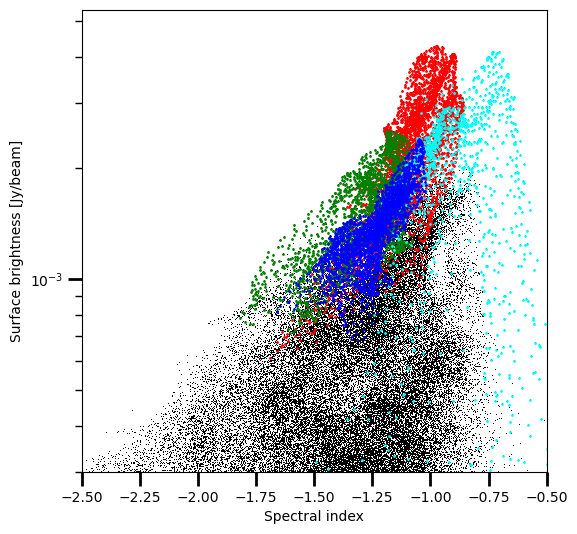} 
    \makebox[0pt][r]{
      \raisebox{14em}{
        \includegraphics[width=.3\linewidth]{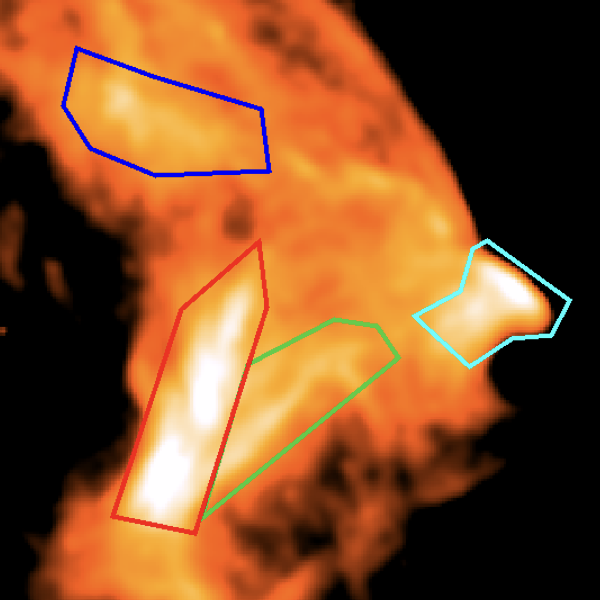}
      }\hspace*{12em}
    }%
    \caption{Similar to the bottom panel of Fig.~\ref{fig:simulation2}, but here we show the spectral index versus the surface brightness of all pixels in the NW relic. Regions with prominent filaments are coloured to underline the different spectral properties.}
    \label{fig:simulation_comparison}
\end{figure} 

In this section we explore the origin of the radio surface brightness and spectral index variations on smaller spatial scales to see if simulations can replicate the flattening of the spectra due to an increase in the magnetic field strength in the downstream region. We analysed a simulation from Dom\'inguez-Fern\'andez et al. (in prep.) where the same set-up as in \cite{Dominguez-Fernandez2021, Dominguez-Fernandez2021a} was used. In this simulation, a $\mathcal{M}=2.5$ initial shock is propagated through a turbulent ICM, and cosmic ray electrons are accelerated once at the shock front via DSA. Furthermore, in this simulation cosmic ray electrons lose energy via synchrotron and inverse Compton radiation as well as adiabatic expansion after being accelerated.

This simulation was produced with the PLUTO code \citep{Mignone2007, Vaidya2018} where a computational domain of 400 kpc $\times$ 200 kpc $\times$ 200 kpc ($256 \times 128 \times 128$ cells) was considered and a total of 3,145,728 Lagrangian particles were used. The right-hand half of the computational domain is filled with a turbulent medium with typical values of the peripheral regions of galaxy clusters: a mean density of $10^{-4}\ m_p\,\mathrm{cm}^{-3}$, an rms Mach number of $\mathcal{M}_s \sim$0.7 and a plasma beta of $\beta \sim$110. The initial magnetic field distribution ranges from 0.01 $\mu$G to $\sim 2\ \mu$G. We refer the reader to \cite{Dominguez-Fernandez2021, Dominguez-Fernandez2021a} for a detailed description of the MHD and Lagrangian numerical set-up and for the synchrotron emissivity computation.

In the top panel of Fig.~\ref{fig:simulation2} we show an example of the synchrotron emissivity at 880 MHz produced by an initial $\mathcal{M}=2.5$ shock. We chose this value to be comparable to the average X-ray derived Mach number for Abell 3667 \citep{Sarazin2016}. At this epoch, the downstream radio-emitting region has a size of $\sim 90$ kpc. The emissivity was computed at 1.2 GHz and 880 MHz. We produced spectral index maps and projected magnetic field maps where the magnetic field was weighted with the 880 MHz emission along the line of sight. In these maps, we isolated the regions corresponding to the shock front and the downstream, considering `downstream' the emission $>20$ kpc from the shock front. In the bottom panels of Fig.~\ref{fig:simulation2} we show the phase plots of the spectral index (considering 1.2 GHz and 880 MHz) and surface brightness for both of these regions. The shock front is characterised by electrons freshly injected from the thermal pool by shocks with a distribution of Mach numbers that translates into variations in the injection spectral index and surface brightness of the accelerated particles. Furthermore, the higher the compression, the larger the perpendicular magnetic field component with respect to the shock normal and this can lead to a higher total magnetic field strength and consequently a higher radio surface brightness.

 In the lower panel of Fig.~\ref{fig:simulation2} we show the corresponding surface brightness variations in the downstream region. We note that some steep spectrum regions trace comparatively high values of the magnetic field. For example, some regions with a spectral index of $\sim -2$ have similar magnetic field values as those close to the shock front with a spectral index of $\sim -1$ (mid panel of Fig.~\ref{fig:simulation2}). This can happen when an old cosmic ray electron population enters a region with a higher magnetic field. In this scenario, the critical radio frequency $\nu_c \propto \gamma^2 B$ for electrons with a fixed $\gamma$ is boosted to a higher frequency. Hence, the spectral index measured between two constant frequencies is set by less energetic electrons. As the low-energy electrons tend to have a flatter spectral index than the high-energy electrons, the observed spectral index will become flatter. An increase in observed brightness would go along with the flattening of the spectral index. We can conclude that in simulations with a single DSA injection we see bright regions with flatter-than-expected spectra in the downstream region, such as the (relatively) steep filament discussed in Sect.~\ref{sec:filaments} (see Fig.~\ref{fig:simulation_comparison}). However, we note that even just 20 kpc away from the shock front it is unlikely to find spectral indices close to $\sim -1$, which is the spectral index found in many of the filaments detected in the NW relic. This instead could be explained by extra re-acceleration sites, for instance due to repeated DSA events either at work in the downstream region or seen in projection. More detailed works dedicated to the study of radio filaments will be necessary to further assess their origin.

%%%%%%%%%%%%%%%%%%%%%%%%%%%%%%%%%%%%%%%%%%%%%%%%%%%%%%%%%%%%%%%%%%%
\section{Conclusions}
\label{sec:conclusions}

In this paper we have analysed new MeerKAT data of the galaxy cluster Abell 3667. The data have been used to obtain radio images, as well as spectral and polarisation maps. We also carried out Faraday rotation analysis and spectral tomography. We summarise here our main results:

\begin{enumerate}
 \item We produced the most detailed images of the famous double radio relic present in Abell 3667. The northern relic is much thicker than what is expected by assuming simple acceleration at the shock front and synchrotron plus inverse Compton ageing. Both relics present filamentary structures all across their bodies. Filaments have generally a flatter spectra compared to the surrounding region but with a variety of values. While the filament delimiting the edge of the NW relic is highly polarised, with a fractional polarisation of $\sim 70\%$, the other filaments appear depolarised compared to the surroundings. The nature of the filaments in the relic is still debated and might be due to substructures in the shock front with complex shapes, such that they effectively trace regions of particle acceleration, or to a local enhancement of magnetic fields combined with a curved energy spectrum of the cosmic ray electron population. Our observations and simulations show a good agreement with the first scenario. We have also shown that the second scenario would predict unrealistically high magnetic fields to produce the filaments with the flattest spectrum located downstream of the shock front.
 \item In addition to large-scale gradients in RM across the relic, we find that different filamentary structures show different Faraday depths. This suggests that the filaments are embedded within a thermal, magnetised medium, with magnetic fields of $<1 \mu$G varying on scales similar to those of the filaments, $\gtrsim 100$~kpc. However, X-ray observations put a limit of $B>3~\mu$G on the maximum inverse Compton emission from the relic region \citep{Finoguenov2010}. Radio relics therefore appear to be made by filamentary regions of high magnetic field strengths embedded in a less magnetised medium.
 \item We identified the presence of an elongated radio halo, previously noticed by \cite{Carretti2013}. While some elongated radio halos are bounded by the merging shock fronts, in the case of Abell 3667 the halo is bounded from one side by the cold front at the edge of the surviving core of the main sub-cluster and on the other side by one of the two merging shocks. The region occupied by the halo is coincident with the wake of the bullet-like sub-cluster that merged from the SE. If the halo traces the turbulence in the wake of the merging sub-cluster, its narrow size implies a large Reynolds number ($>800$) for the ICM. Both the luminosity and the emissivity of the radio halo are a factor of two to three lower than what is expected from scaling relations \citep{Cuciti2021a}.
 \item X-ray data show what is possibly the remnant of the cool core of the merging bullet (the mushroom) right behind the southern arm of the NW radio relic. Radio images show the presence of polarised radio emission surrounding the mushroom region, which we interpreted as an observational consequence of magnetic draping. In this scenario, the cool core of the merging bullet, which moves at super-Alfvénic speed in the magnetised ICM, sweeps up and compresses a significant magnetic layer that is bent over the projectile. While the X-ray and radio signature of magnetic draping is usually hard to detect in clusters, in this case the ICM crossed by the projectile is polluted by cosmic rays accelerated by the shock front, therefore having all the ingredients to produce synchrotron radiation around the mushroom cap. The thickness of the magnetic layer provides an upper limit on the local magnetic field, $B \lesssim 5~\mu$G.
 \item We compared our observations with both MHD cosmological and Lagrangian simulations. The simulations were able to reproduce several of the qualitative and quantitative properties of radio relics, such as the filamentary structure, including spectral and polarisation properties. According to the simulations, filaments can be interpreted as sites of particle acceleration due to the shock front seen at an oblique angle by the observer. The filaments are located at different depths into the relic structure, resulting in a different amount of Faraday rotation, which is visible in both the simulated data and the observations.
\end{enumerate}

These observations give a glimpse of the power of next-generation radio telescopes, where with a few hours of data we could sample very large sources at high resolution (a well-known challenge for standard radio interferometers), also deriving polarisation and spectral information. The combination of these capabilities is perfectly suited for the study of nearby radio relics, whose large apparent size can be exploited to study in detail the processes involved in the acceleration of cosmic rays.

\begin{acknowledgements}

We thank Matt Owers for providing the exact locations of the spectroscopically identified galaxies.

MGCLS data products were provided by the South African Radio Astronomy Observatory and the MGCLS team and were derived from observations with the MeerKAT radio telescope. The MeerKAT telescope is operated by the South African Radio Astronomy Observatory, which is a facility of the National Research Foundation, an agency of the Department of Science and Innovation.

FdG and MB acknowledge support from the Deutsche Forschungsgemeinschaft under Germany's Excellence Strategy - EXC 2121 “Quantum Universe” - 390833306. Basic research in radio astronomy at the U.S. Naval Research Laboratory is supported by 6.1 Base funding. KK is supported by the New Scientific Frontiers grant of the South African Radio Astronomy Observatory.
DW is funded by the Deutsche Forschungsgemeinschaft (DFG, German Research Foundation) - 441694982. P.D.F was supported by the National Research Foundation (NRF) of Korea through grants 2016R1A5A1013277 and 2020R1A2C2102800.
% DS9
This research has made use of SAOImage DS9, developed by Smithsonian Astrophysical Observatory.
% ADS
This research has made use of NASA's Astrophysics Data System.
% NRAO
The National Radio Astronomy Observatory (NRAO) is a facility of the National Science Foundation operated by Associated Universities Inc.
% cosm. sim.
The authors gratefully acknowledge the Gauss Centre for Super-computing e.V. (www.gauss-centre.eu) for supporting this project by providing computing time through the John von Neumann Institute for Computing (NIC) on the GCS Supercomputer JUWELS at Jülich Supercomputing Centre (JSC), under project no. hhh44.

\end{acknowledgements}

%---------------------------------------------------------------------
% bibliography

\bibliographystyle{aa}
\bibliography{references_HA,library}

%---------------------------------------------------------------------

\onecolumn
\begin{appendix}

\section{Radio galaxies in Abell 3667}
\label{app:rg}

\begin{figure*}[ht!]
\centering
\includegraphics[width=.32\textwidth]{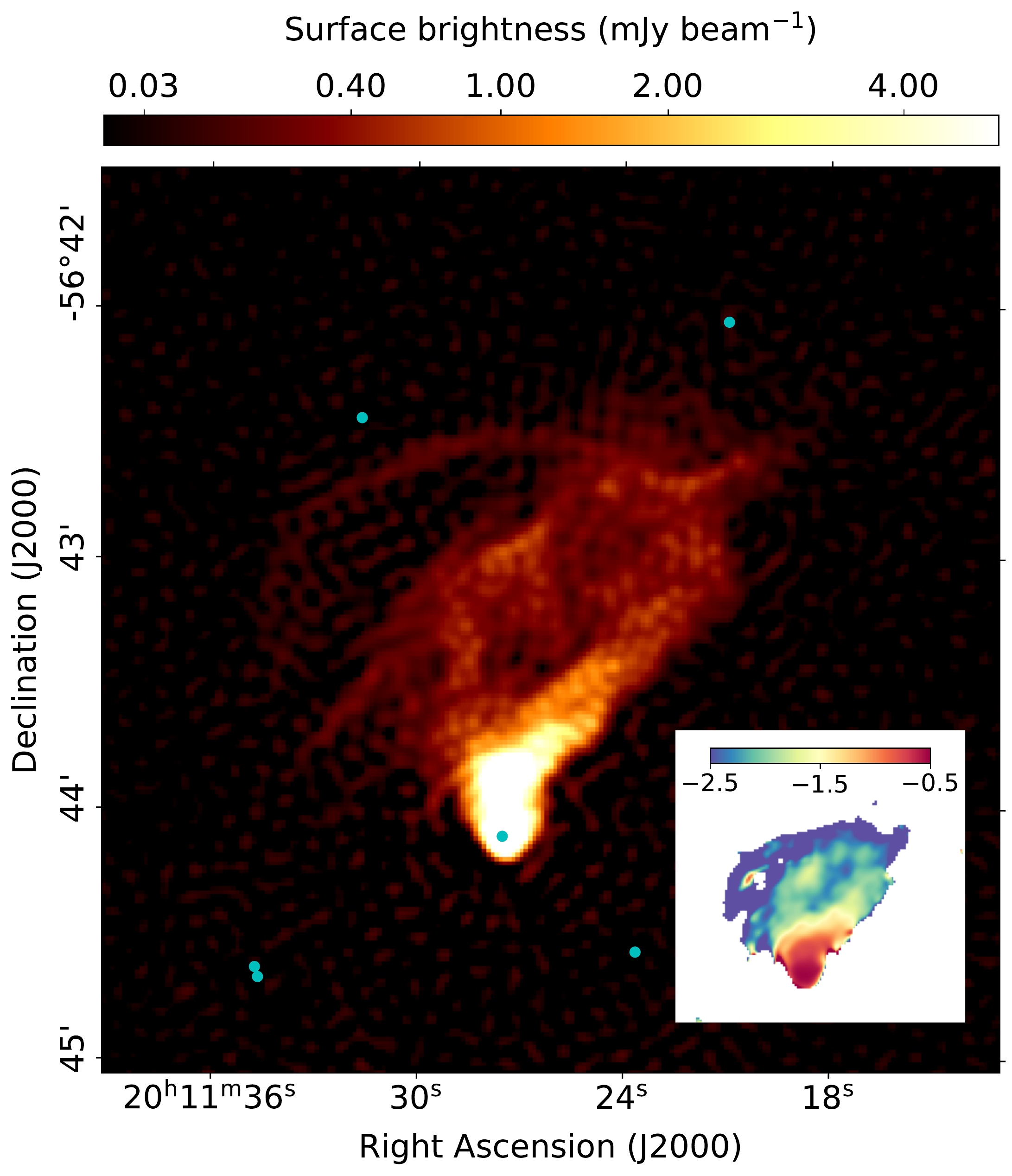}
\includegraphics[width=.34\textwidth]{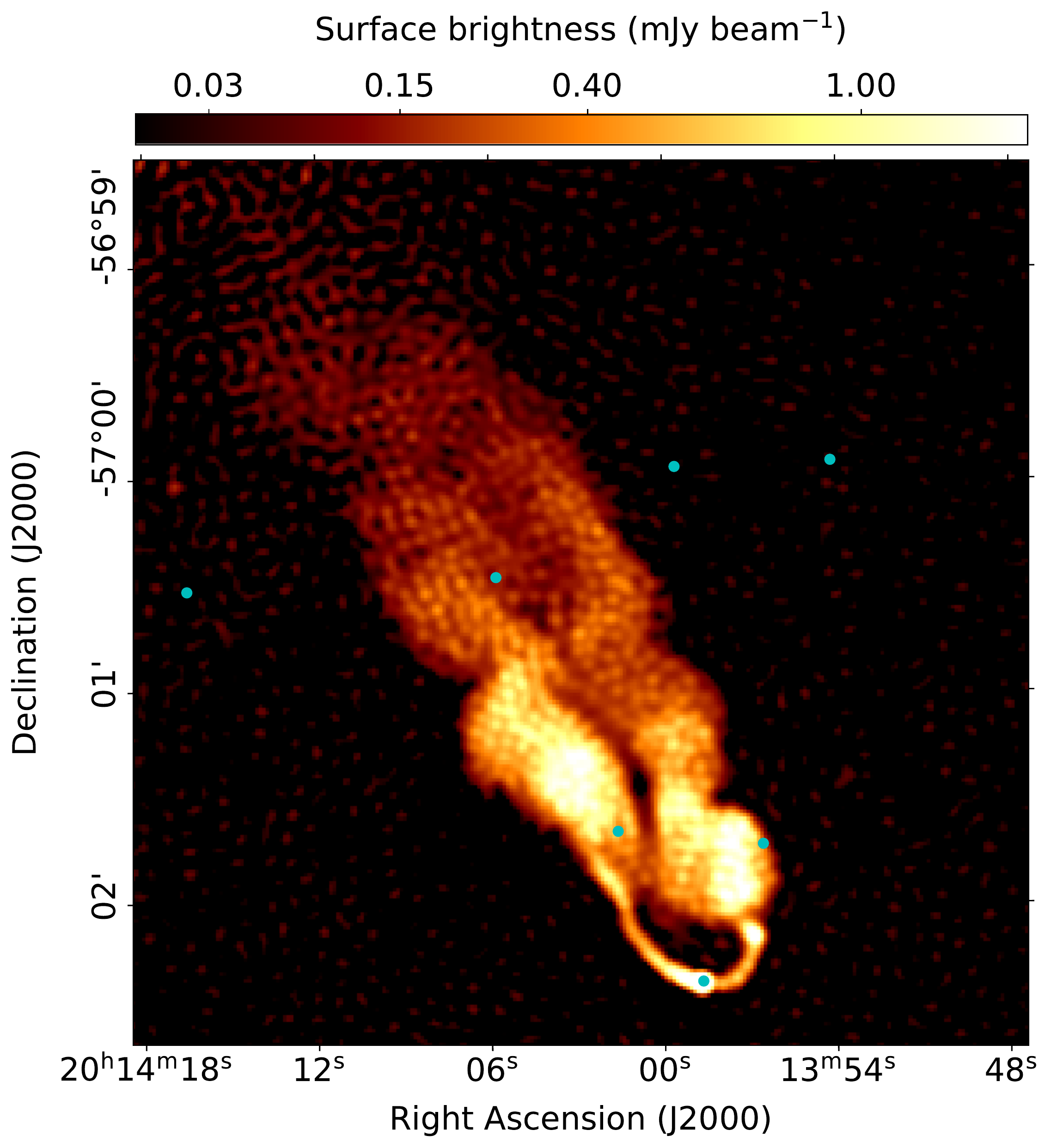}
\includegraphics[width=.33\textwidth]{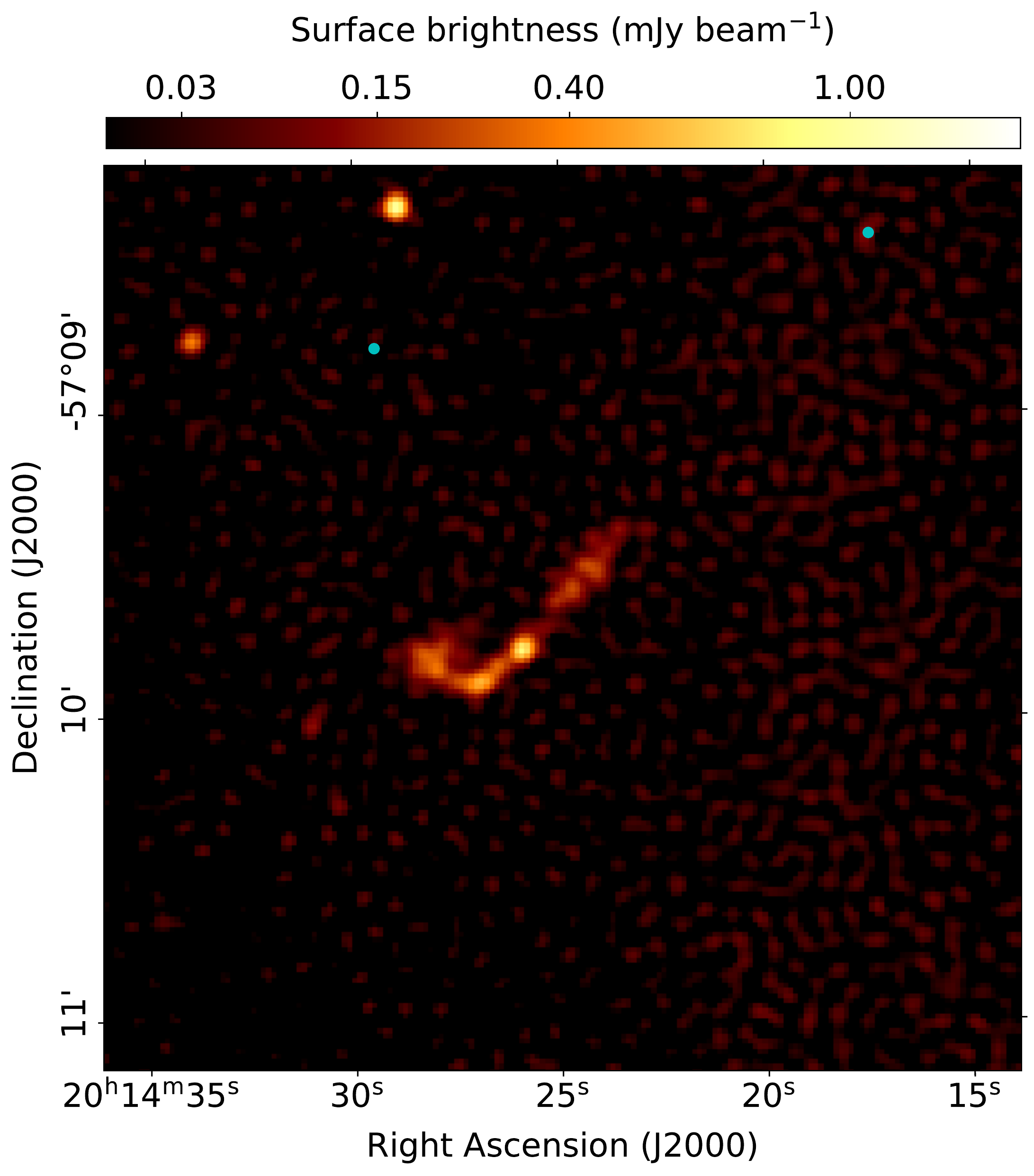}
\\
\caption{Radio emission from the two largest radio galaxies present in Abell 3667 and the double lobed radio galaxy in front of the SE relic. RG1 and RG2 have an optically confirmed counterpart. The map resolution is \beam{3.3}{3.3} and the local rms noise is 20~\mujybeam{} in all cases. Light blue dots show the positions of known cluster members. In the insert of the left panel, the colours correspond to the spectral index of the source (for the spectral index of the other radio galaxies, see Fig.~\ref{fig:spidx}).}\label{fig:wat}
\end{figure*}

In Fig.~\ref{fig:wat} we present high-resolution images of three interesting radio galaxies present in the cluster. The images in this case were obtained using a uniform weighting that provided a resolution of \beam{3}{3}. In the first two cases, the radio emission is associated with a spectroscopically identified galaxy cluster member from the sample of \cite{Owers2009}. The first radio galaxy (RG1) is located within the linear radio halo and the shape of the radio lobes reflects their violent interaction with the ICM that bent the radio tail in the direction of the merger axis. The second source (RG2) is a head-tail radio galaxy with the tail perpendicular to the merging axis that extends for a projected length of $\sim 800$~kpc. The radio flux densities and luminosities of RG1 and RG2 are in Table~\ref{tab:fluxlum}. The third radio galaxy (RG3) is a double lobed radio galaxy located at the edge of the SE radio relic, close to the expected location of the shock front. RG3 shows radio properties of FR-II galaxies and it is not associated with any of the spectroscopically identified cluster members of \cite{Owers2009}. The spectral index of the hotspots appear inverted, with $\alpha = 0.45$ in the eastern lobe and $\alpha = 0.17$ in the western lobe. The source lack an obvious counterpart in the optical, infrared, X-ray and gamma ray bands. With no $z$ information, RG3 may or may not be a member of the cluster; however, the radio emission of the radio galaxy bridged into the relic and it is spatially coincident with a break in the filamentary structure of the relic. This could be an indication of an interaction between RG3 and the eastern relic. Such a configuration is similar to the one present in the radio relic of Coma \citep{Bonafede2021}, where a radio galaxy present in front of the relic is aligned with a break in the radio relic filaments. In the same table we report the flux density and luminosity of the radio emission associated with the brightest cluster galaxy (BCG; RG4) of sub-cluster B. The positions of the radio galaxies are shown in Fig.~\ref{fig:wide}.

\FloatBarrier

\section{Spectral index error maps}
\label{app:spidx_err}

Here we show the spectral index error maps relative to the spectral index maps presented in Sect.~\ref{sec:spidx}.

\begin{figure*}
\centering
\includegraphics[width=\textwidth]{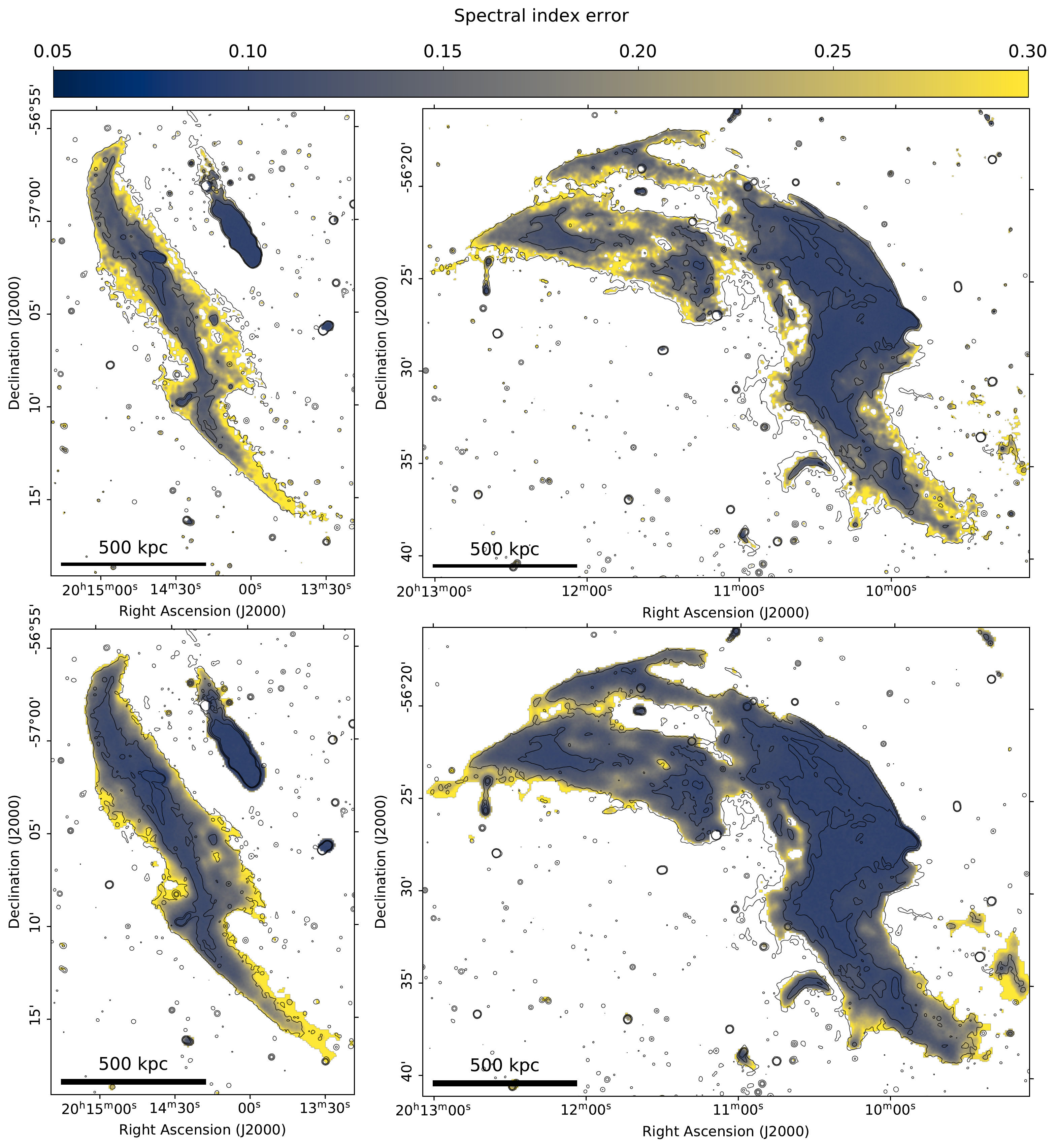}
\caption{Spectral index error maps for the spectral index maps shown in Fig.~\ref{fig:spidx}. The top images have a resolution of \beam{12}{12}, the bottom images  \beam{26}{26}. The values shown in the images represent the $1\sigma$ uncertainties estimated as described in Sect.~\ref{sec:spidx}. Contours are the same as in Fig.~\ref{fig:spidx}.} \label{fig:spidx-err}
\end{figure*}

\FloatBarrier

\section{Alternative visualisation of polarisation maps}
\label{app:pol}

Here we add an alternative visualisation of Fig.~\ref{fig:pola}. The image has been prepared starting from the total intensity maps (Figs.~\ref{fig:relicN} and~\ref{fig:relicS}) and applying the line integral convolution method using for the local direction the one of the B vectors corrected for Faraday rotation to their intrinsic orientation at zero wavelength.

\begin{figure*}
\centering
\includegraphics[width=\textwidth]{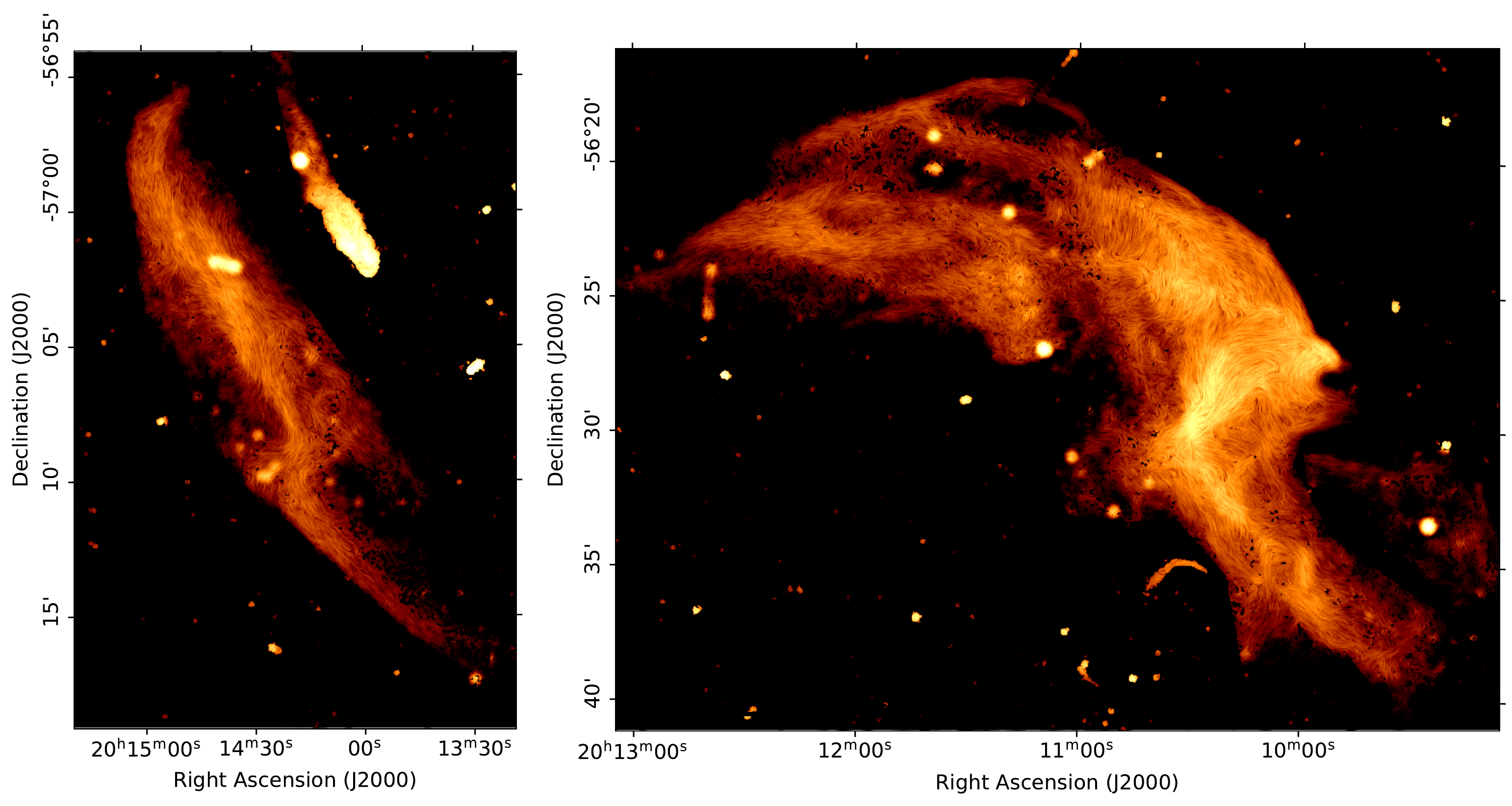}
\caption{B-mode vectors de-rotated to their intrinsic orientation at zero wavelength using the observed Faraday rotation and represented using line integral convolution.} \label{fig:pola2}
\end{figure*}

\end{appendix}

%---------------------------------------------------------------------

\end{document}